\documentclass[reprint,aps,pra,superscriptaddress,amsmath,amssymb,floatfix]{revtex4-2}
\usepackage[margin=1in]{geometry}
\usepackage{graphicx}
\graphicspath{{figures/}}
\usepackage{booktabs}
\usepackage{tikz}
\usetikzlibrary{arrows.meta,positioning}
\usepackage{pgfplots}
\pgfplotsset{compat=1.18}
\usepackage[colorlinks=true,linkcolor=blue!50!black,
 citecolor=blue!50!black,urlcolor=blue!50!black]{hyperref}
\newcommand{\ket}[1]{\lvert #1\rangle}
\newcommand{\bra}[1]{\langle #1\rvert}

\begin{document}
\raggedbottom
\title{Parallel Teleportation via AME(5,2) States on IBM Quantum Hardware}
\author{Zheng Liang Lim}
\email{e0001040@u.nus.edu} % This creates the asterisk (*)
\affiliation{Centre for Quantum Technologies, National University of Singapore, 3 Science Drive 2, 117543, Singapore}
\affiliation{Department of Physics, National University of Singapore, 2 Science Drive 3, Singapore 117551, Singapore}

\author{Hoi-Kwong Lo}
\thanks{hoikwonglo@nus.edu.sg} % This creates the dagger symbol (†)
\affiliation{Centre for Quantum Technologies, National University of Singapore, 3 Science Drive 2, 117543, Singapore}
\affiliation{Department of Physics, National University of Singapore, 2 Science Drive 3, Singapore 117551, Singapore}
\affiliation{Quantum Bridge Technologies Inc., 108 College St., Toronto, ON, Canada}
\affiliation{Department of Electrical and Computer Engineering,
University of Toronto, 10 King’s College Road, Toronto, ON, Canada}

\begin{abstract}
The five-qubit  Absolutely Maximally Entangled AME(5,2) state, whose every pair of qubits is maximally mixed so that all its information lives in non-local correlations, could in principle serve as a single shared quantum networking resource for teleporting two unknown qubits in parallel (each via its own Bell measurement and Pauli correction). Nonetheless, its hardware implementations remain scarce despite its theoretical importance. Here, we demonstrate parallel
teleportation based on a five-qubit absolutely maximally entangled
state, AME$(5,2)$, on the superconducting processor
\texttt{ibm\_pittsburgh}. We construct three-qubit Clifford decoders
that extract two Bell pairs using three logical two-qubit gates for
each of 60 inequivalent sender-receiver assignments. Using parallelization and classical
Pauli frame post-processing, we retain reference-receiver entanglement
on both channels after three teleportation rounds. With XY4 dynamical
decoupling and the lowest-cost decoder, the final entanglement fidelities
are $0.704\pm0.028$ and $0.680\pm0.022$, where uncertainties are
pointwise $99\%$ confidence-interval half-widths over eight trials.
Their lower bounds exceed the separability threshold of $1/2$.
The decoder with lower circuit depth generally retains higher mean fidelity, and
XY4 generally improves on XX. These results establish a hardware
demonstration of two-channel entanglement transmission through
repeated AME-assisted teleportation and provide compact decoding
circuits for reconfigurable multipartite communication.
\end{abstract}
\maketitle

\section{Introduction}
Entanglement is a central resource for quantum communication and
distributed quantum information processing~\cite{Horodecki:2009zz}.
Quantum teleportation transfers an unknown state using shared
entanglement, a Bell measurement and classical
communication~\cite{PhysRevLett.70.1895}. Its experimental development
has extended from photonic demonstrations~\cite{Bouwmeester:1997slj}
to long distance optical links~\cite{Yin:2012vzv,Ma:2012bgk,Ren:2017ivf} and
superconducting implementations~\cite{Baumer:2023vrf,Kang:2024jqx}.
Entanglement swapping has also been demonstrated over a
100 km optical-fibre link~\cite{Sun:2017ylw}. More recently, Lou
et al. demonstrated hundred-channel reconfigurable continuous-variable
teleportation using spatial optical modes~\cite{rfz9-3prw}.
For multichannel communication, an important question is whether
several usable channels can be obtained from a single shared
multipartite state, rather than allocated as independent Bell pairs.
Such resource choices are relevant to quantum network and repeater
architectures~\cite{Azuma:2022nuy}.

Absolutely maximally entangled (AME) states offer this possibility.
An AME state is pure and maximally entangled across every
bipartition~\cite{Helwig:2012nha,Helwig:2013ckb}. For instance, in AME$(5,2)$, every 
reduced state with one and two qubits is maximally mixed. Across any partition of
two qubits against the remaining three, the state therefore contains
two ebits. Helwig et al. realised that a suitable joint operation on the three-qubit side can
convert this entanglement into two Bell pairs, enabling two qubits
to be teleported in parallel~\cite{Helwig:2012nha}. The choice of
senders can be made after the resource is prepared, whereas the
partners of separately distributed Bell pairs are fixed in advance.
This connection between multipartite entanglement and flexible
communication also relates AME states to quantum secret
sharing~\cite{Cleve:1999qg,Helwig:2012nha,Lim:2026cuk,Helwig:2013ckb}.
Moreover, AME states are fundamentally important as holographic quantum error-correction codes, which are studied as toy models in black hole information paradox~\cite{Pastawski:2015qua} and they have recently been studied in the black hole interior constructions~\cite{w7nc-l9s5}.
AME resources cannot be enlarged arbitrarily at fixed local
dimension; in particular, seven-qubit AME states do not
exist~\cite{PhysRevLett.118.200502}.

Despite this theoretical promise, experimental tests of AME-assisted
parallel teleportation remain limited, particularly over successive
rounds. We note that there has been an experimental demonstration of a related quantum secret sharing scheme with IBM hardware~\cite{Graves:2024wlh}. Nonetheless,~\cite{Graves:2024wlh} was limited to a single round of application and had a rather low fidelity. In summary, a high-fidelity and multiple-round experimental demonstration of parallel teleportation is missing. Here, we fill this important gap. Such tests are motivated by quantum repeaters, where preserving
entanglement through noisy processing at intermediate nodes is
essential for long-distance quantum communication~\cite{Azuma:2022nuy}.
Realising Helwig et al.'s proposal~\cite{Helwig:2012nha} on hardware
therefore offers a way to examine whether a shared multipartite
resource can supply multiple teleportation channels whose entanglement
survives repeated use.

In this paper, we present an experimental demonstration of  AME$(5,2)$-assisted parallel
teleportation on the 156-qubit superconducting processor
\texttt{ibm\_pittsburgh}. We construct and verify compact three-qubit
Clifford decoders for all 60 inequivalent sender-receiver assignments,
with three logical two-qubit gates per decoder. Using classical
Pauli frame post-processing, we teleport one half of each of two
Bell pairs through three successive rounds while retaining the other
halves as references. Employing the reduced-depth decoder alongside XY4 dynamical decoupling yields final M3-mitigated entanglement fidelities of approximately 0.70 and 0.68, with pointwise 99\% confidence intervals remaining above 1/2, confirming surviving entanglement across both channels~\cite{Horodecki:1998zh}. We compare two decoder constructions and dynamical decoupling (DD) settings to evaluate how circuit depth and idle-time protection influence fidelity retention. The Supplementary Material details the decoder derivations, implementation diagnostics, full setting-by-setting results, and a separate real-time sequential feed-forward method.

\section{Parallel teleportation and compact AME decoders}
\label{sec:protocol}

\begin{figure}[tbp]
\centering
\includegraphics[width=\linewidth,keepaspectratio]{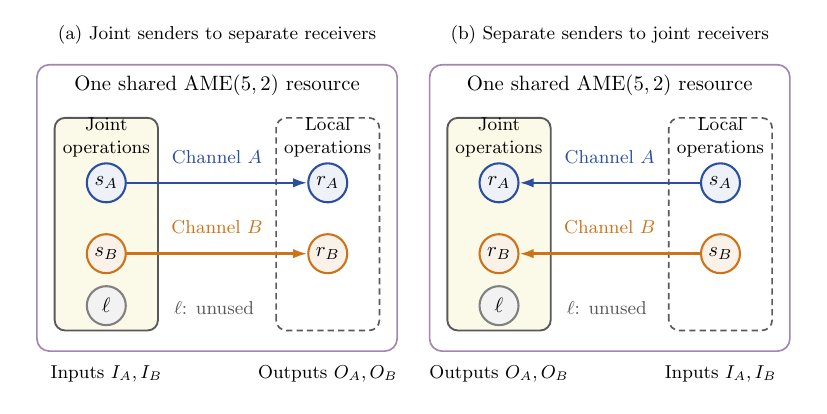}
\caption{Two directions of AME$(5,2)$-assisted parallel teleportation
($n=5$, $d=2$, $m=2$), following
Helwig et al.~\cite{Helwig:2012nha}.
(a) Joint senders to separate receivers; (b) separate senders to
joint receivers. The shaded three-qubit group performs joint
decoding; the dashed group acts locally. Resource qubits $s_A,s_B$
are Bell-measured with inputs $I_A,I_B$, yielding outputs $O_A,O_B$
on receivers $r_A,r_B$, respectively; $\ell$ is disentangled and
unused. Colours distinguish channels; arrows indicate input-state
transfer. Bell measurements, classical communication and Pauli
corrections or frame updates are implicit.}
\label{fig:parallel-teleportation-scenarios}
\end{figure}

While the concept is general, for concreteness, in our experimental demonstration, we consider the $n=5$, $d=2$, $m=2$ case of the parallel-teleportation
protocol proposed by Helwig et al.~\cite{Helwig:2012nha}, where $n$
denotes the number of parties sharing the AME resource, $d$ the
dimension of each party's local quantum system, and $m$ the size
of the smaller group in the chosen bipartition, which determines
the number of systems teleported in parallel. Here, preparing the shared resource state and decoding it jointly set up the two teleportation channels, after which two separate Bell measurements, each followed by a Pauli correction based on its outcome, complete the parallel teleportation (see Figure~\ref{fig:parallel-teleportation-scenarios}).

Let $S=\{s_A,s_B\}$ be the two source qubits of the AME resource and
$T=\{r_A,r_B,\ell\}$ the remaining three. The AME property gives
\begin{equation}
 \ket{\mathrm{AME}(5,2)}
 =\frac12\sum_{a,b=0}^{1}\ket{a}_{s_A}\ket{b}_{s_B}
   \ket{\phi_{ab}}_T,
 \label{eq:resource}
\end{equation}
where the four states $\ket{\phi_{ab}}$ are orthonormal.
A decoder $V_T$ satisfying
$V_T\ket{\phi_{ab}}=\ket{a}_{r_A}\ket{b}_{r_B}\ket{0}_\ell$
therefore produces
\begin{equation}
 (\mathbb I_S\otimes V_T)\ket{\mathrm{AME}(5,2)}
 =\ket{\Phi^+}_{s_A r_A}\ket{\Phi^+}_{s_B r_B}
   \ket{0}_{\ell},
 \label{eq:decoder-verification}
\end{equation}
with $\ket{\Phi^+}=(\ket{00}+\ket{11})/\sqrt2$.
Only the three qubits in $T$ are decoded; the source qubits bypass
$V_T$. The decoder converts the shared five-qubit resource into two
Bell pairs, each supporting one teleportation channel, while $\ell$
is disentangled and no longer participates.

For each channel $c\in{A,B}$, a Bell measurement on the input
$I_c$ and resource qubit $s_c$ yields two classical bits that
specify the Pauli byproduct on receiver $r_c$~\cite{PhysRevLett.70.1895}. Communicating these
outcomes allows the byproduct to be corrected physically or tracked
in a classical Pauli frame~\cite{Baumer:2023vrf}. The two Bell measurements act on
disjoint qubit pairs and can be performed concurrently, transferring
both inputs through the same decoded AME resource. Ideally, after
accounting for the Pauli byproducts, the joint input state is
recovered on $(r_A,r_B)$, including any entanglement between the
inputs or with external reference qubits. The inputs are transferred,
not copied. This preservation of reference entanglement is the
property tested by the cascaded experiment below.

\begin{figure}[tbp]
\centering
\includegraphics[width=\linewidth]{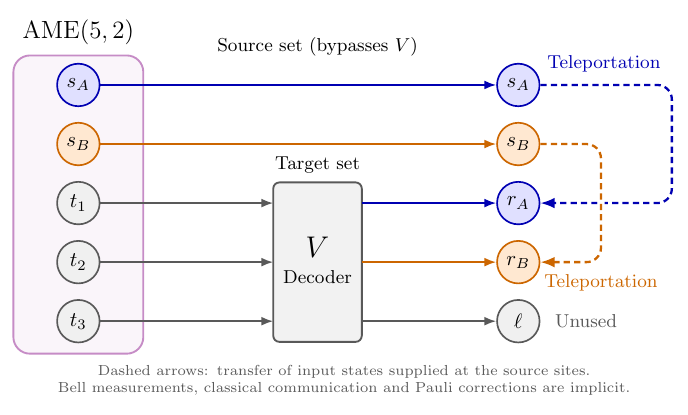}
\caption{Illustration of the role assignment
$(s_A\rightarrow r_A,s_B\rightarrow r_B)$.
The source set $\{s_A,s_B\}$ bypasses $V$; the decoder acts jointly
on the remaining three AME qubits to produce $r_A$, $r_B$ and the
disentangled qubit $\ell$. Dashed arrows indicate teleportation
of inputs supplied at the source sites to the assigned receivers.
Bell measurements, classical communication and Pauli corrections
are implicit. Assigning the five qubits to the ordered roles
$(s_A,r_A,s_B,r_B,\ell)$ gives $5!=120$ labelled assignments;
identifying assignments that differ only by exchanging the two
channel labels leaves $120/2=60$.}
\label{fig:assignment-illustration}
\end{figure}

Figure~\ref{fig:assignment-illustration} makes the role assignment
explicit: two resource qubits bypass the decoder, while the other
three become the assigned receivers and an unused qubit. The dashed
arrows indicate the logical transfer of input states supplied at the
source sites to their corresponding receivers, not physical motion
or copying of the resource qubits.

The resource is prepared with five CNOTs and three Hadamards using
the circuit of Ref.~\cite{Cervera-Lierta:2019juk}. We remark on passing that, up to local Clifford transformations, this AME$(5,2)$ resource
is mathematically equivalent to the codeword of the perfect five-qubit $[[5,1,3]]$ quantum
error-correcting code~\cite{Helwig:2012nha,Laflamme:1996iw}. Assigning the
five resource qubits to the ordered roles
$(s_A,r_A,s_B,r_B,\ell)$ gives $5!=120$ labelled assignments.
Exchanging the complete channels,
$(s_A,r_A)\leftrightarrow(s_B,r_B)$, changes only their labels,
leaving $120/2=60$ inequivalent role assignments.
For each assignment, we enumerate 64 Clifford completions of the
decoder's action on the four-dimensional subspace in
Eq.~\eqref{eq:resource}, synthesise them using the Bravyi-Maslov
Clifford decomposition~\cite{Bravyi:2021ijy}, and verify
Eq.~\eqref{eq:decoder-verification} numerically. Candidates are
ranked lexicographically by logical two-qubit-gate count,
single-qubit gate count and depth.

\begin{figure*}[tp]
\centering
\begin{minipage}[c]{0.38\textwidth}
 \centering
 \includegraphics[width=\linewidth]{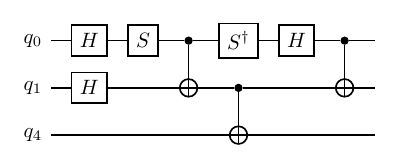}
 \par\smallskip\textbf{(a)} Rank 1, $V_{014}$.
\end{minipage}%
\hfill%
\begin{minipage}[c]{0.59\textwidth}
 \centering
 \includegraphics[width=\linewidth]{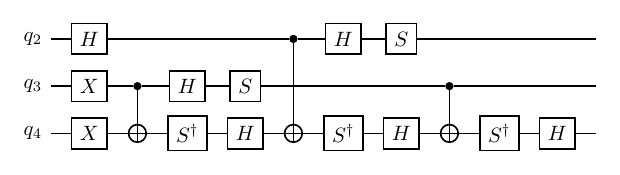}
 \par\smallskip\textbf{(b)} Rank 58, $V_{234}$.
\end{minipage}
\caption{The logical decoder circuits, read from
left to right. Only the three active AME qubits are displayed;
the two source qubits bypass the decoder. Filled circles and
$\oplus$ denote CNOT controls and targets. Both decoders contain
three CNOTs. Rank~1 contains three Hadamards, one $S$ and one
$S^\dagger$ gate; rank~58 contains six Hadamards, two $S$,
three $S^\dagger$ and two $X$ gates. Their depths are six and
ten, respectively, counting sequential layers of all gates
while allowing independent operations on disjoint qubits
to share a layer.}
\label{fig:decoder-circuits}
\end{figure*}

Every assignment admits a decoder with three logical two-qubit
gates in this search. We benchmark
rank 1, $V_{014}$, and rank 58, $V_{234}$
(Figure~\ref{fig:decoder-circuits}). They have the same entangling-gate
count but different single-qubit counts, depths and role assignments.
The rank labels reflect a logical-cost ordering, in which lower rank means lower circuit depth and fewer single-qubit gates. Figure~\ref{fig:decoder-circuits} shows two such decoders, which we implement experimentally. Supplementary Material, Sec.~I, describes AME resource
preparation and derives the ideal teleportation identities.
Sec.~II presents the decoder construction, completion procedure
and logical cost ranking. Sec.~III verifies the preparation and
both decoder circuits by explicit state vector algebra, confirming
that preparation produces the AME resource and each decoder extracts
its assigned Bell pairs with the unused qubit disentangled.

\section{Experimental design}
\label{sec:experiment}
We study three rounds of applications of parallel teleportation. Therefore, we construct 4 checkpoints corresponding the states at various steps of the parallel teleportation. The first checkpoint is $R0$, where no parallel teleportation has been done yet. Here, we simply prepare two copies of the Bell pair $\ket{\Phi^+}$, one for each channel $c$, each between a reference qubit $R_c$ and an input qubit $I_c$, using the same $H$-CNOT construction. The reference qubits are retained throughout the cascade. Checkpoints $R1$, $R2$ and $R3$ correspond to one, two and three rounds of parallel teleportation, respectively. Each round uses a fresh five-qubit AME register and the chosen decoder, and its logical output $O_c^{(R)}$ becomes the input to the next round, $I_c^{(R+1)}\equiv O_c^{(R)}$.

\begin{figure*}[tp]
\centering
\includegraphics[width=\textwidth]{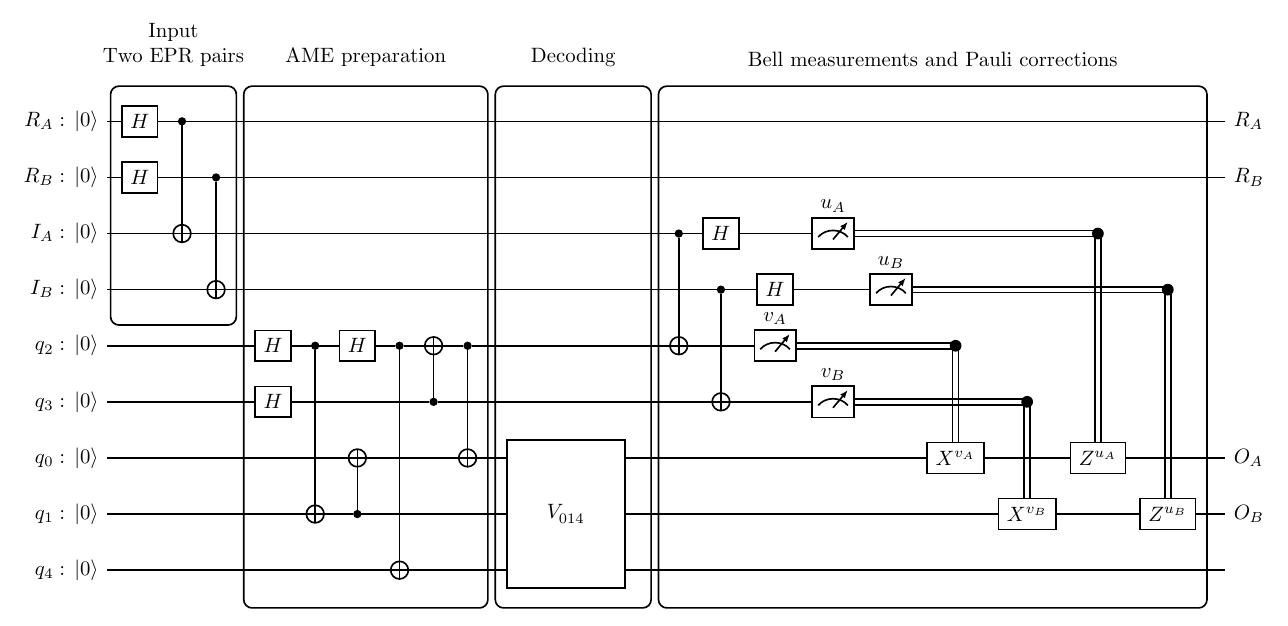}
\caption{Circuit diagram of one logical round of parallel teleportation, using the rank 1 decoder. At $R0$, the $H$-CNOT circuits prepare
$(R_A,I_A)$ and $(R_B,I_B)$ in $\ket{\Phi^+}$. The references
$R_A,R_B$ are retained while $V_{014}$ decodes the AME resource.
Bell outcomes $(u_c,v_c)$ determine the Pauli correction on $O_c$.
The displayed correction boxes are implemented through classical
Pauli frame post-processing in the main experiment. In the cascade,
$I_c^{(R+1)}\equiv O_c^{(R)}$. The same reference qubits are retained through the cascade. The diagram shows logical operations,
not their parallelised hardware schedule.}
\label{fig:rank1-parallel-teleportation}
\end{figure*}

Figure~\ref{fig:rank1-parallel-teleportation} illustrates a single round with the rank-1 decoder, including AME preparation, decoding and the two Bell measurements. The figure also shows the outcome-dependent Pauli boxes that the protocol requires. In our classical-post-processing implementation, these corrections are tracked in the analysis frame rather than executed on the device.

For the measured reference-receiver state $\rho_c(R)$, our primary
metric is the entanglement fidelity~\cite{Schumacher:1996dw},
evaluated from Pauli correlations~\cite{Siddhu:2025jsc}:
\begin{equation}
 \begin{split}
 F_c(R)&=\bra{\Phi^+}\rho_c(R)\ket{\Phi^+}\\
 &=\frac{1+\langle XX\rangle-\langle YY\rangle+\langle ZZ\rangle}{4}.
 \end{split}
 \label{eq:bell-correlation-fidelity}
\end{equation}
A value above $1/2$ certifies entanglement of that two-qubit
state~\cite{Horodecki:1998zh,GuhneToth2009}. Because the input Bell pair preparation and reference readout are imperfect, our entanglement-fidelity estimates include state preparation and measurement errors. So, they do not isolate the intrinsic performance of the channels. We mitigate readout errors using matrix-free measurement error mitigation (M3)~\cite{Nation:2021kye}, which is based on calibrated assignment models~\cite{PhysRevA.103.042605}, and we quantify fidelity loss relative to the measured $R0$ baselines. Backend calibration conditions,
the shot budget, trial selection and uncertainty calculations are
specified in Supplementary Material, Sec.~IV. While definitions of the performance metrics and the layout selection,
DD and readout mitigation methods are given in Supplementary
Material, Sec.~V. For completeness, we also characterise AME preparation
(Stage~1), the decoded Bell pair marginals (Stage~2), and
teleportation of $\ket{+}_A\ket{-}_B$ (Stage~3) using separate
circuits. Input preparation is not reported,
since $\ket{+}$ and $\ket{-}$ require only a few single-qubit
gates and therefore add little circuit overhead.\footnote{The circuit setup is slightly different from Fig.~4 and is shown in Fig.~S6 of Sec.~VI of the Supplementary Material.}
Tomography fidelities, stabilizer-generator fidelity
lower bounds, and SWAP-test pass rates are presented in the
Supplementary Material, Sec.~VI. These measurements check resource
quality, Bell pair extraction, and transfer of a known input,
providing a perspective different from the entanglement fidelity and
complementing our experiment on the multi-round entanglement fidelity.

With entanglement fidelity as the performance metric, we next describe two implementation strategies, namely the parallelization of independent operations and the classical post-processing of Pauli byproducts, aimed at preserving reference-receiver correlations. These strategies reduce avoidable sequential execution and feedback-related waiting, limiting decoherence during the cascade.

\subsection{Parallelization}

Each teleportation round uses a separate AME register, allowing
independent resource preparation and decoding to overlap across
rounds. Bell measurements on disjoint qubit pairs can likewise
overlap once their prerequisite operations are complete. We
explicitly construct these layers in Qiskit without barriers
that enforce sequential execution of complete rounds.
Transpilation, routing and backend-aware scheduling then determine
the permitted concurrency, following the layered approach
illustrated in IBM's long-range-entanglement
tutorial~\cite{IBMLongRangeEntanglement}.

To quantify the additional benefit of parallelization, we compare
sequential and parallelized implementations that both use
classical post-processing (to be discussed in the next subsection). For three rounds on
\texttt{ibm\_pittsburgh} at optimisation level~2 without DD,
parallelization reduces the scheduled duration from approximately
$21.3$ to $14.5\,\mu\mathrm{s}$ for rank~1 and from
$23.3$ to $16.2\,\mu\mathrm{s}$ for rank~58.
These correspond to reductions of $31.8\%$ and $30.2\%$,
respectively. All durations reported here are obtained from
hardware-compiler timing traces, including reported initialization
and finalization.

These shorter schedules can reduce idle-time exposure to
relaxation and dephasing, helping preserve entanglement
fidelity~\cite{Baumer:2023vrf}. The savings reflect the complete
parallelized implementation, including any changes in routing
and gate cancellation. Each round-count circuit is compiled
and scheduled independently, rather than obtained by appending
operations to an earlier compiled checkpoint. 

\subsection{Classical Post-Processing}

Mid-circuit measurement and classical feedback introduce a timing
bottleneck because conditional Pauli corrections must wait for
Bell-measurement outcomes to be acquired and processed. Readout
and classical control can take longer than two-qubit
gates~\cite{IBMExecuteDynamicCircuits}, leaving reference and
receiver qubits exposed to decoherence during this
waiting period~\cite{Baumer:2023vrf}.

We instead use Pauli frame tracking to record the required
corrections classically rather than applying them
physically~\cite{Chamberland:2017mdd,PennyLane-PauliFrameTracking}.
Because the cascade uses Clifford operations, Pauli byproducts
remain Pauli operators as they propagate through subsequent
rounds. Intermediate physical corrections are therefore
unnecessary. After execution, the Bell-measurement history
determines sign corrections to the reference-receiver Pauli
correlations used to calculate entanglement fidelity.
This removes intermediate correction gates and feedback-related
waiting while enabling the overlapping layers described above.
Physical measurements remain necessary, and frame tracking
does not reverse unknown preparation, gate, readout or
decoherence errors.

To separate this benefit from parallelization, we compare the
two sequential implementations using the same routed quantum
backbone. Replacing physical feed-forward with classical
post-processing reduces the scheduled duration from approximately
$23.9$ to $21.3\,\mu\mathrm{s}$ for rank~1 and from
$25.3$ to $23.3\,\mu\mathrm{s}$ for rank~58.
The corresponding savings are $2.67\,\mu\mathrm{s}$ ($11.2\%$)
and $2.09\,\mu\mathrm{s}$ ($8.24\%$).

Combining classical post-processing with parallelization
therefore shortens the schedule from approximately
$23.9$ to $14.5\,\mu\mathrm{s}$ for rank~1 and from
$25.3$ to $16.2\,\mu\mathrm{s}$ for rank~58,
relative to sequential feed-forward. The net savings are
$9.42\,\mu\mathrm{s}$ ($39.4\%$) and
$9.11\,\mu\mathrm{s}$ ($35.9\%$), respectively \footnote{
For context, Baumer et al. report a feed-forward time of
approximately $0.7\,\mu\mathrm{s}$ on
\texttt{ibm\_sherbrooke}~\cite{Baumer:2023vrf}.
In our no DD, optimisation level 2, timing benchmark on
\texttt{ibm\_pittsburgh}, hardware-returned compiler traces give
scheduled feedback waits of $(0.956,0.956,0.944)\,\mu\mathrm{s}$
for rank~1 and $(1.01,0.948,0.964)\,\mu\mathrm{s}$ for
rank~58 at $R1$-$R3$, respectively.
We define this interval from the end of the last Bell-readout
capture in each round to the start of the first conditional
$X$ correction. These intervals
are distinct from the complete circuit durations.
}.\\

The circuits run on the 156-qubit \texttt{ibm\_pittsburgh} processor.
We compare no dynamical decoupling (DD), XX and XY4~\cite{Souza:2011vcv}
at Qiskit optimisation levels 2 and 3~\cite{IBMTranspilerOptimization}.
DD pulses are inserted into eligible idle intervals after
scheduling~\cite{QiskitPadDD}. Mapomatic~\cite{PRXQuantum.4.010327}
chooses a layout for the whole circuit, subject to a channel-balance
criterion. This criterion keeps the calibrated readout errors of the two channels closely matched, so that their entanglement fidelities can be compared fairly. Each tomography
setting and readout calibration uses 5000 shots. Reported results
are means over eight selected trials per setting, with pointwise
two-sided $99\%$ confidence intervals. Note that Supplementary Material, Sec.~VII, details classical Pauli frame
tracking, overlapping operations and the complete hardware fidelity
results summarised in the next section. Section~VIII compares DD
sequences, fidelity retention relative to $R0$ and Bell correlations
using raw and M3 data, testing whether the decoder and DD preferences
depend on readout mitigation or unequal starting fidelities.
Lastly, Sec.~IX reports the real-time sequential feed-forward experiment,
in which Bell outcomes trigger physical Pauli corrections before
each receiver becomes the next input.

\section{Main results}
\label{sec:main-results}
\subsection{Entanglement survives three rounds}

\begin{figure*}[tp]
\centering
\includegraphics[width=\textwidth]{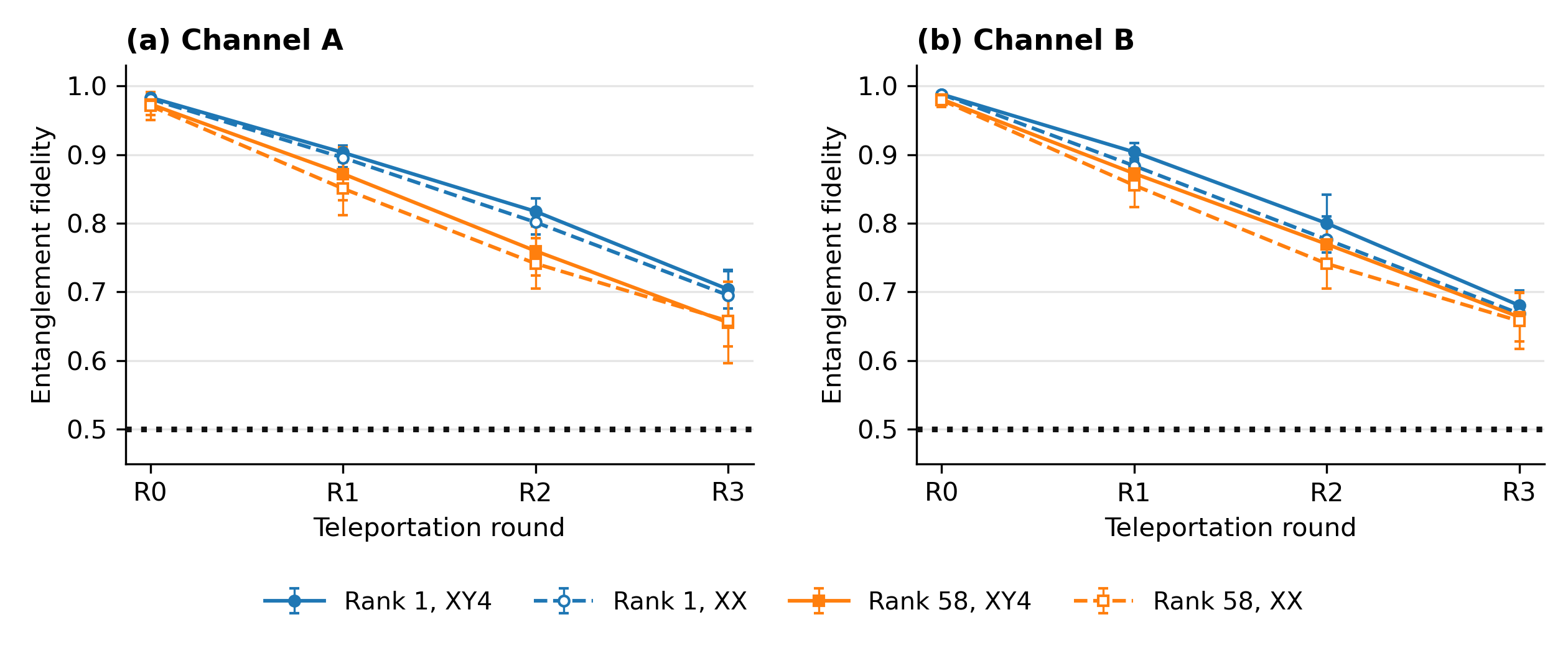}
\caption{Reference-receiver entanglement fidelities after repeated
AME-assisted parallel teleportation on hardware, with optimisation
level 2 and classical Pauli frame post-processing. Panels (a) and
(b) show channels $A$ and $B$. Blue circles and orange squares
identify ranks 1 and 58; solid filled-marker curves show XY4,
and dashed open-marker curves show XX. Points are M3-mitigated
means over eight trials; error bars are pointwise two-sided $99\%$
confidence intervals. The thick black dotted line marks the $1/2$ separability
threshold, and $R0$ is the input Bell pair baseline. The data are obtained from the full grouped-bar results in Supplementary
Material, Sec.~VII.}
\label{fig:cascade-summary}
\end{figure*}

Figure~\ref{fig:cascade-summary} compares the fidelity decay with
XX and XY4 at optimisation level 2. For rank~1 with XY4, the
M3-mitigated channel means $(F_A,F_B)$ decrease from
$(0.983,0.988)$ at $R0$ to $(0.903,0.904)$ at $R1$,
$(0.817,0.800)$ at $R2$, and
$(0.704\pm0.028,0.680\pm0.022)$ at $R3$.
Rank~58 follows the same declining trend, from $(0.973,0.981)$
at $R0$ to $(0.872,0.873)$ at $R1$, $(0.760,0.770)$ at $R2$,
and $(0.655\pm0.059,0.663\pm0.036)$ at $R3$.

The corresponding raw $R3$ fidelities are
$(0.703\pm0.029,0.678\pm0.022)$ for rank~1 and
$(0.654\pm0.059,0.662\pm0.036)$ for rank~58.
All quoted uncertainties are pointwise $99\%$ confidence-interval
half-widths over eight trials. The close agreement between raw
and mitigated endpoints shows that the entanglement-survival
result does not depend on a large M3 correction, especially at the last round.

Despite the accumulated fidelity loss, both channels retain
reference-receiver entanglement after three rounds. The
M3-mitigated lower interval bounds at $R3$ are $(0.676,0.658)$
for rank~1 and $(0.596,0.628)$ for rank~58, all above the
separability threshold of $1/2$; the raw lower bounds also exceed
this threshold. With XY4, this remains true for both decoders
at both optimisation levels. Each interval applies to the
trial-mean fidelity for one combination of decoder, channel,
optimisation level and processing method.

\iffalse
\begin{table}[tbp]
\centering
\caption{$R3$ fidelities with XY4 and optimisation level 2.
Uncertainties are pointwise $99\%$ interval half-widths over eight
trials. All four raw and mitigated lower bounds exceed $1/2$.}
\label{tab:endpoints}
\begin{tabular}{cccc}
\toprule
Rank & Channel & Raw & M3\\
\midrule
1 & $A$ & $0.703\pm0.029$ & $0.704\pm0.028$\\
1 & $B$ & $0.678\pm0.022$ & $0.680\pm0.022$\\
58 & $A$ & $0.654\pm0.059$ & $0.655\pm0.059$\\
58 & $B$ & $0.662\pm0.036$ & $0.663\pm0.036$\\
\bottomrule
\end{tabular}
\end{table}
\fi

\subsection{Comparison of decoders and dynamical-decoupling sequences}

Rank 1 has higher mean XY4 fidelity on both channels at
$R1$-$R3$ for both tested optimisation levels. To account for
the different input-pair fidelities at $R0$, we compare each rank's change
from its own $R0$ baseline. At level 2, the baseline-adjusted
rank 1 advantage (percentage points of entanglement fidelities of rank 1 minus rank 58) on channel $A$ is $2.12$, $4.78$ and $3.88$
percentage points at $R1$, $R2$ and $R3$. On channel $B$, it is
$2.39$, $2.31$ and $0.96$ percentage points. Rank 1 is therefore
favoured in mean fidelity retention on both channels, with a
larger later-round separation on $A$. The $R3$ intervals for
these between-rank differences include zero. Thus, the observed ordering does not establish a resolved advantage at every round. Figure~\ref{fig:baseline-comparisons} shows these comparisons in detail, together with the channel balance and the baseline-adjusted dynamical-decoupling (DD) advantage.

DD provides a second, practically useful improvement. For rank 1
at level 2, the M3-mitigated $R3$ means increase from
$(0.639,0.643)$ without DD to $(0.695,0.668)$ with XX and
$(0.704,0.680)$ with XY4. XY4 generally gives higher mean fidelity
than XX across the tested settings, including at both optimisation
levels (see Sec. VIII of Supplementary materials). Its advantage over no DD at this endpoint is
$6.51\pm6.39$ percentage points on $A$ and
$3.73\pm6.08$ on $B$. The magnitude and uncertainty differ
between channels; the latter interval includes zero.
Overall, XY4 outperforms XX in this campaign, so we adopt it as our working sequence; the one near-exception is rank-58 channel $A$ at $R3$, where XX gives $0.658$ and XY4 $0.655$.

However, moving from optimisation level 2 to 3 does not give a uniform
fidelity improvement. For XY4, the level-3 M3 endpoints are
$(0.693,0.676)$ for rank 1 and $(0.643,0.668)$ for rank 58.
We focus on level 2 because it is slightly better in most of
these comparisons. 

\section{Discussion}
The results establish two operational properties of the AME resource.
A joint decoder on only three resource qubits supplies two Bell
channels, and those channels can preserve reference entanglement
through three successive teleportation rounds on present hardware.
The first property gives flexibility in assigning communication
roles; the second tests whether that ideal entanglement remains
usable after repeated noisy operations. 

\begin{figure*}[tp]
\centering
\includegraphics[width=\textwidth]{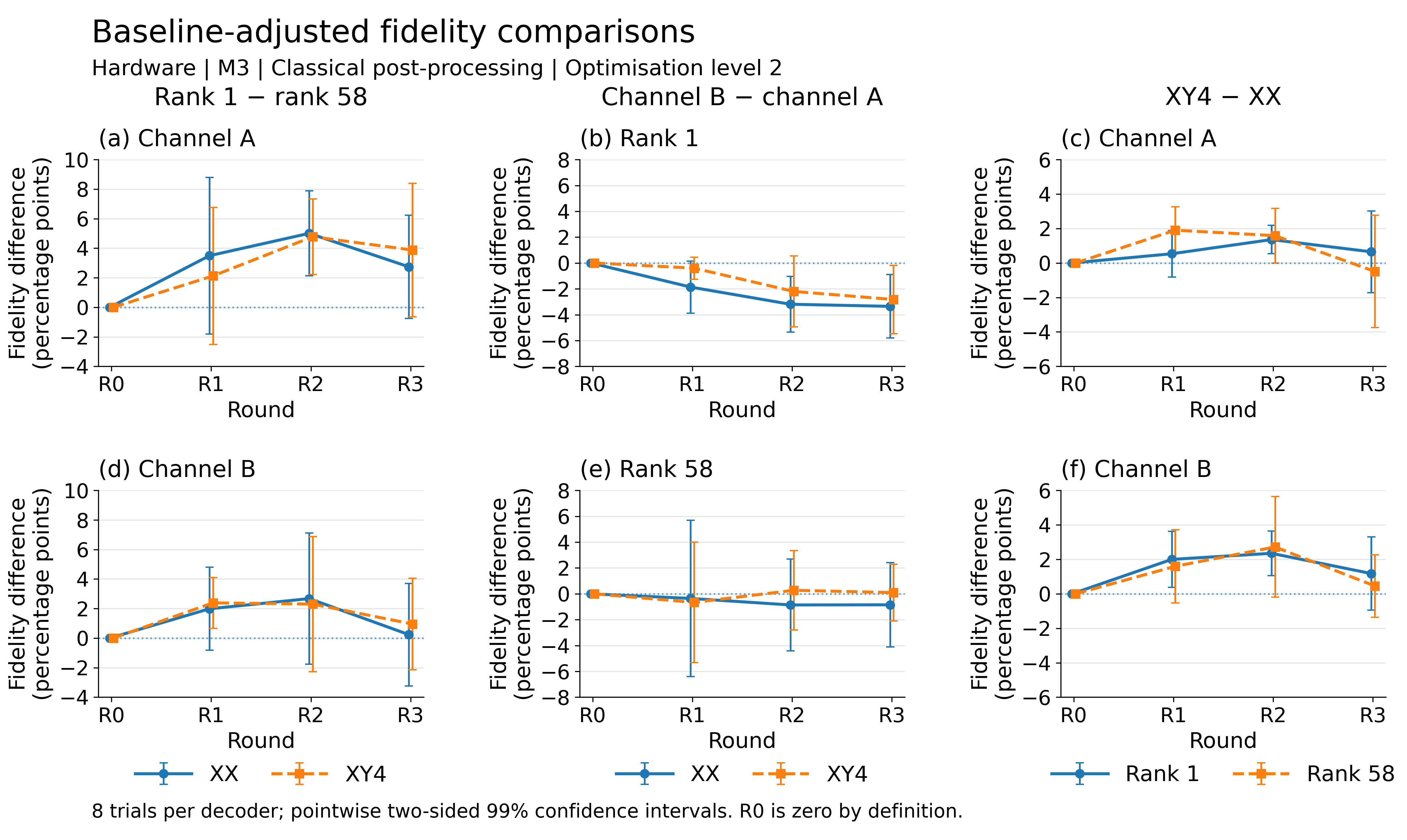}
\caption{Baseline-adjusted hardware entanglement-fidelity comparisons
with M3 mitigation, classical Pauli frame post-processing and
optimisation level 2. Panels (a) and (d) compare rank 1 minus rank 58
on channels $A$ and $B$, respectively, for XX and XY4. Panels (b)
and (e) compare channel $B$ minus channel $A$ for ranks 1 and 58,
respectively, for both DD sequences. Panels (c) and (f) compare XY4
minus XX on channels $A$ and $B$, respectively, for both ranks.
Each comparison subtracts its corresponding within-trial $R0$
baseline before averaging, so $R0$ is zero by construction.
Positive values indicate better fidelity retention by rank 1 in
the left column, channel $B$ in the middle column, or XY4 in the
right column. Differences are in fidelity percentage points.
Error bars denote pointwise two-sided $99\%$ confidence intervals,
calculated separately for each plotted comparison using eight
trials per decoder.}
\label{fig:baseline-comparisons}
\end{figure*}

Figure~\ref{fig:baseline-comparisons} distinguishes decoder choice,
channel balance and DD performance after removing the initial
EPR-pair differences. For rank $r$, channel $c$ and DD sequence $p$,
define the trial-level change
$\delta F_{r,c}^{p}(R)=F_{r,c}^{p}(R)-F_{r,c}^{p}(R0)$.
The figure compares these changes between ranks, channels or DD
sequences, with every $R0$ value zero by construction. Subtracting
the baseline within each trial retains its correlations with later
rounds; the two rank-specific trial pools remain independent.

The positive rank 1 minus rank 58 mean fidelity difference (percentage points) in panels (a) and (d)
show that rank 1 retains fidelity better on both channels with
either DD sequence at $R1$-$R3$. The clearest separation is on
channel $A$ at $R2$, where both XX and XY4 intervals lie above
zero. At $R2$, the difference from the XY4 values is $4.78$ percentage points on channel $A$ and
$2.31$ on $B$. Panels (b) and (e) distinguish
this decoder preference from channel balance. For rank 1, both
DD sequences give greater baseline-relative loss on $B$ than on
$A$; the XY4 difference reaches $-2.83\pm2.65$ percentage points
at $R3$. For rank 58, the corresponding $0.09\pm2.20$ is
consistent with comparable losses on the two channels.

Panels (c) and (f) directly support the preference for XY4 over XX.
After subtracting each sequence's own $R0$ baseline, XY4 has
better mean fidelity retention in eleven of the twelve
rank-channel comparisons at $R1$-$R3$. For rank 1 at $R2$,
the gains are $1.36\pm0.82$ percentage points on $A$ and
$2.35\pm1.29$ on $B$, with both $99\%$ intervals entirely
above zero. XY4 therefore provides an experimentally supported
improvement on both channels, rather than merely a higher initial
EPR fidelity. The exception is rank 58 channel $A$ at $R3$,
where the mean difference is $-0.49\pm3.27$ percentage points.
Taken together, the results favour rank 1 and XY4 for classical post-processing, although not every individual comparison is statistically resolved.

The advantage of the rank 1 against the rank 58 decoder has a natural circuit-cost interpretation.
Both decoders use three logical CNOTs, but rank 1 uses five
single-qubit gates and has depth six, compared with thirteen
single-qubit gates and depth ten for rank 58
(Figure~\ref{fig:decoder-circuits}). Fewer operations and a shallower
executed circuit can reduce accumulated gate errors and decoherence
exposure, motivating circuit optimisation on noisy
hardware~\cite{IBMTranspilerOptimization}. The logical savings of rank 1 may contribute to its measured advantage, but there are other considerations. On the hardware, some phase gates become virtual frame updates, while the native $SX=\sqrt X$ gate~\cite{IBMQuantumSXGate}, routing and gate cancellation change the physical operation count. Fidelity also depends on which qubits and couplers are selected, as calibration-aware circuit mapping shows~\cite{PRXQuantum.4.010327}.
%make sure this is discuss in outlook

The XY4 preference is likewise consistent with its alternating
control axes. Unlike single axis XX, XY4 can average a broader
set of system-environment interactions and compensate some pulse
imperfections~\cite{Souza:2011vcv}. This provides a physical
rationale for the observed retention gains, but the benefit still
depends on the idle intervals, noise spectrum and pulse errors.
While classical Pauli frame tracking removes intermediate physical
corrections and their real-time dependencies, it does not remove
resource-preparation, decoder, Bell-measurement or readout errors.

The supporting correlation analysis also shows that the rank 1
advantage is channel dependent. At $R3$, its advantage on $A$
comes mainly from larger $-YY$ and $ZZ$ components; on $B$, it
comes mainly from $XX$. These profiles identify measured
contributions to the fidelity, not a unique noise mechanism.
Residual phase noise is plausible, and measurement-induced phase
errors and $ZZ$ crosstalk have been reported in superconducting
dynamic circuits~\cite{Shirizly:2024hre}. Our data do not show
uniform protection of $ZZ$, so a pure dephasing explanation is
insufficient. 

\iffalse
The component values and plots are in Supplementary
Material, Sec.~VI.
\fi

In addition, the experiment is limited to eight trials on one processor, with
data acquired over several days. The pointwise intervals quantify
trial-to-trial uncertainty but do not capture all systematic
calibration and mitigation errors. 

\iffalse
Stage-resolved
tomography~\cite{James:2001klt,Kanazawa2023}, stabilizer
witnesses~\cite{Toth:2005qcv} and SWAP
tests~\cite{Barenco:1996kz,Buhrman:2001rma} provide complementary
diagnostics, but
separately compiled stages cannot be subtracted to obtain an
additive error budget; their targets, subsystems and layouts differ.
Mitigated tomography combines readout calibration with state
reconstruction in Qiskit Experiments~\cite{QiskitMSTDocs}.
Those diagnostics are described in Supplementary Material, Secs.~IV
and VIII.
\fi

\section{Conclusion and outlook}
\label{sec:conclusion-outlook}
In this work, we demonstrate AME-assisted parallel teleportation on IBM Quantum
hardware using compact three-qubit decoders. A shared AME$(5,2)$ state enables two qubits to be teleported
in parallel. For each of the 60 inequivalent sender-receiver
assignments, we construct and verify a decoder circuit containing
three two-qubit gates, in addition to single-qubit gates, before
hardware compilation. This provides flexibility
in choosing communication roles after resource preparation, while
the cascaded experiment tests whether the resulting channels
preserve entanglement under repeated noisy operations.

With rank 1, XY4 and optimisation level 2, the M3-mitigated
entanglement fidelities after three rounds are $0.704\pm0.028$
and $0.680\pm0.022$, where uncertainties are pointwise $99\%$
confidence-interval half-widths. The lower bounds exceed $1/2$
for both channels, as do those of the corresponding raw estimates.
The hardware means favour rank 1 over rank 58 and generally favour
XY4 over XX. Classical Pauli frame tracking permits overlapping
operations without intermediate physical corrections. Together,
these results connect efficient multipartite decoding with a
quantitative demonstration of entanglement survival through
repeated parallel teleportation.

In the future work, decoder selection should account for both
logical circuit complexity and hardware properties, including
qubit connectivity, gate durations and the idle time experienced
by each channel. Different decoder, layout and DD choices should
be tested in interleaved runs within the same calibration sessions
to reduce the influence of hardware drift on their comparison~\cite{Magesan:2012xfe}. Experiments with more teleportation rounds and larger AME resources
could examine how entanglement retention scales with circuit length
and the number of parallel channels. Such extensions should also
assess the sampling requirements and readout-mitigation overhead
of longer Bell-measurement histories, using consistent input
baselines and calibration procedures.

Moving beyond the present single-processor demonstration, a key
goal is parallel teleportation between remote nodes. Optically
linked trapped-ion modules offer a promising
platform~\cite{Main:2024iqh}, although distributing the AME
resource while preserving its entanglement and enabling joint
receiver-side decoding remains challenging. Such experiments could
compare the flexibility of shared AME resources with fixed Bell pair
allocations in terms of memory time, distribution cost and
communication latency.

Connections with fusion-based quantum computation motivate exploring
whether AME resources joined by Bell measurements can support
loss-tolerant communication~\cite{Bartolucci:2021wut}. Quantum
secret sharing offers complementary tests of authorised
reconstruction and protection against unauthorised
subsets~\cite{Helwig:2012nha,Cleve:1999qg}. More broadly, nonlocal
information distribution and reconstruction could provide accessible
experimental toy models relevant to holographic error correction
and black hole information, offering educational and conceptual
insights~\cite{Pastawski:2015qua,w7nc-l9s5}.\\

\begin{acknowledgments}

We thank Alexander C.B. Greenwood for insightful discussions on error mitigation and dynamical-decoupling techniques on IBM Heron r3, as well as offering valuable recommendations for the implementation. We would also like to thank Chen-Xun Weng for the insightful discussion during the initial phase of the project. The code for the IBM quantum experiments was developed with assistance from Claude 5.0 Sonnet and OpenAI's Codex, and verified by the authors. This project is supported by the National Research Foundation, Singapore through the National Quantum Office, hosted in A*STAR, under its Centre for Quantum Technologies Funding Initiative. Hoi-Kwong Lo acknowledges research support from the CQT PI core funding and the
National University of Singapore Start-up grant. Zheng Liang Lim thanks the National University of Singapore Physics Department for providing access to the superconducting processor used in this work and acknowledges support from the National Quantum Scholarship Scheme.

\end{acknowledgments}

\section*{Code and data availability}
The data that support the findings of this study are available in the Supplementary Materials.
Both the data and codes used to generate the results of this study are
also available from the authors upon reasonable request \footnote{The main data and codes are available in github: \url{https://github.com/LimZhengLiang/ibm_parallel_teleportation}.}.

% This boundary makes the main-text page count unambiguous.
\label{sec:references}
\bibliographystyle{unsrt}
\bibliography{Ref,additional_refs}

@article{Horodecki:2009zz,
    author = "Horodecki, Ryszard and Horodecki, Pawel and Horodecki, Michal and Horodecki, Karol",
    title = "{Quantum entanglement}",
    eprint = "quant-ph/0702225",
    archivePrefix = "arXiv",
    doi = "10.1103/RevModPhys.81.865",
    journal = "Rev. Mod. Phys.",
    volume = "81",
    pages = "865--942",
    year = "2009"
}

@article{Sun:2017ylw,
    author = "Sun, Qi-Chao and others",
    title = "{Entanglement swapping over 100{\,}{\,}km optical fiber with independent entangled photon-pair sources}",
    eprint = "1704.03960",
    archivePrefix = "arXiv",
    primaryClass = "quant-ph",
    doi = "10.1364/OPTICA.4.001214",
    journal = "Optica",
    volume = "4",
    number = "10",
    pages = "1214--1218",
    year = "2017"
}

@article{PhysRevLett.70.1895,
  title = {Teleporting an unknown quantum state via dual classical and Einstein-Podolsky-Rosen channels},
  author = {Bennett, Charles H. and Brassard, Gilles and Cr{\'e}peau, Claude and Jozsa, Richard and Peres, Asher and Wootters, William K.},
  journal = {Phys. Rev. Lett.},
  volume = {70},
  issue = {13},
  pages = {1895--1899},
  numpages = {0},
  year = {1993},
  month = {Mar},
  publisher = {American Physical Society},
  doi = {10.1103/PhysRevLett.70.1895},
  url = {https://link.aps.org/doi/10.1103/PhysRevLett.70.1895}
}

@article{Bouwmeester:1997slj,
    author = "Bouwmeester, Dik and Pan, Jian-Wei and Mattle, Klaus and Eibl, Manfred and Weinfurter, Harald and Zeilinger, Anton",
    title = "{Experimental quantum teleportation}",
    eprint = "1901.11004",
    archivePrefix = "arXiv",
    primaryClass = "quant-ph",
    doi = "10.1038/37539",
    journal = "Nature",
    volume = "390",
    pages = "575--579",
    year = "1997"
}

@article{Yin:2012vzv,
    author = "Yin, Juan and others",
    title = "{Quantum teleportation and entanglement distribution over 100-kilometre free-space channels}",
    eprint = "1205.2024",
    archivePrefix = "arXiv",
    primaryClass = "quant-ph",
    doi = "10.1038/nature11332",
    journal = "Nature",
    volume = "488",
    number = "7410",
    pages = "185--188",
    year = "2012"
}

@article{Ma:2012bgk,
    author = "Ma, Xiao-Song and others",
    title = "{Quantum teleportation over 143 kilometres using active feed-forward}",
    eprint = "1205.3909",
    archivePrefix = "arXiv",
    primaryClass = "quant-ph",
    doi = "10.1038/nature11472",
    journal = "Nature",
    volume = "489",
    number = "7415",
    pages = "269--273",
    year = "2012"
}

@article{Ren:2017ivf,
    author = "Ren, Ji-Gang and others",
    title = "{Ground-to-satellite quantum teleportation}",
    eprint = "1707.00934",
    archivePrefix = "arXiv",
    primaryClass = "quant-ph",
    doi = "10.1038/nature23675",
    journal = "Nature",
    volume = "549",
    pages = "70--73",
    year = "2017"
}

@article{rfz9-3prw,
  title = {Hundred-Channel Reconfigurable Quantum Teleportation},
  author = {Lou, Yanbo and Wang, Jiabin and Zou, Yuyan and Hou, Lingyue and Liu, Shengshuai and Jing, Jietai},
  journal = {Phys. Rev. Lett.},
  volume = {137},
  issue = {8},
  pages = {080801},
  numpages = {7},
  year = {2026},
  month = {Aug},
  publisher = {American Physical Society},
  doi = {10.1103/rfz9-3prw},
  url = {https://link.aps.org/doi/10.1103/rfz9-3prw}
}

@article{Azuma:2022nuy,
    author = "Azuma, Koji and Economou, Sophia E. and Elkouss, David and Hilaire, Paul and Jiang, Liang and Lo, Hoi-Kwong and Tzitrin, Ilan",
    title = "{Quantum repeaters: From quantum networks to the quantum internet}",
    eprint = "2212.10820",
    archivePrefix = "arXiv",
    primaryClass = "quant-ph",
    doi = "10.1103/RevModPhys.95.045006",
    journal = "Rev. Mod. Phys.",
    volume = "95",
    number = "4",
    pages = "045006",
    year = "2023"
}

@article{Baumer:2023vrf,
    author = {B{\"a}umer, Elisa and Tripathi, Vinay and Wang, Derek S. and Rall, Patrick and Chen, Edward H. and Majumder, Swarnadeep and Seif, Alireza and Minev, Zlatko K.},
    title = "{Efficient Long-Range Entanglement Using Dynamic Circuits}",
    eprint = "2308.13065",
    archivePrefix = "arXiv",
    primaryClass = "quant-ph",
    doi = "10.1103/PRXQuantum.5.030339",
    journal = "PRX Quantum",
    volume = "5",
    number = "3",
    pages = "030339",
    year = "2024"
}

@article{Kang:2024jqx,
    author = "Kang, Haiyue and Kam, John F. and Mooney, Gary J. and Hollenberg, Lloyd C. L.",
    title = "{Teleporting two-qubit entanglement across 19 qubits on a superconducting quantum computer}",
    eprint = "2407.02858",
    archivePrefix = "arXiv",
    primaryClass = "quant-ph",
    doi = "10.1103/PhysRevApplied.23.014057",
    journal = "Phys. Rev. Applied",
    volume = "23",
    number = "1",
    pages = "014057",
    year = "2025"
}

@article{Helwig:2012nha,
    author = "Helwig, Wolfram and Cui, Wei and Riera, Arnau and Latorre, Jos{\'e} I. and Lo, Hoi-Kwong",
    title = "{Absolute Maximal Entanglement and Quantum Secret Sharing}",
    eprint = "1204.2289",
    archivePrefix = "arXiv",
    primaryClass = "quant-ph",
    doi = "10.1103/PhysRevA.86.052335",
    journal = "Phys. Rev. A",
    volume = "86",
    pages = "052335",
    year = "2012"
}

@article{PhysRevLett.118.200502,
  title = {Absolutely Maximally Entangled States of Seven Qubits Do Not Exist},
  author = {Huber, Felix and G{\"u}hne, Otfried and Siewert, Jens},
  journal = {Phys. Rev. Lett.},
  volume = {118},
  issue = {20},
  pages = {200502},
  numpages = {5},
  year = {2017},
  month = {May},
  publisher = {American Physical Society},
  doi = {10.1103/PhysRevLett.118.200502},
  url = {https://link.aps.org/doi/10.1103/PhysRevLett.118.200502}
}

@article{Pastawski:2015qua,
    author = "Pastawski, Fernando and Yoshida, Beni and Harlow, Daniel and Preskill, John",
    title = "{Holographic quantum error-correcting codes: Toy models for the bulk/boundary correspondence}",
    eprint = "1503.06237",
    archivePrefix = "arXiv",
    primaryClass = "hep-th",
    doi = "10.1007/JHEP06(2015)149",
    journal = "JHEP",
    volume = "06",
    pages = "149",
    year = "2015"
}

@article{w7nc-l9s5,
  title = {Holographic Absolutely Maximally Entangled States in Black Hole Interiors},
  author = {Anegawa, Takanori and Tamaoka, Kotaro},
  journal = {Phys. Rev. Lett.},
  volume = {135},
  issue = {26},
  pages = {261601},
  numpages = {6},
  year = {2025},
  month = {Dec},
  publisher = {American Physical Society},
  doi = {10.1103/w7nc-l9s5},
  url = {https://link.aps.org/doi/10.1103/w7nc-l9s5}
}

@article{Cleve:1999qg,
    author = "Cleve, Richard and Gottesman, Daniel and Lo, Hoi-Kwong",
    title = "{How to share a quantum secret}",
    eprint = "quant-ph/9901025",
    archivePrefix = "arXiv",
    reportNumber = "LA-UR-98-5842, LAUR98-5842",
    doi = "10.1103/PhysRevLett.83.648",
    journal = "Phys. Rev. Lett.",
    volume = "83",
    pages = "648--651",
    year = "1999"
}

@article{Graves:2024wlh,
    author = "Graves, Jay and Nelson, Mike and Chitambar, Eric",
    title = "{Implementing Quantum Secret Sharing on Current Hardware}",
    eprint = "2410.11640",
    archivePrefix = "arXiv",
    primaryClass = "quant-ph",
    doi = "10.3390/e27100993",
    journal = "Entropy",
    volume = "27",
    number = "10",
    pages = "993",
    year = "2025"
}

@article{Cervera-Lierta:2019juk,
    author = "Cervera-Lierta, Alba and Latorre, Jos{\'e} Ignacio and Goyeneche, Dardo",
    title = "{Quantum circuits for maximally entangled states}",
    eprint = "1904.07955",
    archivePrefix = "arXiv",
    primaryClass = "quant-ph",
    doi = "10.1103/PhysRevA.100.022342",
    journal = "Phys. Rev. A",
    volume = "100",
    number = "2",
    pages = "022342",
    year = "2019"
}

@article{Schumacher:1996dw,
    author = "Schumacher, Benjamin",
    title = "{Sending entanglement through noisy quantum channels}",
    eprint = "quant-ph/9604023",
    archivePrefix = "arXiv",
    doi = "10.1103/PhysRevA.54.2614",
    journal = "Phys. Rev. A",
    volume = "54",
    number = "4",
    pages = "2614",
    year = "1996"
}

@article{James:2001klt,
    author = "James, Daniel F. V. and Kwiat, Paul G. and Munro, William J. and White, Andrew G.",
    title = "{Measurement of qubits}",
    eprint = "quant-ph/0103121",
    archivePrefix = "arXiv",
    reportNumber = "LA-UR-01-1143",
    doi = "10.1103/PhysRevA.64.052312",
    journal = "Phys. Rev. A",
    volume = "64",
    number = "5",
    pages = "052312",
    year = "2001"
}

@article{Kanazawa2023, doi = {10.21105/joss.05329}, url = {https://doi.org/10.21105/joss.05329}, year = {2023}, publisher = {The Open Journal}, volume = {8}, number = {84}, pages = {5329}, author = {Kanazawa, Naoki and Egger, Daniel J. and Ben-Haim, Yael and Zhang, Helena and Shanks, William E. and Aleksandrowicz, Gadi and Wood, Christopher J.}, title = {Qiskit Experiments: A Python package to characterize and calibrate quantum computers}, journal = {Journal of Open Source Software} }

@article{Barenco:1996kz,
    author = "Barenco, Adriano and Berthiaume, Andr{\'e} and Deutsch, David and Ekert, Artur and Jozsa, Richard and Macchiavello, Chiara",
    title = "{Stabilization of Quantum Computations by Symmetrization}",
    eprint = "quant-ph/9604028",
    archivePrefix = "arXiv",
    doi = "10.1137/S0097539796302452",
    journal = "SIAM J. Comput.",
    volume = "26",
    number = "5",
    pages = "1541--1557",
    year = "1997"
}

@article{Buhrman:2001rma,
    author = "Buhrman, Harry and Cleve, Richard and Watrous, John and de Wolf, Ronald",
    title = "{Quantum Fingerprinting}",
    eprint = "quant-ph/0102001",
    archivePrefix = "arXiv",
    doi = "10.1103/PhysRevLett.87.167902",
    journal = "Phys. Rev. Lett.",
    volume = "87",
    number = "16",
    pages = "167902",
    year = "2001"
}

@article{Toth:2005qcv,
    author = "T{\'o}th, G{\'e}za and Guehne, Otfried",
    title = "{Entanglement detection in the stabilizer formalism}",
    eprint = "quant-ph/0501020",
    archivePrefix = "arXiv",
    doi = "10.1103/PhysRevA.72.022340",
    journal = "Phys. Rev. A",
    volume = "72",
    number = "2",
    pages = "022340",
    year = "2005"
}

@article{PRXQuantum.4.010327,
  title = {Suppressing Quantum Circuit Errors Due to System Variability},
  author = {Nation, Paul D. and Treinish, Matthew},
  journal = {PRX Quantum},
  volume = {4},
  issue = {1},
  pages = {010327},
  numpages = {9},
  year = {2023},
  month = {Mar},
  publisher = {American Physical Society},
  doi = {10.1103/PRXQuantum.4.010327},
  url = {https://link.aps.org/doi/10.1103/PRXQuantum.4.010327}
}

@misc{QiskitMSTDocs,
  author       = {{IBM Qiskit Development Team}},
  title        = {\texttt{MitigatedStateTomography}},
  howpublished = {\url{https://qiskit-community.github.io/qiskit-experiments/stubs/qiskit_experiments.library.tomography.MitigatedStateTomography.html}},
  note         = {Accessed 13 October 2026}
}

@article{PhysRevA.103.042605,
  title = {Mitigating measurement errors in multiqubit experiments},
  author = {Bravyi, Sergey and Sheldon, Sarah and Kandala, Abhinav and Mckay, David C. and Gambetta, Jay M.},
  journal = {Phys. Rev. A},
  volume = {103},
  issue = {4},
  pages = {042605},
  numpages = {12},
  year = {2021},
  month = {Apr},
  publisher = {American Physical Society},
  doi = {10.1103/PhysRevA.103.042605},
  url = {https://link.aps.org/doi/10.1103/PhysRevA.103.042605}
}

@article{Souza:2011vcv,
    author = "Souza, Alexandre M. and {\'A}lvarez, Gonzalo A. and Suter, Dieter",
    title = "{Robust dynamical decoupling}",
    eprint = "1110.6334",
    archivePrefix = "arXiv",
    primaryClass = "quant-ph",
    doi = "10.1098/rsta.2011.0355",
    journal = "Phil. Trans. Roy. Soc. Lond. A",
    volume = "370",
    pages = "4748--4769",
    year = "2012"
}

@misc{QiskitPadDD,
  author       = {{IBM Qiskit Development Team}},
  title        = {\texttt{PadDynamicalDecoupling}},
  howpublished = {\url{https://quantum.cloud.ibm.com/docs/en/api/qiskit/qiskit.transpiler.passes.PadDynamicalDecoupling}},
  note         = {Accessed 13 October 2026}
}

@article{Bravyi:2021ijy,
    author = "Bravyi, Sergey and Shaydulin, Ruslan and Hu, Shaohan and Maslov, Dmitri",
    title = "{Clifford Circuit Optimization with Templates and Symbolic Pauli Gates}",
    eprint = "2105.02291",
    archivePrefix = "arXiv",
    primaryClass = "quant-ph",
    doi = "10.22331/q-2021-11-16-580",
    journal = "Quantum",
    volume = "5",
    pages = "580",
    year = "2021"
}

@article{Shirizly:2024hre,
    author = "Shirizly, Liran and Govia, Luke C. G. and McKay, David C.",
    title = "{Randomized benchmarking protocol for dynamic circuits}",
    eprint = "2408.07677",
    archivePrefix = "arXiv",
    primaryClass = "quant-ph",
    doi = "10.1103/PhysRevA.111.012611",
    journal = "Phys. Rev. A",
    volume = "111",
    number = "1",
    pages = "012611",
    year = "2025"
}

@article{Horodecki:1998zh,
    author = "Horodecki, Michal and Horodecki, Pawel and Horodecki, Ryszard",
    title = "{General teleportation channel, singlet fraction, and quasidistillation}",
    eprint = "quant-ph/9807091",
    archivePrefix = "arXiv",
    doi = "10.1103/PhysRevA.60.1888",
    journal = "Phys. Rev. A",
    volume = "60",
    number = "3",
    pages = "1888",
    year = "1999"
}

@article{Nielsen:2002obk,
    author = "Nielsen, Michael A.",
    title = "{A simple formula for the average gate fidelity of a quantum dynamical operation}",
    eprint = "quant-ph/0205035",
    archivePrefix = "arXiv",
    doi = "10.1016/S0375-9601(02)01272-0",
    journal = "Phys. Lett. A",
    volume = "303",
    number = "4",
    pages = "249--252",
    year = "2002"
}

@article{Siddhu:2025jsc,
    author = "Siddhu, Vikesh and Winston, Erick and McKay, David C. and Javadi-Abhari, Ali",
    title = "{Basic entanglement distillation with realistic noise}",
    eprint = "2504.06175",
    archivePrefix = "arXiv",
    primaryClass = "quant-ph",
    doi = "10.1103/xx9g-pkht",
    journal = "Phys. Rev. Res.",
    volume = "7",
    number = "4",
    pages = "043069",
    year = "2025"
}

@misc{IBMQuantumSXGate,
  author       = {{IBM Qiskit Development}},
  title        = {{SXGate}: Qiskit API Documentation},
  howpublished = {\url{https://quantum.cloud.ibm.com/docs/en/api/qiskit/qiskit.circuit.library.SXGate}},
  note         = {Accessed 9 October 2026}
}

@article{Nation:2021kye,
    author = "Nation, Paul D. and Kang, Hwajung and Sundaresan, Neereja and Gambetta, Jay M.",
    title = "{Scalable Mitigation of Measurement Errors on Quantum Computers}",
    eprint = "2108.12518",
    archivePrefix = "arXiv",
    primaryClass = "quant-ph",
    doi = "10.1103/PRXQuantum.2.040326",
    journal = "PRX Quantum",
    volume = "2",
    number = "4",
    pages = "040326",
    year = "2021"
}

@misc{IBMLongRangeEntanglement,
  author       = {{IBM Quantum}},
  title        = {Long-range entanglement with dynamic circuits},
  howpublished = {IBM Quantum Documentation},
  url          = {https://quantum.cloud.ibm.com/docs/en/tutorials/long-range-entanglement},
  note         = {Accessed 12 October 2026}
}

@article{Magesan:2012xfe,
    author = "Magesan, Easwar and others",
    title = "{Efficient Measurement of Quantum Gate Error by Interleaved Randomized Benchmarking}",
    eprint = "1203.4550",
    archivePrefix = "arXiv",
    primaryClass = "quant-ph",
    doi = "10.1103/PhysRevLett.109.080505",
    journal = "Phys. Rev. Lett.",
    volume = "109",
    number = "8",
    pages = "080505",
    year = "2012"
}

@article{Helwig:2013ckb,
    author = "Helwig, Wolfram and Cui, Wei",
    title = "{Absolutely Maximally Entangled States: Existence and Applications}",
    eprint = "1306.2536",
    archivePrefix = "arXiv",
    primaryClass = "quant-ph",
    month = "6",
    year = "2013"
}

@article{Laflamme:1996iw,
    author = "Laflamme, Raymond and Miquel, Cesar and Paz, Juan Pablo and Zurek, Wojciech Hubert",
    title = "{Perfect Quantum Error Correcting Code}",
    eprint = "quant-ph/9602019",
    archivePrefix = "arXiv",
    doi = "10.1103/PhysRevLett.77.198",
    journal = "Phys. Rev. Lett.",
    volume = "77",
    number = "1",
    pages = "198",
    year = "1996"
}

@article{Bartolucci:2021wut,
    author = "Bartolucci, Sara and others",
    title = "{Fusion-based quantum computation}",
    eprint = "2101.09310",
    archivePrefix = "arXiv",
    primaryClass = "quant-ph",
    doi = "10.1038/s41467-023-36493-1",
    journal = "Nature Commun.",
    volume = "14",
    number = "1",
    pages = "912",
    year = "2023"
}

@article{Main:2024iqh,
    author = "Main, D. and Drmota, P. and Nadlinger, D. P. and Ainley, E. M. and Agrawal, A. and Nichol, B. C. and Srinivas, R. and Araneda, G. and Lucas, D. M.",
    title = "{Distributed quantum computing across an optical network link}",
    eprint = "2407.00835",
    archivePrefix = "arXiv",
    primaryClass = "quant-ph",
    doi = "10.1038/s41586-024-08404-x",
    journal = "Nature",
    volume = "638",
    number = "8050",
    pages = "383--388",
    year = "2025"
}

@article{Chamberland:2017mdd,
    author = "Chamberland, Christopher and Iyer, Pavithran and Poulin, David",
    title = "{Fault-tolerant quantum computing in the Pauli or Clifford frame with slow error diagnostics}",
    eprint = "1704.06662",
    archivePrefix = "arXiv",
    primaryClass = "quant-ph",
    doi = "10.22331/q-2018-01-04-43",
    journal = "Quantum",
    volume = "2",
    pages = "43",
    year = "2018"
}

@misc{PennyLane-PauliFrameTracking,
  author = {Ren, David D. W.},
  title = {Pauli Frame Tracking},
  year = {2026},
  howpublished = {\url{https://pennylane.ai/compilation/pauli-frame-tracking}}
}

@misc{IBMExecuteDynamicCircuits,
  author = {{IBM Quantum}},
  title = {Execute dynamic circuits},
  howpublished = {\url{https://quantum.cloud.ibm.com/docs/en/guides/execute-dynamic-circuits}},
  note = {Accessed 16 October 2026}
}

@article{GuhneToth2009,
 author={G{\"u}hne, Otfried and T{\'o}th, G{\'e}za},
 title={Entanglement detection},
 journal={Physics Reports},
 volume={474},number={1--6},pages={1--75},year={2009},
 doi={10.1016/j.physrep.2009.02.004},
 eprint={0811.2803},archivePrefix={arXiv},primaryClass={quant-ph}
}

@article{Lim:2026cuk,
 author={Lim, Zheng Liang and Lo, Hoi-Kwong},
 title={Encrypted Cloning, Absolute Maximal Entanglement and Quantum Secret Sharing},
 journal={arXiv preprint arXiv:2605.26866},
 eprint={2605.26866},archivePrefix={arXiv},primaryClass={quant-ph},
 month={5},year={2026}
}

@misc{IBMTranspilerOptimization,
 author={{IBM Quantum}},
 title={Set transpiler optimization level},
 howpublished={IBM Quantum Documentation},
 url={https://quantum.cloud.ibm.com/docs/en/guides/set-optimization},
 note={Accessed 12 October 2026}
}

@article{Gottesman:1997zz,
    author = "Gottesman, Daniel",
    title = "{Stabilizer codes and quantum error correction}",
    eprint = "quant-ph/9705052",
    archivePrefix = "arXiv",
    month = "5",
    year = "1997"
}

@misc{IBMCommutativeCancellation,
  author = {{IBM Quantum}},
  title = {{CommutativeCancellation}},
  howpublished = {\url{https://quantum.cloud.ibm.com/docs/en/api/qiskit/qiskit.transpiler.passes.CommutativeCancellation}},
  note = {Accessed 19 October 2026}
}

@misc{IBMSabreSwap,
  author = {{IBM Quantum}},
  title = {{SabreSwap}},
  howpublished = {\url{https://quantum.cloud.ibm.com/docs/en/api/qiskit/qiskit.transpiler.passes.SabreSwap}},
  note = {Accessed 19 October 2026}
}

@misc{IBMRuntimeScheduling,
  author = {{IBM Quantum}},
  title = {Scheduling passes, Qiskit IBM Runtime 0.43},
  howpublished = {\url{https://quantum.cloud.ibm.com/docs/en/api/qiskit-ibm-runtime/0.43/transpiler-passes-scheduling}},
  note = {Accessed 19 October 2026}
}
\end{document}

\begin{center}
\Large\bfseries
Supplementary Information: Parallel Teleportation via AME(5,2)
States on IBM Quantum Hardware
\end{center}
\vspace{0.8\baselineskip}

This Supplementary Material describes the AME resource and decoder
construction and their algebraic derivations (Secs.~I, II and III), hardware and diagnostic methods
(Secs.~IV and V), stage-resolved characterisation
(Sec.~VI), complete classical post-processing results
and supporting analyses (Secs.~VII and VIII), and a separate sequential feed-forward
experiment (Sec.~IX).
Unless otherwise stated, uncertainty
bars denote pointwise two-sided $99\%$ confidence intervals.
%add back stage-resolved interpretation after experiment

\section{Resource preparation and teleportation}
\label{sec:supp-resource}
This section explains how one shared absolutely maximally entangled
five-qubit state, $\mathrm{AME}(5,2)$, provides two parallel teleportation
channels. We describe resource preparation, relate the logical gates
to the processor's native gates, and derive how joint decoding,
Bell measurements and Pauli corrections transfer the two input states
to their assigned receivers.

\subsection*{AME states}

An absolutely maximally entangled (AME) state $\ket{\mathrm{AME}(n,d)}$ is a pure state of $n$ qudits of dimension $d$ that is maximally entangled across every bipartition~\cite{Helwig:2012nha}. Consider any subset $B$ of $m\le\lfloor n/2\rfloor$ parties, and let $A$ be the remaining $n-m$ parties, so that $A$ is the larger side. The state is AME if the reduced state of every such $B$ is maximally mixed,
\begin{equation}
\rho_B=\mathrm{Tr}_A\ket{\Psi}\!\bra{\Psi}=\frac{1}{d^{m}}\,\mathbb{I}_{d^{m}},\qquad 1\le m\le\Big\lfloor\frac n2\Big\rfloor ,
\label{supp-eq:ame-def}
\end{equation}
which means that its von Neumann entropy takes the largest possible value, $S(\rho_B)=m\log_2 d$. In practice, it is enough to check the $\binom{n}{\lfloor n/2\rfloor}$ marginals of size $\lfloor n/2\rfloor$, since a partial trace of the maximally mixed state is again maximally mixed, so all smaller marginals follow automatically. Equivalently, a state is AME if and only if it can be written as
\begin{equation}
\ket{\mathrm{AME}}=\frac{1}{\sqrt{d^{m}}}\sum_{k\in\mathbb Z_d^{m}}\ket{k_1}_{B_1}\cdots\ket{k_m}_{B_m}\,\ket{\phi( k)}_{A},
\label{supp-eq:ame-form}
\end{equation}
where $\braket{\phi(k)|\phi(k')}=\delta_{k,k'}$, for every choice of $B$ with $m=|B|\le n/2$. Equation~(\ref{supp-eq:ame-form}) says that the parties in $B$ are in a maximally entangled state with a $d^m$-dimensional subspace of the larger side $A$. It is this subspace that a joint operation on $A$ can later convert into $m$ Bell pairs~\cite{Helwig:2012nha}.

For $n=2$, the definition requires only that each qubit be maximally mixed, so the AME qubit state is the Bell (EPR) pair. For $n=3$ it is the GHZ state. AME qubit states also exist for $n=5$ and $n=6$, but not for $n=4$, and not for any $n\ge7$~\cite{PhysRevLett.118.200502}. This non-existence for $n \ge 7$ is specific to qubits ($d=2$); for qudit systems, $\text{AME}(n,d)$ states exist for arbitrary $n$ provided the local dimension $d$ is sufficiently large. The five-qubit state, AME$(5,2)$, is therefore the smallest AME qubit state beyond the textbook Bell and GHZ examples. Furthermore, $\text{AME}(5,2)$ forms the underlying state representation for the minimal, perfect $[[5,1,3]]$ quantum error-correcting code, serving as a bridge between quantum error correction and multi-party communication~\cite{Helwig:2012nha,Laflamme:1996iw}. This makes it a natural first candidate for a parallel teleportation experiment.

\subsection*{General Encoding}

A stabilizer state of \(n\) qubits is the unique common \(+1\)
eigenstate of \(n\) independent commuting Pauli stabilizer generators
\(g_1,\ldots,g_n\)~\cite{Gottesman:1997zz}. The reference state \(\ket{0}^{\otimes n}\) is
stabilized by the single-qubit Pauli operators
\begin{equation}
 Z_i=\mathbb I^{\otimes(i-1)}\otimes Z
     \otimes\mathbb I^{\otimes(n-i)},
 \qquad
 Z=\begin{pmatrix}1&0\\0&-1\end{pmatrix}.
\end{equation}
A target stabilizer state can therefore be prepared by a Clifford
circuit \(U\) satisfying
\begin{equation}
    UZ_iU^\dagger=g_i,\qquad i=1,\ldots,n.
\end{equation}

For an \([[n,m,d]]\) stabilizer code, take the first \(n-m\) qubits
as ancillas initialized in \(\ket{0}\) and the remaining \(m\) qubits
as the input. An encoding Clifford \(U_{\mathrm{enc}}\) satisfies
\begin{align}
 &U_{\mathrm{enc}}Z_iU_{\mathrm{enc}}^\dagger
   =g_i, \qquad i=1,\ldots,n-m,\\
 &U_{\mathrm{enc}}Z_{n-m+j}U_{\mathrm{enc}}^\dagger
   =\bar Z_j,\\
 &U_{\mathrm{enc}}X_{n-m+j}U_{\mathrm{enc}}^\dagger
   =\bar X_j,
   \qquad j=1,\ldots,m,
\end{align}
where \(\bar X_j\) and \(\bar Z_j\) are the logical Pauli operators of
the \(j\)th encoded qubit. Such circuits can be synthesized using
Hadamard, phase, and CNOT gates.

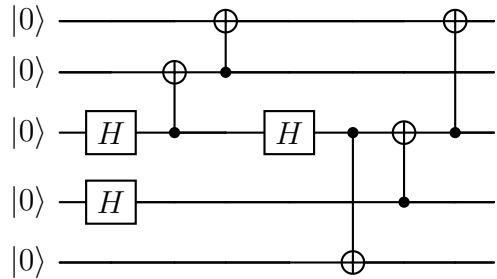
\begin{figure}[H]
    \centering
    \begin{quantikz}[
        row sep=0.35cm,
        column sep=0.35cm
    ]
        \lstick{\(\ket{0}\)}
            & \qw      & \qw       & \targ{}
            & \qw      & \qw       & \qw
            & \targ{}  & \qw \\
        \lstick{\(\ket{0}\)}
            & \qw      & \targ{}   & \ctrl{-1}
            & \qw      & \qw       & \qw
            & \qw      & \qw \\
        \lstick{\(\ket{0}\)}
            & \gate{H} & \ctrl{-1} & \qw
            & \gate{H} & \ctrl{2}  & \targ{}
            & \ctrl{-2}& \qw \\
        \lstick{\(\ket{0}\)}
            & \gate{H} & \qw       & \qw
            & \qw      & \qw       & \ctrl{-1}
            & \qw      & \qw \\
        \lstick{\(\ket{0}\)}
            & \qw      & \qw       & \qw
            & \qw      & \targ{}   & \qw
            & \qw      & \qw
    \end{quantikz}
    \caption{Circuit for preparing the AME\((5,2)\) state. The circuit
    contains five CNOT gates and three Hadamard gates and is optimized
    with respect to the minimisation of the number of entangling gates (obtained from figure 16 of~\cite{Cervera-Lierta:2019juk}).}
    \label{supp-fig:ame}
\end{figure}

The resource AME$(5,2)$ state used for teleportation is a limiting case of this
construction. The four stabilizer generators of the \([[5,1,3]]\)
code may be chosen as four cyclic shifts of \(XZZXI\). Adjoining the
logical operator
\begin{equation}
    \bar Z=ZZZZZ
\end{equation}
as a fifth stabilizer selects the logical state \(\ket{\bar 0}\) and
removes the remaining logical degree of freedom. The resulting one-dimensional stabilizer space is an AME\((5,2)\)
state. The circuit below prepares the chosen resource representative;
its relation to a standard five-qubit-code state is understood up to
local Clifford transformations. The
encoding circuit in Figure~\ref{supp-fig:ame} contains five CNOT and three Hadamard gates and has logical depth six.

It is optimal with respect to the minimisation of the number of
entangling gates.

On the \texttt{ibm\_pittsburgh} Heron r3
processor, the relevant native gate set is
\(\{R_z(\theta),\sqrt X,X,\mathrm{CZ}\}\). To make the relation between these
gates and rotations on the Bloch sphere explicit, define
\begin{equation}
 R_\alpha(\theta)
   =\exp\!\left(-\frac{i\theta}{2}\sigma_\alpha\right),
 \qquad \alpha\in\{x,y,z\}.
\end{equation}
Their single-qubit representations are
\begin{align}
 R_x(\theta)
 &=\begin{pmatrix}
 \cos(\theta/2)&-i\sin(\theta/2)\\
 -i\sin(\theta/2)&\cos(\theta/2)
 \end{pmatrix},\\
 R_y(\theta)
 &=\begin{pmatrix}
 \cos(\theta/2)&-\sin(\theta/2)\\
 \sin(\theta/2)&\cos(\theta/2)
 \end{pmatrix},\\
 R_z(\theta)
 &=\begin{pmatrix}
 e^{-i\theta/2}&0\\
 0&e^{i\theta/2}
 \end{pmatrix}.
\end{align}

Writing \(\sqrt X\) for the native \(SX\) gate \cite{IBMQuantumSXGate} and using
\(\widetilde{=}\) to denote equality up to a global phase, the logical gates
are synthesized as
\begin{align}
 &H\widetilde{=}
 R_z\!\left(\frac{\pi}{2}\right)
 \sqrt X\,
 R_z\!\left(\frac{\pi}{2}\right),\quad
 S\widetilde{=} R_z\!\left(\frac{\pi}{2}\right),\quad
 S^\dagger\widetilde{=} R_z\!\left(-\frac{\pi}{2}\right),\nonumber\\
 &Z\widetilde{=} R_z(\pi), \qquad
 \mathrm{CNOT}_{c\rightarrow t}=(\mathbb I_c\otimes H_t)\,
   \mathrm{CZ}_{c\rightarrow t}\,
   (\mathbb I_c\otimes H_t).
 \label{supp-eq:native}
\end{align}
More general rotations about the \(x\) and \(y\)-axes can be
expressed as
\begin{align}
 R_x(\theta)
 &\widetilde{=}
 R_z\!\left(\frac{\pi}{2}\right)\sqrt X\,
 R_z(\theta+\pi)\sqrt X\,
 R_z\!\left(\frac{\pi}{2}\right),\\
 R_y(\theta)
 &\widetilde{=}
 R_z(\pi)\sqrt X\,
 R_z(\theta+\pi)\sqrt X .
\end{align}

The \(S\), \(S^\dagger\), and \(Z\) gates are implemented as virtual
\(R_z\) frame updates: they require no additional microwave pulse and
are ordinarily assigned zero duration. In contrast, the above
decomposition of \(H\) contains one physical \(\sqrt X\) pulse, while
a general \(R_x\) or \(R_y\) rotation contains two. Before transpiler simplification, a logical CNOT therefore becomes
one native CZ together with two Hadamards on its target qubit.
We use Qiskit optimisation levels 2 and 3, the two highest available
settings~\cite{IBMTranspilerOptimization}. Both optimise single-qubit
gate sequences and cancel eligible self-inverse gate pairs, such as
$HH=\mathbb I$ and $\mathrm{CZ}^2=\mathbb I$, including through
commuting operations~\cite{IBMCommutativeCancellation}. Level 3
additionally resynthesises two-qubit blocks, potentially reducing
the native entangling gate count~\cite{IBMTranspilerOptimization}.
The executed circuit can therefore be simpler than a direct
gate-by-gate expansion of the logical circuit.

\subsection*{General Decoding and AME states for parallel teleportation}

In our setting, decoding means turning a multipartite state into a set of Bell pairs that can be used for teleportation, and then completing the teleportation with the classical outcomes of the senders' measurements. As in the previous subsection, every operation in the decoder is a Clifford gate, specified by how it maps Pauli operators to Pauli operators.

Following Theorem 1 of~\cite{Helwig:2012nha}, a joint unitary operation $U_A$
\begin{equation}
U_A\ket{\phi(k)}_A=\ket{k_1}_{A_1}\cdots\ket{k_m}_{A_m}\ket{0}_{A'},
\label{supp-eq:UA}
\end{equation}
can be used to transform the AME state into $m$ maximally entangled pairs $(B_i,A_i)$ and a leftover register $A'$ in $\ket 0$. Each pair then teleports one qudit in the standard way, where $B_i$ performs a generalised Bell measurement on its half of the pair and the qudit to be sent, and communicates two classical ``dits'' to $A_i$, which applies a local correction. In total, $m$ qudits are teleported by a single shared state.

For the AME\((5,2)\) protocol, choose the two sender qubits as
\(B=\{s_A,s_B\}\) and the three receiver qubits as
\(A=\{r_A,r_B,\ell\}\). Setting \(m=2\) and \(d=2\) in
equation~\eqref{supp-eq:ame-form} gives
\begin{equation}
 \ket{\mathrm{AME}(5,2)}
 =
 \frac{1}{2}
 \sum_{(k_1,k_2)\in\mathbb Z_2^2}
 \ket{k_1}_{s_A}\ket{k_2}_{s_B}
 \ket{\phi(k)}_{A},
 \label{supp-eq:ame-five-schmidt}
\end{equation}
where the four receiver states
\(\ket{\phi(k)}_A\) are orthonormal, as required. For the chosen sender-receiver
assignment, the three-qubit Clifford decoder \(V\) satisfies
\begin{equation}
 V\ket{\phi(k)}_{A}
 =
 \ket{k_1}_{r_A}\ket{k_2}_{r_B}\ket{0}_{\ell}.
 \label{supp-eq:decoder-action}
\end{equation}

It follows that
\begin{align}
 \left(\mathbb I_{B}\otimes V_{A}\right)
 \ket{\mathrm{AME}(5,2)}
 &=
 \frac{1}{2}
 \sum_{(k_1,k_2)\in\mathbb Z_2^2}
 \ket{k_1}_{s_A}\ket{k_1}_{r_A}
 \ket{k_2}_{s_B}\ket{k_2}_{r_B}
 \ket{0}_{\ell}
 \nonumber\\
 &=
 \ket{\Phi^+}_{s_A r_A}
 \ket{\Phi^+}_{s_B r_B}
 \ket{0}_{\ell},
 \label{supp-eq:decoded-ame}
\end{align}
where
\begin{equation}
 \ket{\Phi^+}
 =
 \frac{\ket{00}+\ket{11}}{\sqrt{2}}.
\end{equation}
Thus, \(V_A\) extracts two Bell pairs from the AME state,
while the remaining receiver qubit \(\ell\) is disentangled in
\(\ket{0}\). The decoded AME state therefore provides one Bell channel for each sender-receiver pair, reducing the remainder of the protocol to two parallel instances of standard quantum teleportation.

Let the input qubit \(x_c\) be
prepared in the state \(\ket{\psi_c}\), where \(c\in\{A,B\}\). A
Bell-basis measurement of \(x_c\) and \(s_c\) is implemented by
applying a CNOT from \(x_c\) to \(s_c\), followed by a Hadamard gate
on \(x_c\) and computational basis measurements of both qubits. If
the measurement outcomes of \(x_c\) and \(s_c\) are \(m_1^c\) and
\(m_2^c\), respectively, the receiver qubit is left in the state
\begin{equation}
 X_{r_c}^{m_2^c}Z_{r_c}^{m_1^c}\ket{\psi_c}.
 \label{supp-eq:teleportation-byproduct}
\end{equation}
The conditional Pauli correction
\begin{equation}
 C_c(m_1^c,m_2^c)
 =
 Z_{r_c}^{m_1^c}X_{r_c}^{m_2^c}
 \label{supp-eq:teleportation-correction}
\end{equation}
recovers \(\ket{\psi_c}\) on \(r_c\). Performing these operations for
both \(c=A\) and \(c=B\) transfers the input states from \(x_A,x_B\)
to \(r_A,r_B\), completing the deterministic two-channel
teleportation protocol.

All in all, the joint decoding converts the $\mathrm{AME}(5,2)$ resource into two Bell pairs and an unused qubit in $\ket{0}$. A Bell measurement and the corresponding Pauli correction then complete teleportation on each
channel. These identities establish the ideal two-channel protocol
implemented in the experiments.

\section{Decoder and their circuits}
\label{sec:supp-decoders}
This section describes the construction of compact three-qubit
Clifford decoders for the 60 inequivalent assignments of sender and
receiver roles within a shared five-qubit absolutely maximally
entangled resource, $\mathrm{AME}(5,2)$. We explain the
search over decoder completions, numerical verification and
logical cost ranking, and identify the rank 1 and rank 58 circuits
selected for the hardware experiments.

\begin{figure}[t]
\centering
\includegraphics[width=0.9\linewidth]{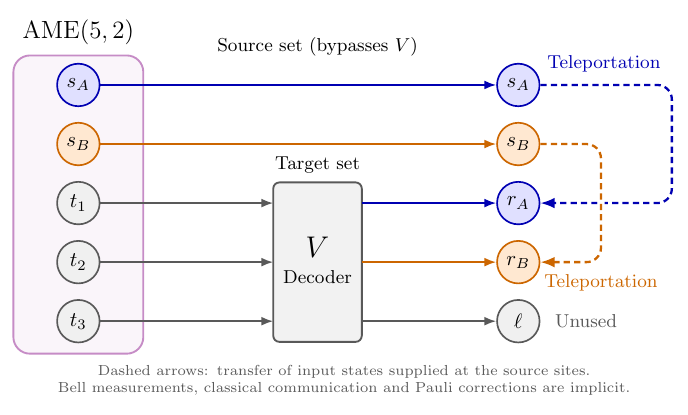}
\caption{Illustration of a specific role assignment \((s_A \rightarrow r_A, s_B \rightarrow r_B)\). The five physical qubits are partitioned into a source set \(\{s_A, s_B\}\), which bypasses the decoder \(V\) unchanged, and a target set consisting of the remaining three AME qubits, which are unitarily transformed by \(V\) into the receiver qubits \(r_A, r_B\) and an unused qubit \(\ell\).}
\label{supp-fig:assignment-illustration}
\end{figure}

To synthesize the decoders, we first establish the explicit allocation of sender and receiver roles for communication channels A and B within the five-qubit AME state. Specifically, an assignment is denoted by the mapping
\[
(s_A\rightarrow r_A,\;s_B\rightarrow r_B),
\]
where the source qubits \(s_A\) and \(s_B\) act as the senders for channels A and B, respectively, and are left unchanged. As illustrated in figure~\ref{supp-fig:assignment-illustration}, this assignment partitions the five physical qubits such that the source qubits bypass the decoding circuit entirely, while the decoder acts exclusively on the remaining three AME qubits to generate the two receiver qubits (\(r_A\) and \(r_B\)) and an unused qubit \(\ell\).

The four conditional states associated with the computational basis values of \(s_A\) and \(s_B\) constrain only a four-dimensional subspace of this three-qubit register. Consequently, the action on the orthogonal subspace is not unique. For each channel assignment, we enumerated all \(64\) Clifford completions \footnote{The four logical Pauli generators each have two equivalent representatives differing by the code stabilizer, giving \(2^4\) choices. Completing the Clifford tableau for the remaining qubit admits four signed choices for its complementary Pauli generator, hence \(2^4\times4=64\).} and synthesised the corresponding decoder using Clifford circuit synthesis~\cite{Bravyi:2021ijy}.

There are \(5!=120\) labelled assignments of two disjoint channels among five qubits. Interchanging the labels \(A\) and \(B\) leaves the physical decoder unchanged, reducing the search to \(60\) physically distinct assignments. Every candidate was verified against the full five-qubit relation
\begin{equation}
 V\ket{\mathrm{AME}(5,2)}
 =
 \ket{\Phi^+}_{s_A r_A}
 \ket{\Phi^+}_{s_B r_B}
 \ket{0}_{\ell},
 \label{supp-eq:decoder-verification}
\end{equation}
with unit state fidelity within numerical precision. For each assignment, candidate completions were ranked in order of two-qubit gate count, single-qubit gate count, and circuit depth. In particular, candidates were compared first by their two-qubit gate count, with ties resolved by single-qubit gate count and subsequently by circuit depth. This ordering prioritises the reduction of entangling gates, which are generally the most costly operations on the hardware.

The two decoders compared experimentally are the rank 1 decoder \(V_{014}\), which acts on AME qubits \(q_0,q_1,q_4\), and the rank 58 decoder \(V_{234}\), which acts on \(q_2,q_3,q_4\). Their channel assignments and logical circuit costs are summarised in table~\ref{supp-tab:decoder-comparison}. Here, \(s_c\) and \(r_c\) denote the source and receiver qubits of channel \(c\in\{A,B\}\), while \(\ell\) denotes the remaining qubit, which is disentangled in \(\ket{0}\) by the decoder. The quantities \(N_{\mathrm{2q}}\) and \(N_{\mathrm{1q}}\) are the logical two and single-qubit gate counts. The difference between their logical rankings therefore arises from the additional single-qubit gates and greater depth of \(V_{234}\), rather than from its two-qubit gate count.

\begin{table}[t]
 \centering
 \caption{Logical properties of the rank 1 and rank 58 decoders. The subscript of \(V\) identifies the three AME qubits on which the decoder acts.}
 \label{supp-tab:decoder-comparison}
 \begin{tabular}{c c c c c c c}
  \toprule
  Rank
  & Decoder
  & \((s_A,r_A)\)
  & \((s_B,r_B)\)
  & \(\ell\)
  & \((N_{\mathrm{2q}},N_{\mathrm{1q}})\)
  & Depth \\
  \midrule
  \(1\)
  & \(V_{014}\)
  & \((q_2,q_0)\)
  & \((q_3,q_1)\)
  & \(q_4\)
  & \((3,5)\)
  & \(6\) \\
  \(58\)
  & \(V_{234}\)
  & \((q_0,q_4)\)
  & \((q_1,q_3)\)
  & \(q_2\)
  & \((3,13)\)
  & \(10\) \\
  \bottomrule
 \end{tabular}
\end{table}

\begin{figure*}[t]
 \centering
 \begin{subfigure}[c]{0.38\textwidth}
  \centering
  \includegraphics[width=\linewidth]{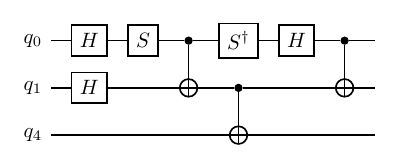}
  \caption{Rank 1 decoder \(V_{014}\).}
 \end{subfigure}
 \hfill
 \begin{subfigure}[c]{0.58\textwidth}
  \centering
  \includegraphics[width=\linewidth]{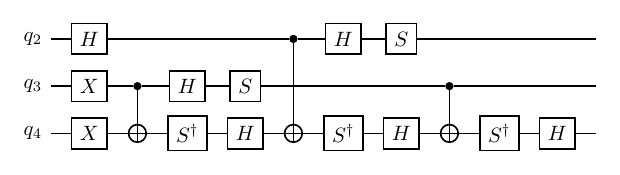}
  \caption{Rank 58 decoder \(V_{234}\).}
 \end{subfigure}
 \caption{Logical circuits for the two decoders compared experimentally. The circuits are read from left to right. A filled circle denotes the control and \(\oplus\) the target of a CNOT gate. The circuits in our figures are expressed in the gate set \(\{H,S,S^\dagger,X,Z,\mathrm{CNOT}\}\). The Pauli \(Y\) gate was not excluded from the synthesis; it simply does not occur in the rank 1 or rank 58 circuits considered here. Only the three active AME qubits are shown.}
 \label{supp-fig:decoder-circuits}
\end{figure*}

The corresponding logical circuits of the decoders are shown in figure~\ref{supp-fig:decoder-circuits}. Only the three qubits acted on by each decoder are displayed; the source qubits remain unchanged. Although both circuits contain three CNOT gates, their entangling patterns and single-qubit gates differ. The rank 1 decoder \(V_{014}\) uses the CNOT sequence
\[
 \mathrm{CNOT}_{0\rightarrow1},\qquad
 \mathrm{CNOT}_{1\rightarrow4},\qquad
 \mathrm{CNOT}_{0\rightarrow1},
\]
whereas the rank~58 decoder \(V_{234}\) uses
\[
 \mathrm{CNOT}_{3\rightarrow4},\qquad
 \mathrm{CNOT}_{2\rightarrow4},\qquad
 \mathrm{CNOT}_{3\rightarrow4},
\]

 where the complete algebraic derivation of equation (\ref{supp-eq:decoder-verification}) of the two decoders are shown in the Supplementary materials.

Having identified the two decoder circuits to be compared, we next evaluate how their performance propagates through repeated uses of the parallel teleportation protocol. This is examined by cascading three teleportation rounds and monitoring the two channels after each round.

\section{Algebraic verification of the preparation and decoders}
\label{sec:supp-algebra}
This section explicitly verifies that the circuit in Figure~S1
prepares the claimed AME$(5,2)$ resource and that the decoder
circuits in Figure~S3 produce the assigned Bell pairs.
Starting from $\ket{0}^{\otimes5}$, we derive the prepared state
and show how $V_{014}$ (rank~1) and $V_{234}$ (rank~58) extract
their respective Bell pairs and disentangle the unused qubit
in $\ket{0}$. Qubit ordering and phase conventions are kept
explicit throughout.

Throughout this section, all binary additions are modulo \(2\), and
\(\oplus\) denotes addition in \(\mathbb{Z}_2\). We use
\begin{equation}
 H\ket{u}
 =
 \frac{1}{\sqrt{2}}
 \sum_{v\in\mathbb{Z}_2}
 (-1)^{uv}\ket{v},
 \qquad
 S\ket{u}=\mathrm{i}^{\,u}\ket{u},
 \label{supp-eq:H-S-actions}
\end{equation}
together with
\begin{equation}
 \sum_{u\in\mathbb{Z}_2}(-1)^{ut}
 =
 2\delta_{t,0},
 \qquad t\in\mathbb{Z}_2.
 \label{supp-eq:binary-character-identity}
\end{equation}

\subsubsection*{AME preparation}

Beginning with
\(\ket{\psi_0}=\ket{0}^{\otimes 5}\), the preparation circuit applies

\begin{align*}
 &H_2H_3,\quad \mathrm{CNOT}_{2\rightarrow1},
 \quad \mathrm{CNOT}_{1\rightarrow0},\\
 &H_2,\quad \mathrm{CNOT}_{2\rightarrow4},
 \quad \mathrm{CNOT}_{3\rightarrow2},
 \quad \mathrm{CNOT}_{2\rightarrow0},
\end{align*}
in the order listed.

Here, the computational-basis kets are ordered as
\(q_0q_1q_2q_3q_4\), and all binary indices belong to
\(\mathbb{Z}_2=\{0,1\}\).

The initial Hadamard gates produce
\begin{align}
\ket{\psi_1}
&=
H_2H_3\ket{\psi_0}
\nonumber\\
&=
\frac{1}{2}
\sum_{i,j\in\mathbb{Z}_2}
\ket{0,0,i,j,0}.
\end{align}
The following two CNOT gates copy \(i\) from \(q_2\) to \(q_1\) and
then from \(q_1\) to \(q_0\):
\begin{align}
\ket{\psi_2}
&=
\mathrm{CNOT}_{2\rightarrow1}\ket{\psi_1}
\nonumber\\
&=
\frac{1}{2}
\sum_{i,j\in\mathbb{Z}_2}
\ket{0,i,i,j,0},
\\[3pt]
\ket{\psi_3}
&=
\mathrm{CNOT}_{1\rightarrow0}\ket{\psi_2}
\nonumber\\
&=
\frac{1}{2}
\sum_{i,j\in\mathbb{Z}_2}
\ket{i,i,i,j,0}.
\end{align}

Using
\[
H\ket{i}
=
\frac{1}{\sqrt{2}}
\sum_{k\in\mathbb{Z}_2}
(-1)^{ik}\ket{k},
\]
the second Hadamard gate on \(q_2\) gives
\begin{align}
\ket{\psi_4}
&=
H_2\ket{\psi_3}
\nonumber\\
&=
\frac{1}{2\sqrt{2}}
\sum_{i,j,k\in\mathbb{Z}_2}
(-1)^{ik}
\ket{i,i,k,j,0}.
\end{align}

The remaining CNOT gates act successively as
\begin{align}
\ket{\psi_5}
&=
\mathrm{CNOT}_{2\rightarrow4}\ket{\psi_4}
\nonumber\\
&=
\frac{1}{2\sqrt{2}}
\sum_{i,j,k\in\mathbb{Z}_2}
(-1)^{ik}
\ket{i,i,k,j,k},
\\[3pt]
\ket{\psi_6}
&=
\mathrm{CNOT}_{3\rightarrow2}\ket{\psi_5}
\nonumber\\
&=
\frac{1}{2\sqrt{2}}
\sum_{i,j,k\in\mathbb{Z}_2}
(-1)^{ik}
\ket{i,i,j\oplus k,j,k},
\\[3pt]
\ket{\psi_7}
&=
\mathrm{CNOT}_{2\rightarrow0}\ket{\psi_6}
\nonumber\\
&=
\frac{1}{2\sqrt{2}}
\sum_{i,j,k\in\mathbb{Z}_2}
(-1)^{ik}
\ket{i\oplus j\oplus k,i,j\oplus k,j,k}.
\end{align}
For the final CNOT, the control qubit \(q_2\) contains
\(j\oplus k\); hence the target bit on \(q_0\) changes from \(i\) to
\(i\oplus j\oplus k\). Therefore,
\begin{equation}
\begin{split}
&\ket{\mathrm{AME}(5,2)}
=
\ket{\psi_7}
\\
&=
\frac{1}{2\sqrt{2}}
\sum_{i,j,k\in\mathbb{Z}_2}
(-1)^{ik}
\ket{i\oplus j\oplus k}_0
\ket{i}_1
\ket{j\oplus k}_2
\ket{j}_3
\ket{k}_4 .
\end{split}
\label{supp-eq:ame52-algebraic}
\end{equation}

\subsubsection*{Rank 1 Decoder}

The rank~1 decoder acts on qubits \((q_0,q_1,q_4)\). Reading its
circuit from left to right gives
\begin{equation}
V_{014}
=
\mathrm{CNOT}_{0\rightarrow1}
H_0S_0^\dagger
\mathrm{CNOT}_{1\rightarrow4}
\mathrm{CNOT}_{0\rightarrow1}
S_0H_0H_1,
\label{supp-eq:V014-product}
\end{equation}
where the rightmost operation acts first, where $S_0\ket{x}_0=e^{i\frac{\pi}{2}x}\ket{x}_0$.

To evaluate its action, introduce binary variables \(x\) and \(y\) for
the first Hadamard gates on \(q_0\) and \(q_1\), and \(z\) for the
second Hadamard gate on \(q_0\). The phases introduced by \(S_0\) and
\(S_0^\dagger\) cancel because \(q_0\) retains the value \(x\) between
these two gates. Direct application of the circuit to
equation \eqref{supp-eq:ame52-algebraic} therefore gives
\begin{align}
V_{014}&\ket{\mathrm{AME}(5,2)}
=
\frac{1}{8}
\sum_{\substack{i,j,k\\x,y,z\in\mathbb{Z}_2}}
(-1)^{
ik+(i\oplus j\oplus k)x+iy+xz
}
\nonumber\\
&\quad\times
\ket{z}_0
\ket{y\oplus x\oplus z}_1
\ket{j\oplus k}_2
\ket{j}_3
\ket{k\oplus y\oplus x}_4 .
\label{supp-eq:rank1-expanded}
\end{align}
The terms containing \(i\) can be collected as
\[
(-1)^{i(k\oplus x\oplus y)}.
\]
Summing over \(i\) and using
equation \eqref{supp-eq:binary-character-identity} imposes
\[
y=x\oplus k.
\]
Equation~\eqref{supp-eq:rank1-expanded} consequently becomes
\begin{align}
&V_{014}\ket{\mathrm{AME}(5,2)}\notag\\
&=
\frac{1}{4}
\sum_{j,k,x,z}
(-1)^{x(j\oplus k\oplus z)}
\ket{z}_0
\ket{k\oplus z}_1
\ket{j\oplus k}_2
\ket{j}_3
\ket{0}_4 .
\end{align}
The sum over \(x\) similarly imposes \(z=j\oplus k\), yielding
\begin{align}
V_{014}\ket{\mathrm{AME}(5,2)}
&=
\frac{1}{2}
\sum_{j,k}
\ket{j\oplus k}_0
\ket{j}_1
\ket{j\oplus k}_2
\ket{j}_3
\ket{0}_4 .
\end{align}
Defining \(a=j\oplus k\) and \(b=j\), this becomes
\begin{align}
V_{014}\ket{\mathrm{AME}(5,2)}
&=
\frac{1}{2}
\sum_{a,b\in\mathbb{Z}_2}
\ket{a}_0\ket{b}_1\ket{a}_2\ket{b}_3\ket{0}_4
\nonumber\\
&=
\ket{\Phi^+}_{q_2q_0}
\ket{\Phi^+}_{q_3q_1}
\ket{0}_{q_4}.
\label{supp-eq:rank1-decoded}
\end{align}
Thus, for rank~1,
\[
(s_A,r_A)=(q_2,q_0),\qquad
(s_B,r_B)=(q_3,q_1),\qquad
\ell=q_4.
\]

\subsubsection*{Rank 58 Decoder}

The rank~58 decoder acts on qubits \((q_2,q_3,q_4)\). Its circuit
corresponds to
\begin{align}
V_{234}
={}&
H_4S_4^\dagger
\mathrm{CNOT}_{3\rightarrow4}
(S_2H_4)
(H_2S_4^\dagger)
\mathrm{CNOT}_{2\rightarrow4}
\nonumber\\
&\times
(S_3H_4)
(H_3S_4^\dagger)
\mathrm{CNOT}_{3\rightarrow4}
(H_2X_3X_4),
\label{supp-eq:V234-product}
\end{align}
again with the rightmost operation applied first.

For this decoder, define
\begin{equation}
a=i\oplus j\oplus k,
\qquad
b=i.
\end{equation}
It follows that
\[
j\oplus k=a\oplus b,
\qquad
k=a\oplus b\oplus j.
\]
Equation~\eqref{supp-eq:ame52-algebraic} can then be regrouped as
\begin{equation}
\ket{\mathrm{AME}(5,2)}
=
\frac{1}{2}
\sum_{a,b\in\mathbb{Z}_2}
\ket{a}_0\ket{b}_1
\ket{\chi_{ab}}_{234},
\label{supp-eq:ame-rank58-decomposition}
\end{equation}
where
\begin{equation}
\ket{\chi_{ab}}_{234}
=
\frac{1}{\sqrt{2}}
\sum_{j\in\mathbb{Z}_2}
(-1)^{b(a\oplus b\oplus j)}
\ket{a\oplus b,j,a\oplus b\oplus j}_{234}.
\label{supp-eq:chi-ab}
\end{equation}
The four states appearing in this decomposition are
\begin{align}
\ket{\chi_{00}}
&=
\frac{\ket{000}+\ket{011}}{\sqrt{2}},
&
\ket{\chi_{01}}
&=
\frac{-\ket{101}+\ket{110}}{\sqrt{2}},
\\
\ket{\chi_{10}}
&=
\frac{\ket{101}+\ket{110}}{\sqrt{2}},
&
\ket{\chi_{11}}
&=
\frac{\ket{000}-\ket{011}}{\sqrt{2}}.
\end{align}
Applying the gate sequence in equation \eqref{supp-eq:V234-product} gives
\begin{align}
V_{234}\ket{\chi_{00}}
&=
e^{-\mathrm{i}\pi/4}\ket{000}_{234},
&
V_{234}\ket{\chi_{01}}
&=
e^{-\mathrm{i}\pi/4}\ket{010}_{234},
\\
V_{234}\ket{\chi_{10}}
&=
e^{-\mathrm{i}\pi/4}\ket{001}_{234},
&
V_{234}\ket{\chi_{11}}
&=
e^{-\mathrm{i}\pi/4}\ket{011}_{234}.
\end{align}
Equivalently,
\begin{equation}
V_{234}\ket{\chi_{ab}}
=
e^{-\mathrm{i}\pi/4}
\ket{0,b,a}_{234}.
\label{supp-eq:rank58-chi-map}
\end{equation}
Substitution into equation (\eqref{supp-eq:ame-rank58-decomposition}) gives
\begin{align}
V_{234}\ket{\mathrm{AME}(5,2)}
&=
\frac{e^{-\mathrm{i}\pi/4}}{2}
\sum_{a,b\in\mathbb{Z}_2}
\ket{a}_0\ket{b}_1\ket{0}_2\ket{b}_3\ket{a}_4
\nonumber\\
&=
e^{-\mathrm{i}\pi/4}
\ket{\Phi^+}_{q_0q_4}
\ket{\Phi^+}_{q_1q_3}
\ket{0}_{q_2}.
\label{supp-eq:rank58-decoded-phase}
\end{align}
After discarding the physically irrelevant global phase
\(e^{-\mathrm{i}\pi/4}\),
\begin{equation}
V_{234}\ket{\mathrm{AME}(5,2)}
=
\ket{\Phi^+}_{q_0q_4}
\ket{\Phi^+}_{q_1q_3}
\ket{0}_{q_2}.
\label{supp-eq:rank58-decoded}
\end{equation}
Thus, for rank~58,
\[
(s_A,r_A)=(q_0,q_4),\qquad
(s_B,r_B)=(q_1,q_3),\qquad
\ell=q_2.
\]

All in all, the calculation recovers Bell pairs $(q_2,q_0)$ and $(q_3,q_1)$
for rank~1, with $q_4$ unused, and Bell pairs $(q_0,q_4)$ and
$(q_1,q_3)$ for rank~58, with $q_2$ unused. In both cases the
unused qubit is disentangled in $\ket{0}$ and the output is two
$\ket{\Phi^+}$ pairs, up to an irrelevant global phase. This
explicitly verifies the two ideal decoder mappings used in the
hardware experiments.

\section{Hardware, acquisition and statistical conventions}
\label{sec:supp-hardware}

This section specifies the hardware acquisition and statistical
procedures used to report the experimental results on
\texttt{ibm\_pittsburgh}. We state the shot budget and eight-trial
sampling convention, explain how confidence intervals are calculated,
and distinguish within-trial comparisons from comparisons between
independent decoder trial pools.

The experiments were run on the 156-qubit superconducting processor
\texttt{ibm\_pittsburgh}. Mid-circuit Bell measurements use the
processor's \texttt{MidCircuitMeasure} instruction
(\texttt{measure\_2}). The main-text classical implementation does
not execute intermediate conditional Pauli corrections. The sequential
feed-forward method in Sec.~IX below executes those corrections.

Each tomography measurement setting was sampled with 5000 shots,
and the M3 calibration circuits used 5000 shots. Each reported
setting contains eight selected trials. For a set of eight trial estimates $f_i$, the reported interval is
 \begin{equation}
 \bar f \pm t_{0.995,7}\frac{s_f}{\sqrt8},
 \qquad
 s_f^2=\frac1{7}\sum_{i=1}^8(f_i-\bar f)^2.
 \label{supp-eq:mean-ci}
\end{equation}
Matched DD comparisons and channel differences use the eight
within-trial differences in place of $f_i$. Between-rank comparisons
use independent trial pools; their intervals use the unequal-variance
standard error and Welch-Satterthwaite degrees of freedom.
Baseline subtraction is always performed within a rank and trial
before comparing the changes. This preserves the covariance with
the same trial's $R0$ estimate. Trial-level intervals are not
shot-level binomial intervals and are not adjusted for simultaneous
testing across all conditions.

The cascade data were collected over several days, with reported median two-qubit-gate and readout errors of approximately $1.70\times10^{-3}$ and $5.00\times10^{-3}$, respectively. The separate stage-resolved acquisition was performed on a single day, with reported medians of $1.62\times10^{-3}$ and $4.39\times10^{-3}$.

The statistical analysis compares dynamical decoupling sequences and channels within
the same trials, subtracts each trial's own $R0$ baseline, and treats
the two decoder trial pools as independent. Each pointwise $99\%$
confidence interval applies to the mean of eight selected trials for
one condition; it does not certify every trial or all conditions
simultaneously. Calibration drift also remains relevant when
interpreting comparisons across the multi-day cascade acquisition.

\section{Diagnostic definitions and Methodologies}
\label{sec:supp-diagnostics}
This section defines what each diagnostic measures and explains how
the circuits are mapped, scheduled and analysed. We distinguish
reference-receiver entanglement fidelity from tomography fidelity,
SWAP-test pass rates and stabilizer-generator fidelity lower bounds,
and describe layout selection, dynamical decoupling (DD), readout
mitigation and classical Pauli frame tracking.

\subsection*{Performance Metrics}

The main text uses entanglement fidelity as the primary measure of teleportation performance, quantifying how well each channel preserves entanglement with a reference qubit. For completeness, the Supplementary Material presents complementary diagnostics using quantum state tomography, SWAP tests and stabilizer-generator witnesses. These respectively reconstruct quantum states, estimate state overlaps and provide fidelity lower bounds, and are applied at the relevant stages of AME preparation, decoding or teleportation.

\subsubsection*{Entanglement fidelity}

The entanglement fidelity of a quantum channel \(\mathcal N\) is
\begin{equation}
 F_{\mathrm e}(\mathcal N)
 =
 \bra{\Phi^+}
 (\mathbb{I}\otimes\mathcal N)
 \left(\ket{\Phi^+}\!\bra{\Phi^+}\right)
 \ket{\Phi^+}.
 \label{supp-eq:Fe}
\end{equation}
It measures how well the channel preserves a maximally entangled
state when acting on one half of it~\cite{Schumacher:1996dw}. Operationally, one prepares a Bell pair, sends one qubit through the channel, and
compares the resulting two-qubit state \(\rho\) with
\(\ket{\Phi^+}\):
\begin{equation}
 F_{\mathrm e}
 =
 \bra{\Phi^+}\rho\ket{\Phi^+}.
 \label{supp-eq:F}
\end{equation}
The ideal value is \(F_{\mathrm e}=1\). An entanglement fidelity greater
than \(1/2\) certifies that the two qubits are entangled, since no
separable state can have a larger overlap with
\(\ket{\Phi^+}\) \cite{Horodecki:1998zh}. For a qubit channel, the average input state
fidelity is related to the entanglement fidelity by
\begin{equation}
 F_{\mathrm{avg}}
 =
 \frac{2F_{\mathrm e}+1}{3}.
\end{equation}
Consequently, \(F_{\mathrm e}=1/2\) corresponds to the classical
teleportation threshold \(F_{\mathrm{avg}}=2/3\) \cite{Horodecki:1998zh,Nielsen:2002obk}. Because $\ket{\Phi^+}$ is the unique simultaneous $+1$ eigenstate of $X\otimes X$ and $Z\otimes Z$, its projector can be expanded via its stabilizer generators as
\begin{equation}
\begin{split}
\ket{\Phi^+}\bra{\Phi^+}& = \frac{1}{4}(I\otimes I + X\otimes X)(I\otimes I + Z\otimes Z) \\
&= \frac{1}{4}(I\otimes I + X\otimes X - Y\otimes Y + Z\otimes Z),
\end{split}
\end{equation}
where $(X\otimes X)(Z\otimes Z) = -Y\otimes Y$. Taking the expectation value over $\rho$ expresses the entanglement fidelity directly in terms of two-body Pauli correlations:
\begin{equation}
F_{\mathrm e} = \frac{1 + \langle XX\rangle_\rho - \langle YY\rangle_\rho + \langle ZZ\rangle_\rho}{4},
\label{supp-eq:bell-correlation-fidelity}
\end{equation}
with $\langle PP\rangle_\rho = \operatorname{Tr}[\rho(P\otimes P)]$ \cite{Siddhu:2025jsc}. 

\subsubsection*{State tomography fidelity}

Quantum state tomography reconstructs a density matrix from
measurements in several bases~\cite{James:2001klt}. Local Pauli tomography of \(n\) qubits
uses the \(3^n\) combinations of the \(X\), \(Y\), and \(Z\)
measurement bases. Once the density matrix \(\rho\) has been
reconstructed, its fidelity with a pure target state \(\ket{T}\) is
\begin{equation}
 F_T
 =
 \bra{T}\rho\ket{T}.
 \label{supp-eq:FT}
\end{equation}
For \(\ket{T}=\ket{\Phi^+}\), this reduces to the same overlap used
in equation~\eqref{supp-eq:F}. However, their physical interpretations differ. Equation \eqref{supp-eq:FT} characterises state preparation accuracy, whereas equation \eqref{supp-eq:Fe} evaluates dynamic channel transmission across an entangled subsystem. Measurement circuits and density matrix fitting were generated using the \texttt{StateTomography} module in Qiskit Experiments~\cite{Kanazawa2023}. Although state tomography provides full information about $\rho$, the required settings scale exponentially as $3^n$. Tomography on reduced subsystems significantly reduces measurement overhead, albeit at the cost of tracing out inter-register correlations.

\subsubsection*{SWAP test}

\begin{figure}[t]
    \centering
    \begin{quantikz}[
        row sep=0.35cm,
        column sep=0.45cm
    ]
        \lstick{\(\ket{0}\)}
            & \gate{H}
            & \ctrl{1}
            & \gate{H}
            & \meter{}
            & \cw \\
        \lstick{\(\ket{\phi}\)}
            & \qw
            & \swap{1}
            & \qw
            & \qw
            & \qw \\
        \lstick{\(\ket{\psi}\)}
            & \qw
            & \targX{}
            & \qw
            & \qw
            & \qw
    \end{quantikz}
    \caption{Circuit for Quantum SWAP test.}
    \label{supp-fig:swap-test}
\end{figure}
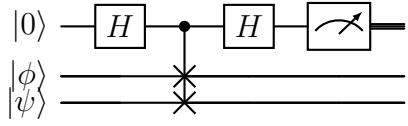

The SWAP test estimates the overlap of two states without
reconstructing their density matrices~\cite{Barenco:1996kz,Buhrman:2001rma}. An ancilla
is prepared in \(\ket{+}\), used to control SWAP operations between
the two registers, and then measured after a final Hadamard gate as shown in Figure~\ref{supp-fig:swap-test}.

If the registers contain independently prepared states \(\rho\) and
\(\sigma\), the probability of measuring the ancilla in \(0\) is
\begin{equation}
 P(0)
 =
 \frac{1}{2}
 \left(1+\operatorname{Tr}(\rho\sigma)\right).
 \label{supp-eq:swap}
\end{equation}
For pure states \(\ket{\psi}\) and \(\ket{\varphi}\), this becomes
\begin{equation}
 P(0)
 =
 \frac{1}{2}
 \left(1+\left|\braket{\psi|\varphi}\right|^2\right).
\end{equation}

We consider using a SWAP test against a reference
register initialized in the ideal target state
\(\sigma=\ket{T}\!\bra{T}\). For an experimentally prepared state
\(\rho\), the probability of measuring the SWAP test ancilla in
\(\ket{0}\) is
\begin{equation}
 P(0)=\frac{1+\operatorname{Tr}(\rho\sigma)}{2}
     =\frac{1+\bra{T}\rho\ket{T}}{2}.
\end{equation}
The target-state fidelity is therefore obtained as
\begin{equation}
 F_T=\bra{T}\rho\ket{T}=2P(0)-1.
\end{equation}
Thus, the SWAP test pass rate directly estimates the overlap between
the experimentally prepared state and its intended target.

The SWAP test requires fewer measurement settings than tomography,
but its controlled-SWAP gates increase the depth and width of the
circuit. Its result may therefore also be affected by errors in the
test circuit itself.

\subsubsection*{Stabilizer-generator witness}

A more economical fidelity bound is available when the target
\(\ket{T}\) is a stabilizer state. Let
\(g_1,\ldots,g_n\) be independent generators of its stabilizer group.
The target projector can be written as
\begin{equation}
 \ket{T}\!\bra{T}
 =
 \prod_{i=1}^{n}\frac{\mathbb I+g_i}{2}.
\end{equation}
Since the corresponding projectors commute, the fidelity satisfies
\begin{equation}
 F_T
 \geq
 1-\sum_{i=1}^{n}
 \frac{1-\langle g_i\rangle}{2}.
 \label{supp-eq:witness}
\end{equation}
This bound requires evaluating expectation values across only $n$ generator settings rather than the $3^n$ settings of full Pauli tomography. Expectation values $\langle g_i\rangle$ are obtained by applying local basis rotations and evaluating the parity of the measurement outcomes. Because individual generator deficits are summed linearly, equation \eqref{supp-eq:witness} provides a conservative lower bound that may fall below the true fidelity. When this lower bound exceeds the maximum fidelity achievable by separable states, it serves as a rigorous, quantitative entanglement witness \cite{Toth:2005qcv}. Together, these four diagnostic metrics provide a complementary characterisation: tomography yields full state information, the SWAP test measures global overlap directly, the stabilizer witness provides an efficient lower bound, and the entanglement fidelity assesses dynamic channel performance.

In the experiment, the reference-receiver entanglement-fidelity
estimate includes imperfections in the input pair, retained reference
and measurements. The formal channel entanglement fidelity assumes
an ideal reference/input preparation. We therefore assess experimental entanglement fidelity losses relative to
the measured $R0$ baselines.
The relation $F_{\mathrm{avg}}=(2F_{\mathrm e}+1)/3$ concerns a
trace-preserving qubit channel with the appropriate ideal channel
entanglement fidelity; it is not an independent SPAM correction
of the measured overlaps.

\subsection*{Error mitigation and suppression techniques}
\label{sec:supp-classical}

To mitigate physical hardware noise, we implement a multi-layered error management framework. Subgraph isomorphism layout ranking (\texttt{mapomatic}) and open-loop control (dynamical decoupling) act during hardware selection and circuit execution to suppress noise~\cite{PRXQuantum.4.010327,Souza:2011vcv}, while matrix-free readout mitigation (\texttt{mthree}) and mitigated state tomography (\texttt{MST}) post-correct measurement distributions~\cite{PRXQuantum.4.010327,PhysRevA.103.042605}.

\subsubsection*{Mapomatic}

Physical qubit and gate error rates vary across superconducting quantum processors. Selecting a layout based solely on isolated single- and two-qubit gate calibrations is often insufficient, as routing an unmapped circuit onto a fixed device coupling map can introduce high-overhead SWAP gates. Mapomatic addresses this routing trade-off after compilation, when the required entangling connectivity graph is established~\cite{PRXQuantum.4.010327}. It constructs the interaction graph of the compiled circuit and identifies all isomorphic subgraphs on the physical processor coupling map. Each matching layout $M$ is then assigned a heuristic error score:
\begin{equation}
\mathcal E(M)=1-\prod_{g\in\mathcal C}\left[1-\epsilon_g(M)\right],\label{supp-eq:mapomatic-score}\end{equation}where $\mathcal C$ represents the set of compiled circuit instructions and $\epsilon_g(M)$ is the calibrated gate or readout error of instruction $g$ on layout $M$. The layout minimizing $\mathcal E(M)$ is selected for execution without altering the compiled routing structure.

\subsubsection*{Dynamical decoupling: XY4 and XX}

Idle qubits accumulate phase noise and crosstalk during execution delays. Open-loop dynamical decoupling suppresses this decoherence by inserting high-frequency control pulse sequences into idle intervals, reducing sensitivity to selected system-environment interactions over a cycle~\cite{Souza:2011vcv}. We compare two pulse sequences:

\begin{equation}
\mathrm{XX}=(X,X),\qquad\mathrm{XY4}=(X,Y,X,Y).
\label{supp-eq:dd-sequences}
\end{equation}

The XX sequence targets slowly varying $Z$-dephasing, whereas the alternating axes of the XY4 sequence compensate for broader system-environment interactions and pulse rotation errors. For an idle duration $T_{\mathrm{idle}}$ and cumulative sequence duration $T_{\mathrm{seq}}$, the remaining slack delay $\Delta = T_{\mathrm{idle}} - T_{\mathrm{seq}}$ is distributed symmetrically around the pulses according to the fraction schedules:

\begin{align}
\mathrm{XX}:&\qquad\left(\frac14,\frac12,\frac14\right),\\
\mathrm{XY4}:&\qquad\left(\frac18,\frac14,\frac14,\frac14,\frac18\right).
\label{supp-eq:dd-spacing}
\end{align}

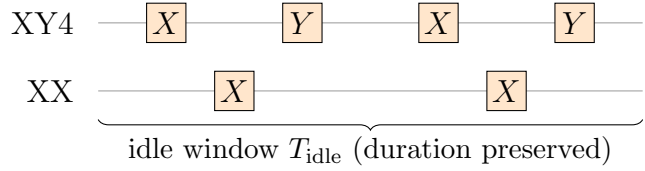
\begin{figure}[t]
\centering
\begin{tikzpicture}[x=0.72cm,y=0.9cm,pulse/.style={draw,fill=orange!20,minimum width=0.52cm,minimum height=0.52cm,inner sep=0pt}]
\node[anchor=east] at (-0.25,1) {$\mathrm{XY4}$};\draw[gray!75] (0,1)--(10,1);\node[pulse] at (1.25,1) {$X$};\node[pulse] at (3.75,1) {$Y$};\node[pulse] at (6.25,1) {$X$};\node[pulse] at (8.75,1) {$Y$}; 
 \node[anchor=east] at (-0.25,0) {\(\mathrm{XX}\)};
 \draw[gray!75] (0,0)--(10,0);
 \node[pulse] at (2.50,0) {\(X\)};
 \node[pulse] at (7.50,0) {\(X\)};

 \draw[
     decorate,
     decoration={brace,mirror,amplitude=4pt}
 ] (0,-0.42)--(10,-0.42);
 \node at (5,-0.88)
     {\small idle window \(T_{\mathrm{idle}}\) (duration preserved)};
\end{tikzpicture}
\caption{Symmetric pulse placement for XX and XY4 dynamical decoupling sequences within an idle window $T_{\mathrm{idle}}$.}
\label{supp-fig:dd-placement}
\end{figure}

The spacing preserves the total idle-window duration
$T_{\mathrm{idle}}$ (Figure~\ref{supp-fig:dd-placement}).
DD is scheduled after compilation using
\texttt{ALAPScheduleAnalysis}, \texttt{PadDelay} and
\texttt{PadDynamical\allowbreak Decoupling}.
In this implementation, eligible top-level delays receive pulses
subject to backend timing alignment. Conditional blocks, terminal
readout and the excluded post-measurement regions are not padded.
A missing native $Y$ is implemented by a physical $X$ pulse and
a virtual update, $Y=R_z(\pi)X$.

We test Qiskit optimisation levels 2 and 3. Level 2 includes
commutation-based cancellation; level 3 performs more extensive
two-qubit block resynthesis~\cite{IBMTranspilerOptimization}.
The level changes compilation effort and can reshape the physical
depth and idle windows. It does not specify a DD sequence or guarantee
higher hardware fidelity. DD can suppress some idle-time noise but
cannot reverse energy relaxation and can add pulse errors.
The no-DD, XX and XY4 comparisons therefore remain empirical.

\subsubsection*{Readout error mitigation: M3 and MST}

Measurement readout noise is modelled by an assignment matrix $A$ with elements $A_{yx} = \Pr(y\mid x)$, mapping ideal probability distributions $p$ to observed distributions $f = Ap$. Matrix-free measurement mitigation (\texttt{mthree} or M3) bypasses full $2^n \times 2^n$ matrix inversion by solving for corrected distributions directly within the subspace of bit strings present in the experimental data~\cite{Nation:2021kye}. Under a tensor-product readout model $A \approx \bigotimes_i A_i$, single-qubit assignment matrices are calibrated as\begin{equation}A_i =\begin{pmatrix}\Pr(0\mid0) & \Pr(0\mid1)\\Pr(1\mid0) & \Pr(1\mid1)\end{pmatrix}.\label{supp-eq:local-assignment}\end{equation}Linear expectation values are evaluated directly from the resulting quasi-probability distribution. When physical density matrix reconstruction requires non-negative probabilities, the distribution is projected onto the nearest probability simplex using \texttt{nearest\_probability\_distribution}.For full state reconstruction, Mitigated State Tomography (\texttt{MST}) integrates single-qubit readout calibration (\texttt{LocalReadoutError}) directly into the tomographic maximum-likelihood fit within Qiskit Experiments~\cite{PhysRevA.103.042605}. Readout calibration requires $2$ calibration circuits ($\ket{0}^{\otimes n}$ and $\ket{1}^{\otimes n}$) alongside $3^n$ Pauli measurement settings, totaling $3^n+2$ configurations~\cite{QiskitMSTDocs}.

\subsection*{Reference tracking and overlapping operations}
At $R0$, the reference-input pairs $(R_A,I_A)$ and $(R_B,I_B)$
are prepared in $\ket{\Phi^+}$ with the same $H$-CNOT circuit.
Each round uses a separate five-qubit AME resource and the selected
decoder. Logical wiring identifies $I_c^{(R+1)}=O_c^{(R)}$,
while the reference qubits are retained.

For one channel and three rounds, the Bell measurements act on
the disjoint pairs
\[
 (I_c,s_c^{(1)}),\qquad
 (o_c^{(1)},s_c^{(2)}),\qquad
 (o_c^{(2)},s_c^{(3)}).
\]
Without intermediate physical corrections, these measurements need
not wait for earlier Bell outcomes. Resource preparation and decoding
can overlap across registers. The final observables use the accumulated
Pauli frame and the necessary Bell history. That history is retained
until the frame-corrected parities have been evaluated.

Each round-count circuit is constructed
and independently compiled using Qiskit's \texttt{transpile}
function. We specify the logical circuit, target backend,
optimisation level and layout-selection criteria; routing and
operation times are computed automatically rather than assigned
gate by gate. An initial layout is selected automatically using Mapomatic's
subgraph-isomorphism search and calibration-based
scoring~\cite{PRXQuantum.4.010327}, supplemented by our readout-balance
constraint. For calibrated single-qubit readout errors $e_i,e_j$,
the pair-risk proxy is $1-(1-e_i)(1-e_j)$. We require the difference
between the worst-case channel $A$ and channel $B$ pair risks
to be at most $0.06$. Qiskit's SABRE heuristic inserts SWAPs where required by the
backend connectivity~\cite{IBMSabreSwap}. The routed circuit is
then optimised in the native gate basis at level 2 or 3, with
further routing disabled~\cite{IBMTranspilerOptimization}.
Scheduling is performed locally using the
\texttt{ALAPScheduleAnalysis} and \texttt{PadDelay} passes from
Qiskit IBM Runtime, using backend instruction durations and circuit
dependencies~\cite{IBMRuntimeScheduling}. Independent operations
can therefore overlap. The resulting circuits are submitted through
IBM Runtime's \texttt{SamplerV2} for hardware execution. Consequently, each checkpoint is compiled and scheduled separately,
not obtained by appending physical gates to an earlier compiled
checkpoint. Note that a common initial layout does not guarantee identical routed mappings or schedules across round counts.

All in all, entanglement fidelity assesses reference-receiver entanglement
preservation, with values above $1/2$ certifying entanglement.
Tomography estimates agreement with a specified target state,
generator witnesses provide fidelity lower bounds, and SWAP test
pass rates probe state overlap. M3 and mitigated state tomography
address readout bias, while DD targets idle time noise and classical
Pauli frame tracking accounts for known Bell measurement byproducts
without physical correction gates. Together, these complementary
metrics and error management techniques support assessment of
resource preparation, decoding and teleportation performance.

\section{Stage-resolved characterisation and interpretation}
\label{sec:supp-stages}

This section uses separate circuits to characterise AME resource
preparation (Stage~1), the decoded Bell pair marginals (Stage~2),
and teleportation of $\ket{+}_A\ket{-}_B$ (Stage~3). The aim is
to assess resource quality, Bell pair extraction and transfer
of a known input, complementing the cascade's test of
reference-receiver entanglement preservation. Tomography
fidelities estimate agreement with the corresponding ideal
targets, stabilizer-generator witnesses provide fidelity lower
bounds, and SWAP-test pass rates probe state overlap. Bell-outcome-conditioned receiver tomography
also examines differences between teleportation branches that
may be hidden by outcome averaging. These diagnostics check resource
quality, Bell pair extraction and transfer of a known input,
providing a different perspective from the entanglement fidelity.

\begin{figure*}[t]
\centering
\includegraphics[width=\textwidth]
 {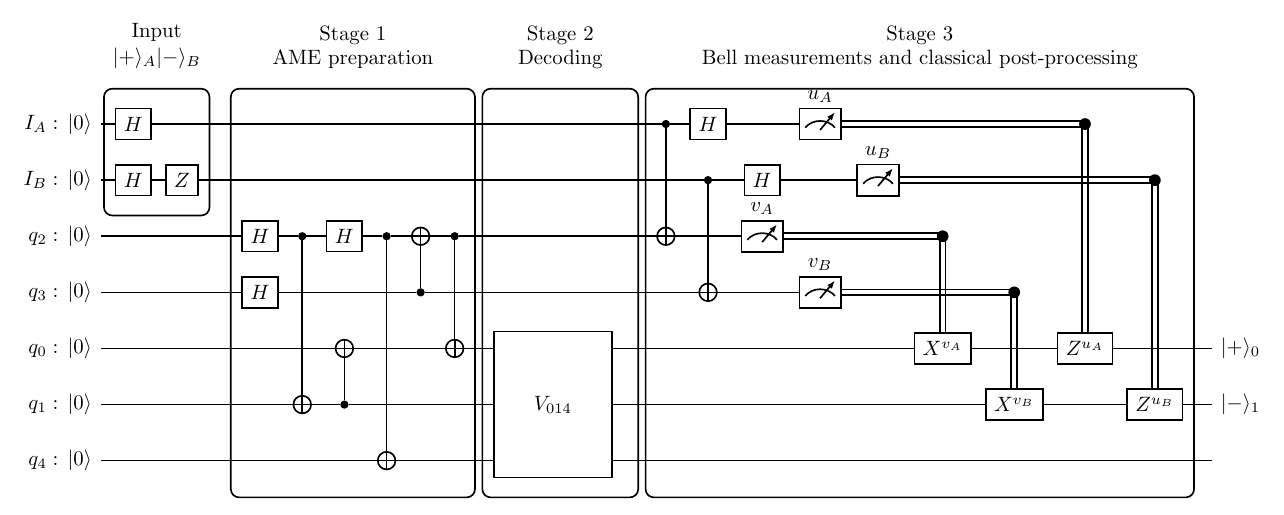}
\caption{Circuit diagram for one round of parallel
teleportation using the rank-1 decoder and inputs
$\ket{+}_A\ket{-}_B$.
Stage~1 prepares the shared AME$(5,2)$ resource.
Stage~2 applies the decoder $V_{014}$ jointly to
$q_0,q_1,q_4$.
Stage~3 performs Bell measurements on $(I_A,q_2)$ and
$(I_B,q_3)$, yielding outcomes $(u_A,v_A)$ and $(u_B,v_B)$.
The corresponding receivers are $q_0$ for channel $A$ and
$q_1$ for channel $B$.
The displayed $X^{v_c}$ and $Z^{u_c}$ boxes indicate
outcome-dependent Pauli corrections accounted for in
classical post-processing. Ideally, the corrected outputs are
$\ket{+}_0\ket{-}_1$.}
\label{supp-fig:rank1-product-teleportation}
\end{figure*}

The stage-resolved experiment treats AME preparation, decoding and
teleportation as separate characterisation circuits. Stage 1 concerns
the full five-qubit AME target; Stage 2 concerns the two decoded
Bell pair marginals; Stage 3 concerns the receiver state after Bell
measurement. Figure~\ref{supp-fig:rank1-product-teleportation} illustrates the
logical circuit underlying the three-stage diagnostics
with product-state inputs, using the rank-1 decoder. Tomography, generator witnesses and SWAP test definitions
are given in Sec.~IV. The available stage protocols include
product-state inputs, distinct from the entangled reference-input
pairs used in the cascade.

For the inputs $\ket{+}_A\ket{-}_B$, the Stage~3 target is
the same product state. Pauli byproducts are accounted for in
classical post-processing, and the corrected conditional
fidelities are averaged with weights given by the joint
Bell-outcome probabilities.\footnote{
For each trial,
$F_3=\sum_{\mathbf{b}\in\mathcal{B}}
p_{\mathbf{b}}F_{3|\mathbf{b}}$,
where $\mathcal{B}=\{0,1\}^4$ contains the sixteen joint
Bell outcomes, $p_{\mathbf{b}}$ is the outcome probability,
and $F_{3|\mathbf{b}}$ is the conditional receiver state
fidelity after classical Pauli correction.
For raw data,
$p_{\mathbf{b}}^{\mathrm{raw}}
=N_{\mathbf{b}}/\sum_{\mathbf{b}'\in\mathcal{B}}N_{\mathbf{b}'}$,
where $N_{\mathbf{b}}$ counts shots with Bell outcome
$\mathbf{b}$, summed over receiver measurement outcomes
and tomography settings.
For M3, the weights are obtained from the corresponding
normalised readout-mitigated distributions.
The resulting fidelities are averaged over eight trials.
}

SWAP test circuits add noisy controlled-SWAP operations, and generator
witnesses provide lower bounds rather than reconstructed-state
fidelities. Readout mitigation does not remove preparation, decoder,
Bell-measurement or idle-time errors. Bell outcome conditioned
tomography can reveal branch dependence, but demonstrating that no
branch differs requires its corresponding numerical data and
uncertainty analysis.
\subsection*{Quantitative stage results}
With XY4, the rank-1 hardware tomography means at Stages 1, 2
and 3 are $0.930$, $0.945$ and $0.905$ raw, and $0.947$,
$0.959$ and $0.926$ with M3. These target-state fidelities are
not the cascade's reference-receiver entanglement fidelities.
At Stage 2, rank 58 gives $0.948$ raw and $0.958$ with M3,
close to rank 1's $0.945$ and $0.959$. Its generator-witness
means are $0.839$ raw and $0.884$ with M3, compared with
$0.786$ and $0.861$ for rank 1. Here, Stage~2 includes AME preparation and decoding, while
Stage~3 includes preparation, decoding and teleportation. However,
Stage~1 tomography assesses the full five-qubit AME state, whereas
Stage~2 reports the mean fidelity of the two decoded Bell pair
marginals. Stage~3 assesses the joint receiver state after classical
Pauli correction, averaged over Bell outcome probabilities.
The increase from Stage~1 to Stage~2 therefore compares different
subsystems and targets. Each endpoint is evaluated in a separate circuit.
The code freezes its preparation and DD sequence before appending
tomography basis rotations, preventing changes to the preparation
across measurement bases. 

For rank-1 Stage 3 with XY4, the displayed Bell outcome conditioned
means range from $0.896$ to $0.918$ raw and $0.905$ to $0.952$
with M3. M3 raises each of the 16 displayed branch means.
These ranges are not confidence bounds or a formal test of
equal branch performance. The figures below present tomography fidelities, generator witness values and SWAP-test pass rates as grouped bar charts, together with heatmaps of joint receiver fidelities conditioned on the Bell-measurement outcomes.

\begin{figure}[!htbp]
\centering
\includegraphics[width=\textwidth,height=0.34\textheight,keepaspectratio]{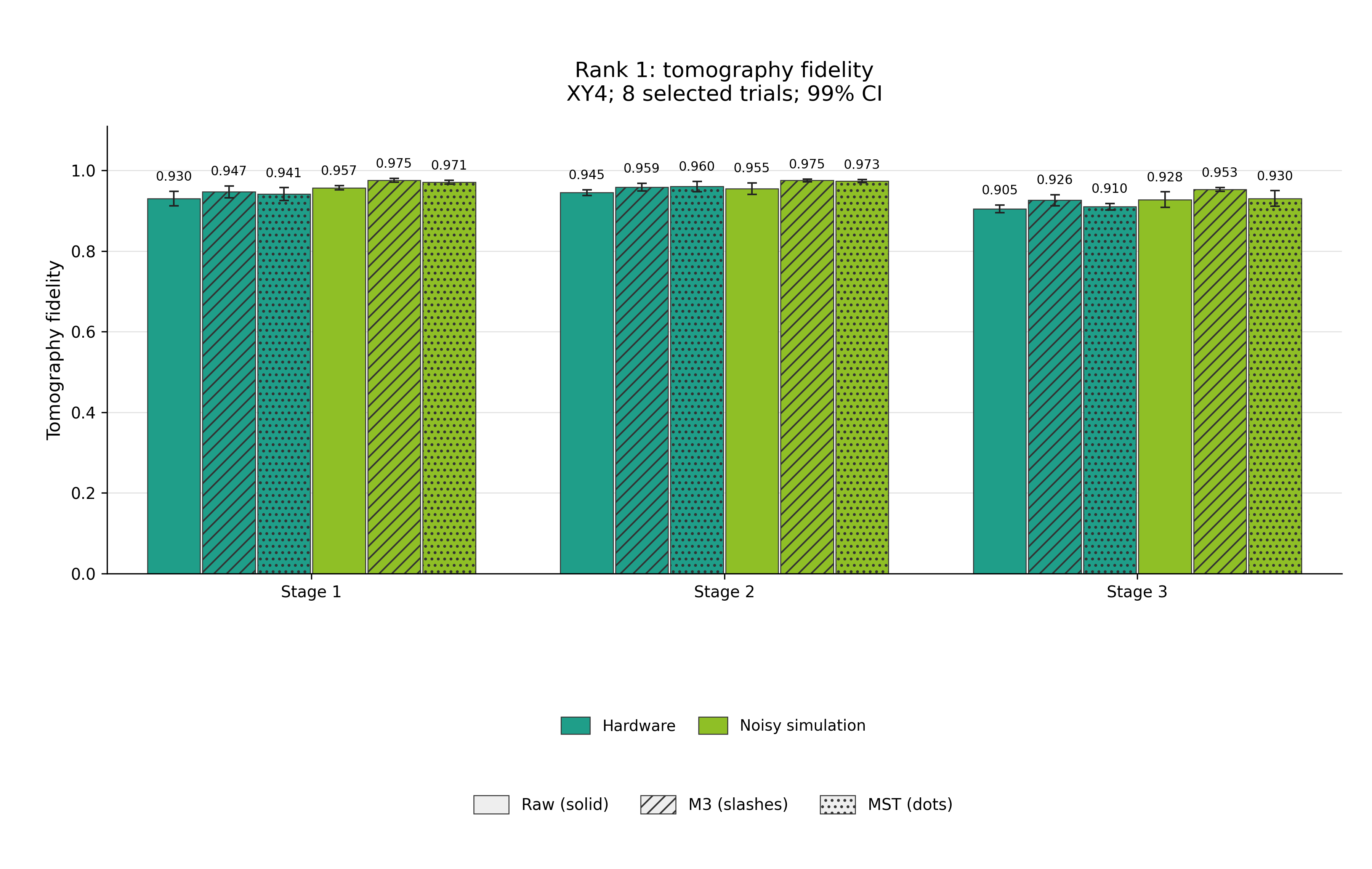}
\par\medskip
\includegraphics[width=\textwidth,height=0.34\textheight,keepaspectratio]{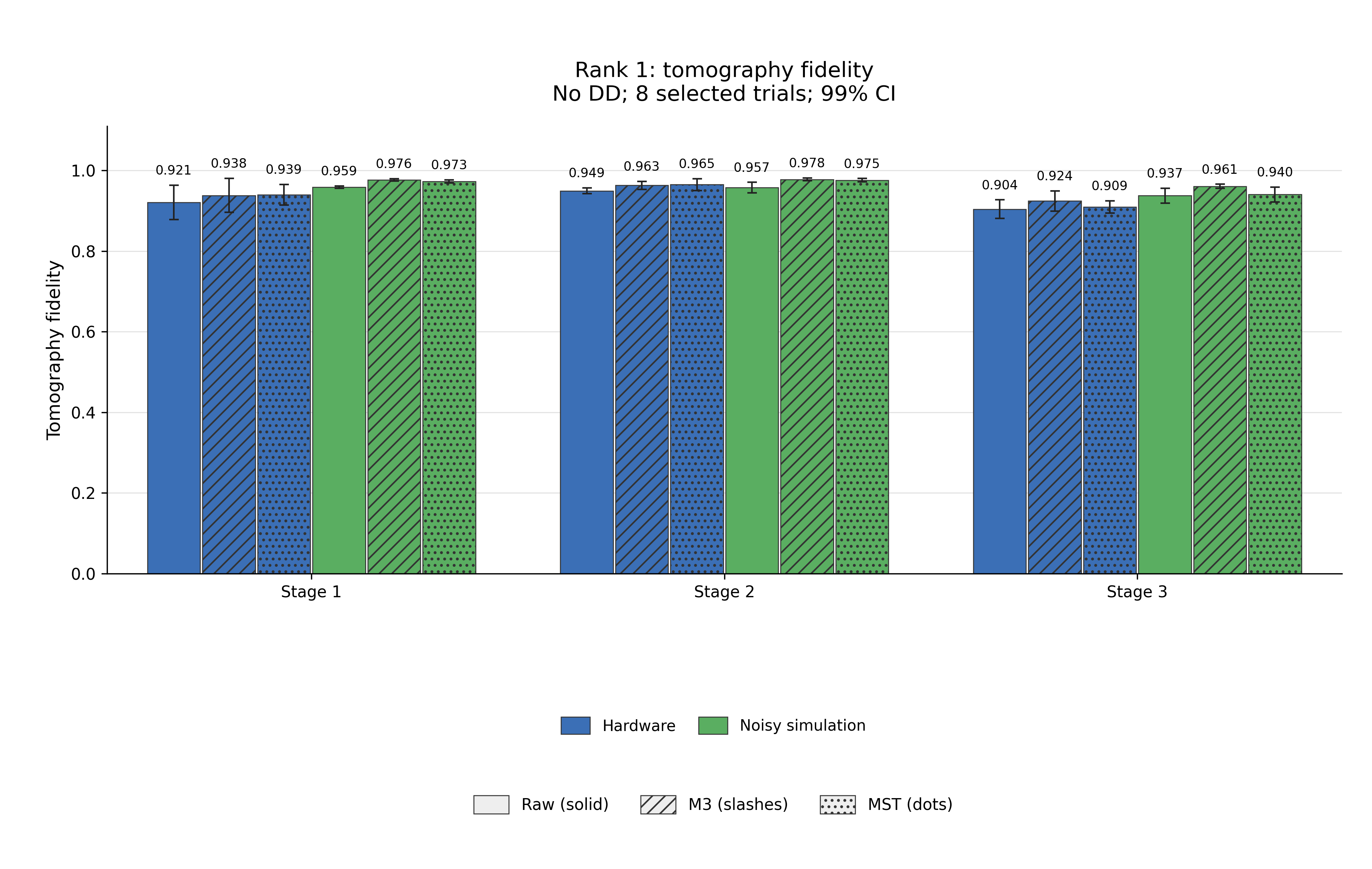}
\caption{Rank-1 tomography fidelities after AME preparation (Stage~1),
decoding (Stage~2) and parallel teleportation (Stage~3), with
XY4 (top) and no DD (bottom). Colours distinguish hardware
from noisy simulations; solid, diagonally hatched and dotted
bars denote raw, M3-mitigated and mitigated state tomography
(MST) estimates. Bars and annotations show eight-trial means,
with pointwise two-sided $99\%$ confidence intervals.
Each stage uses its corresponding ideal target state.}
\label{supp-fig:stage-tomography}
\end{figure}

\begin{figure}[!htbp]
\centering
\includegraphics[width=\textwidth,height=0.35\textheight,keepaspectratio]{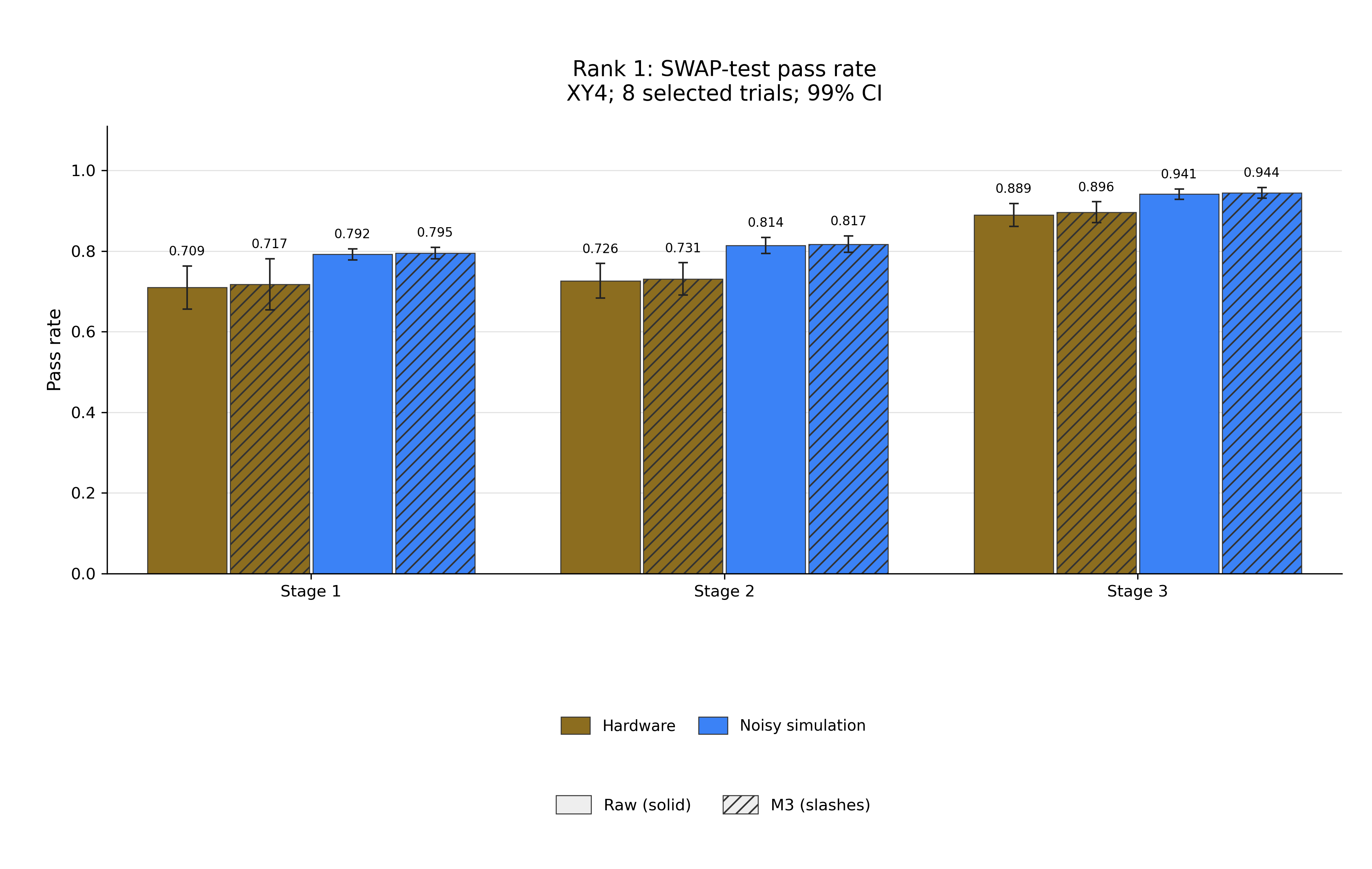}
\par\medskip
\includegraphics[width=\textwidth,height=0.35\textheight,keepaspectratio]{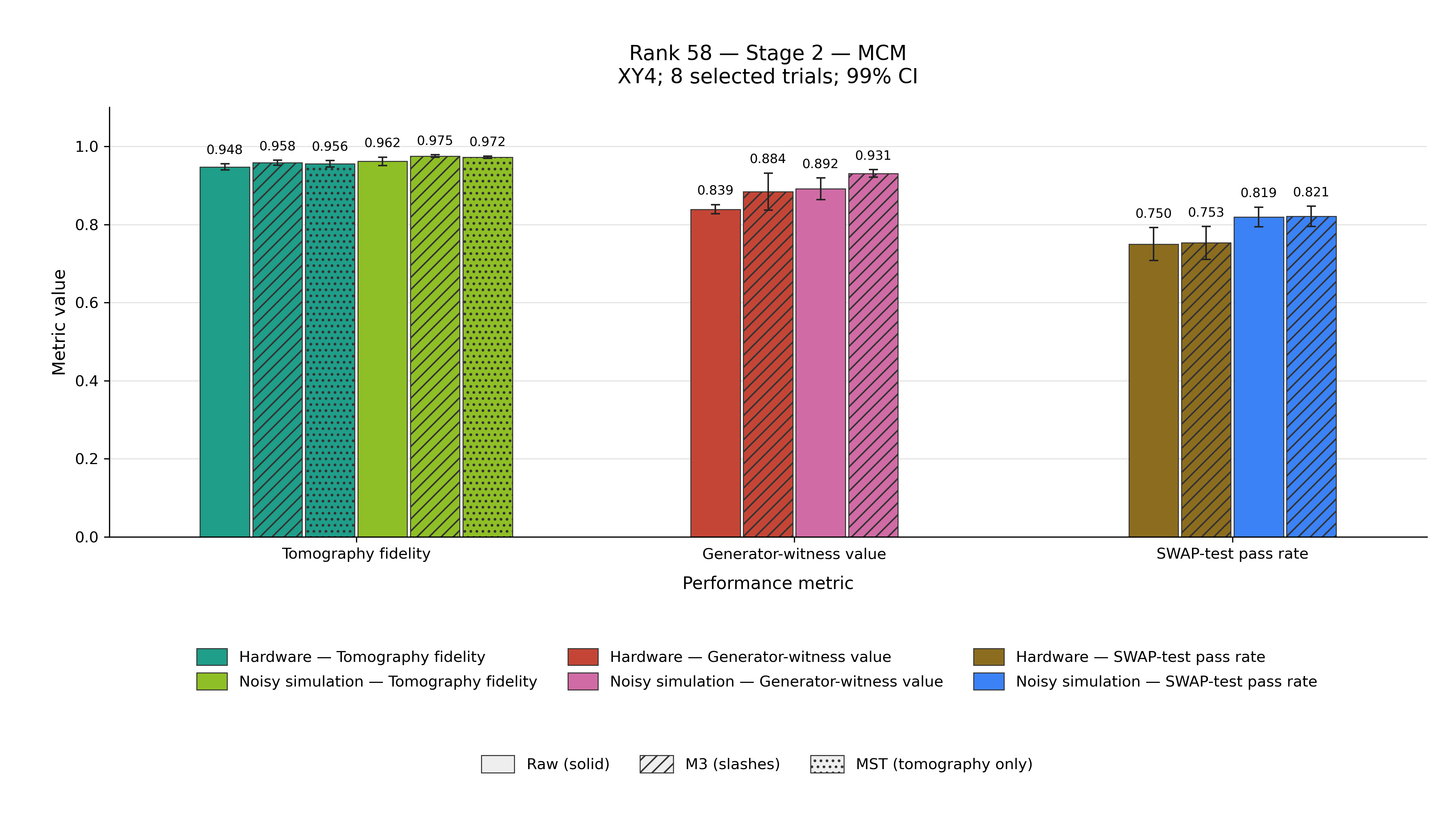}
\caption{Complementary diagnostics with XY4. Top: rank-1
SWAP-test pass rates at the three diagnostic stages.
Bottom: comparison of rank-58 diagnostic metrics after
Stage~2 decoding. SWAP-test pass rates represent test-acceptance
probabilities. The Plotted values are
averages over eight trials, with error bars showing pointwise two-sided 99\% Student-t confidence intervals.}
\label{supp-fig:stage-swap-and-rank58}
\end{figure}

\begin{figure}[!htbp]
\centering
\includegraphics[width=\textwidth,height=0.35\textheight,keepaspectratio]{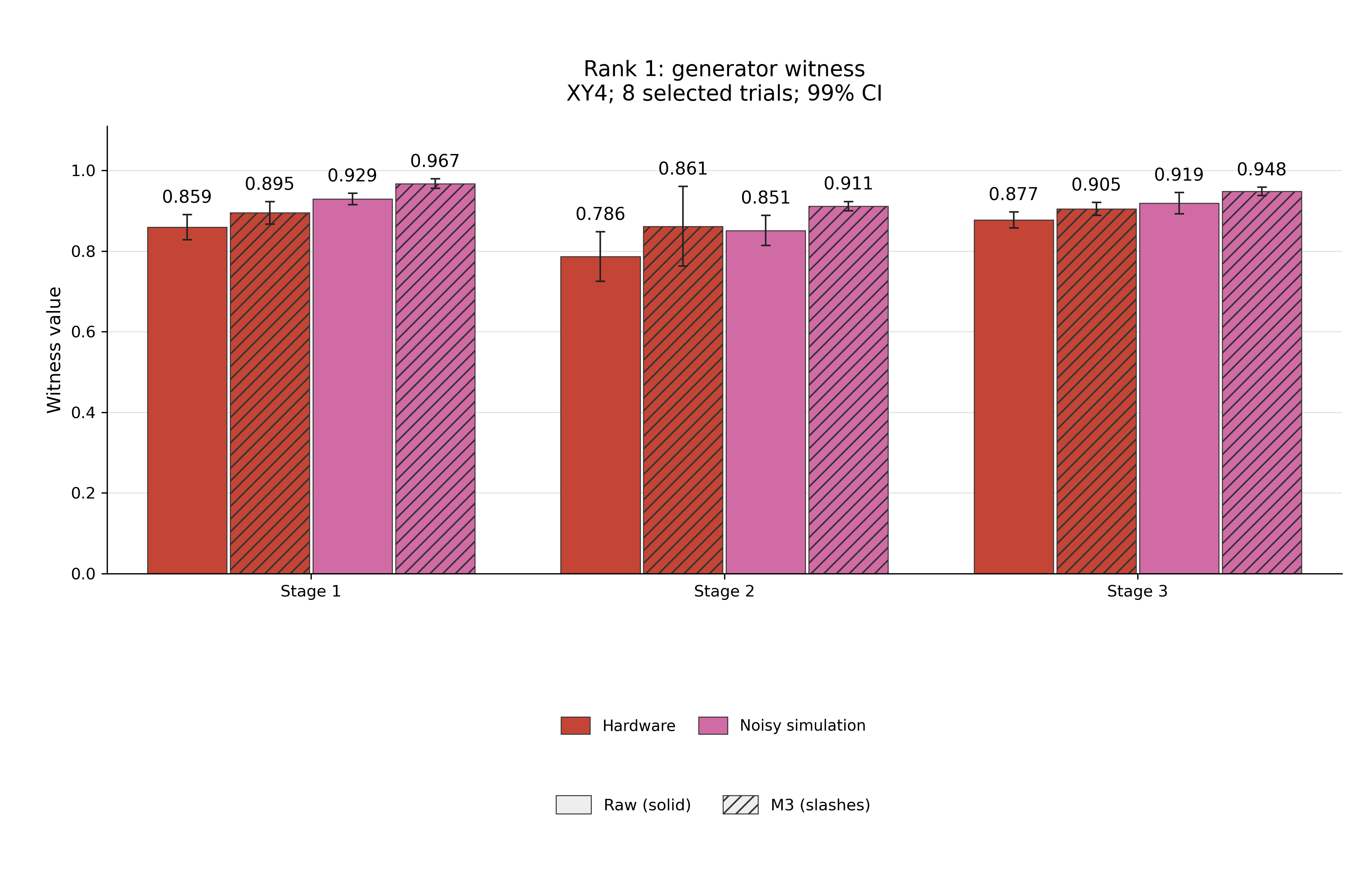}
\par\medskip
\includegraphics[width=\textwidth,height=0.35\textheight,keepaspectratio]{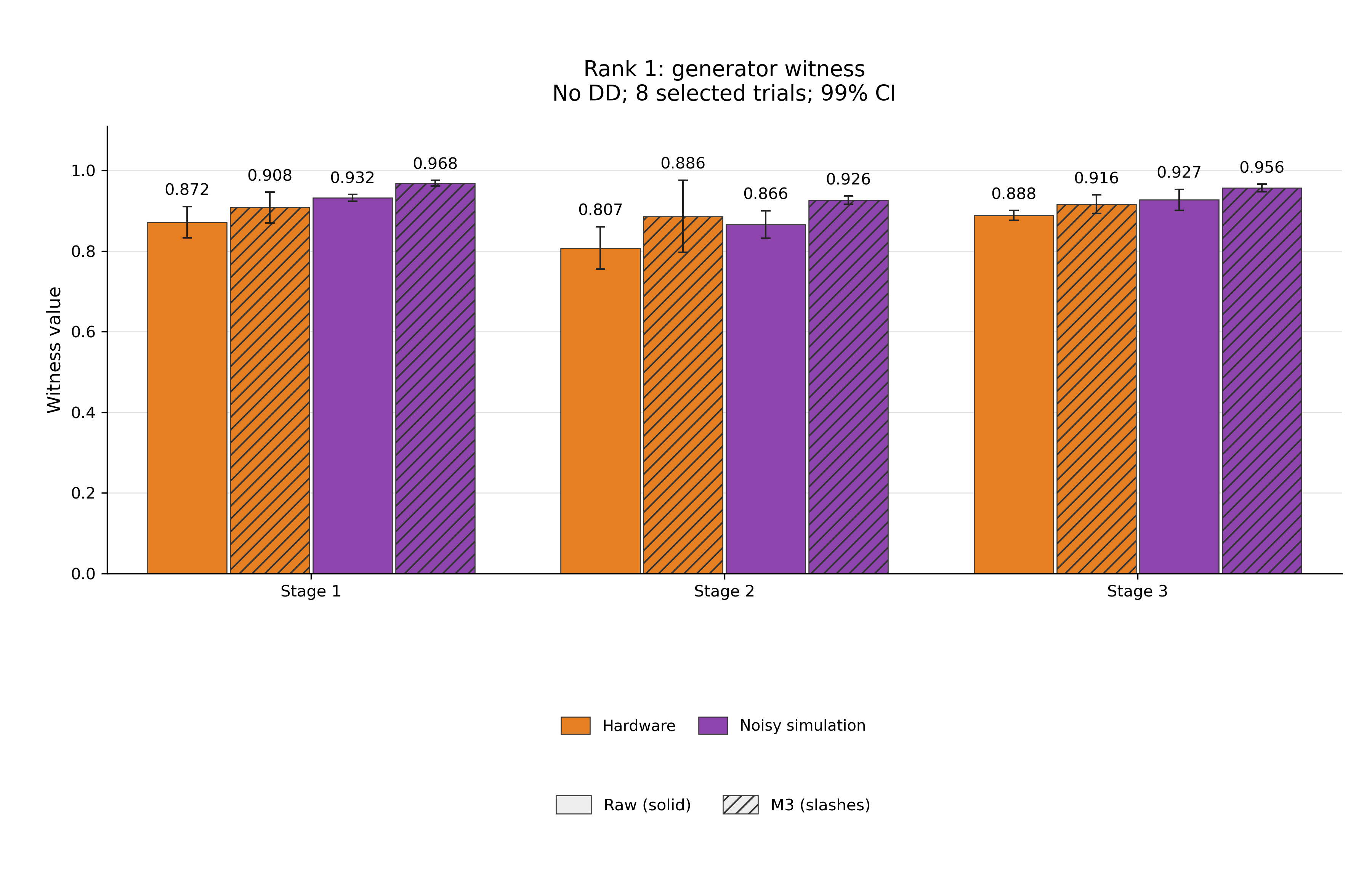}
\caption{Rank-1 fidelity lower bounds obtained from
stabilizer generator measurements after AME state preparation
(Stage~1), decoding (Stage~2) and parallel teleportation
(Stage~3), with XY4 (top) and no DD (bottom). Colours distinguish
hardware results from noisy simulations, while bar patterns
identify raw and mitigated estimates. Bars represent means over
eight trials, with pointwise two-sided $99\%$ confidence intervals.
Each lower bound refers to the corresponding ideal stage target.}
\label{supp-fig:stage-witness}
\end{figure}

\begin{figure}[p]
\centering
\includegraphics[
 width=\textwidth,height=0.35\textheight,keepaspectratio
]{rank1_mcm_swap_passrate_xy4_no_2copy.png}
\par\medskip
\includegraphics[
 width=\textwidth,height=0.35\textheight,keepaspectratio
]{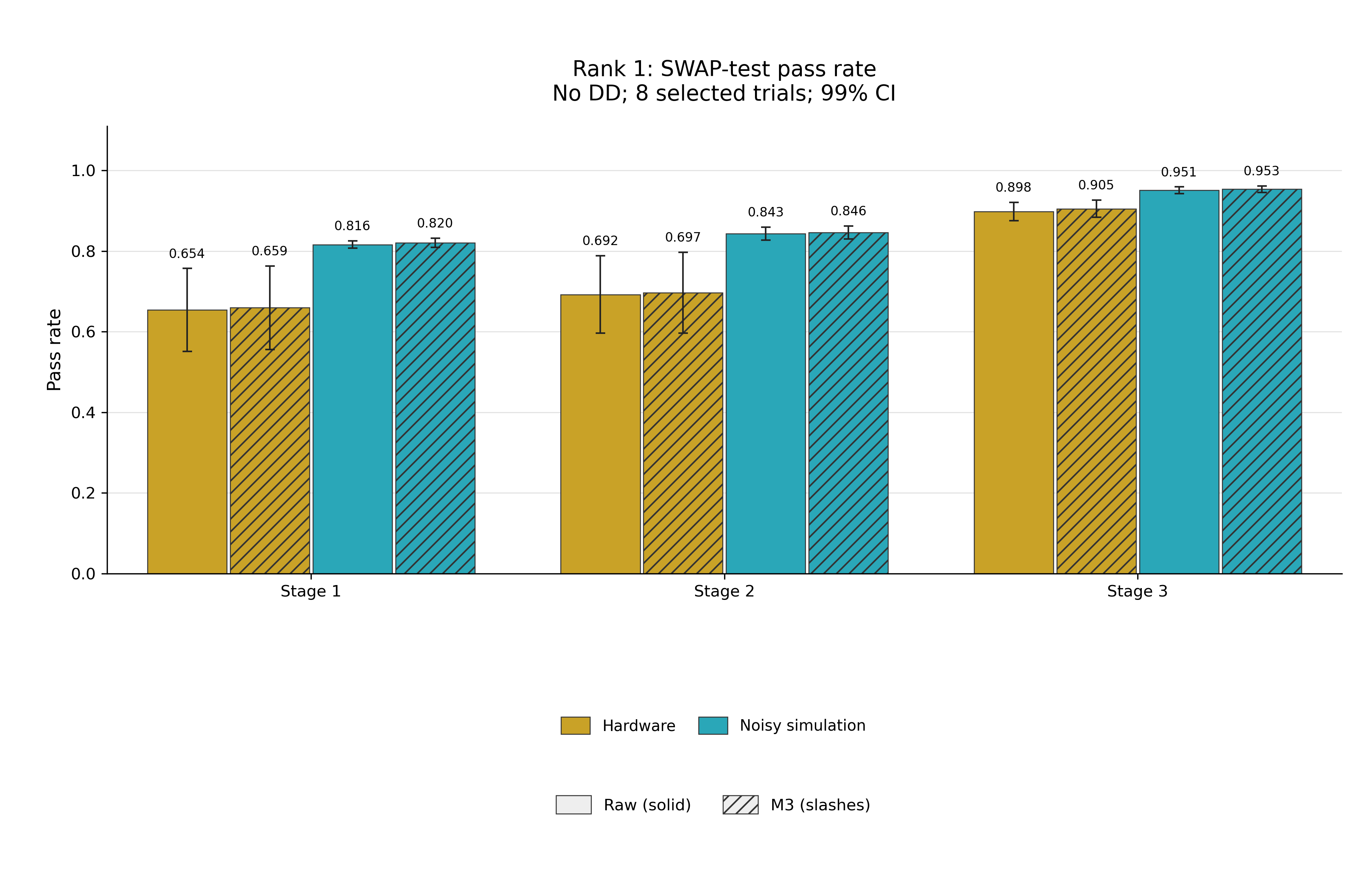}
\caption{Rank-1 SWAP-test pass rates after AME preparation
(Stage~1), decoding (Stage~2) and parallel teleportation
(Stage~3), with XY4 (top) and no DD (bottom).
Colours distinguish hardware results from noisy simulations;
solid and diagonally hatched bars show raw and M3-mitigated
estimates, respectively. Bars represent means over eight
trials, with pointwise two-sided $99\%$ confidence intervals.
Numerical annotations give the mean pass rates.}
\label{supp-fig:stage-swap-variants}
\end{figure}

\begin{figure}[p]
\centering
\includegraphics[width=\textwidth,height=0.31\textheight,keepaspectratio]{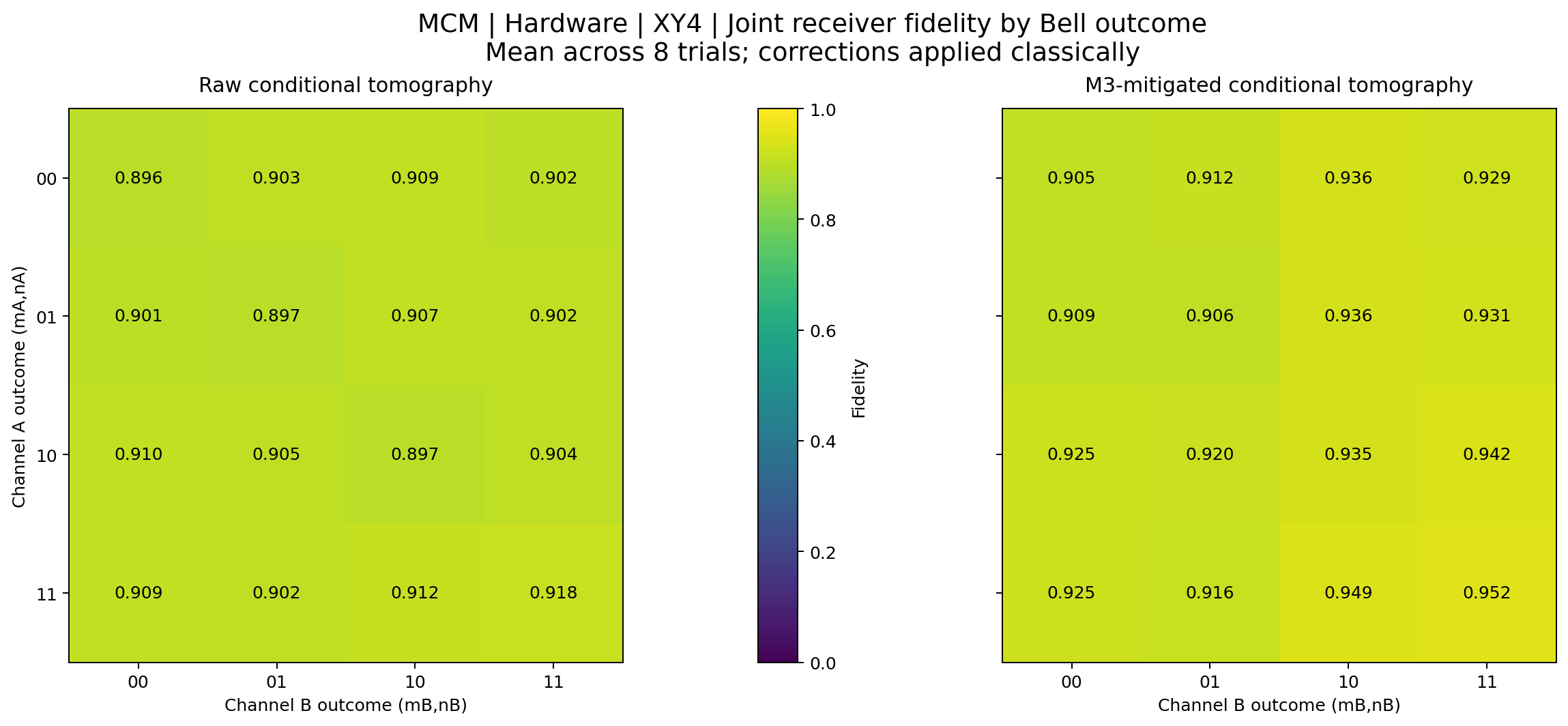}
\par\medskip
\includegraphics[width=\textwidth,height=0.31\textheight,keepaspectratio]{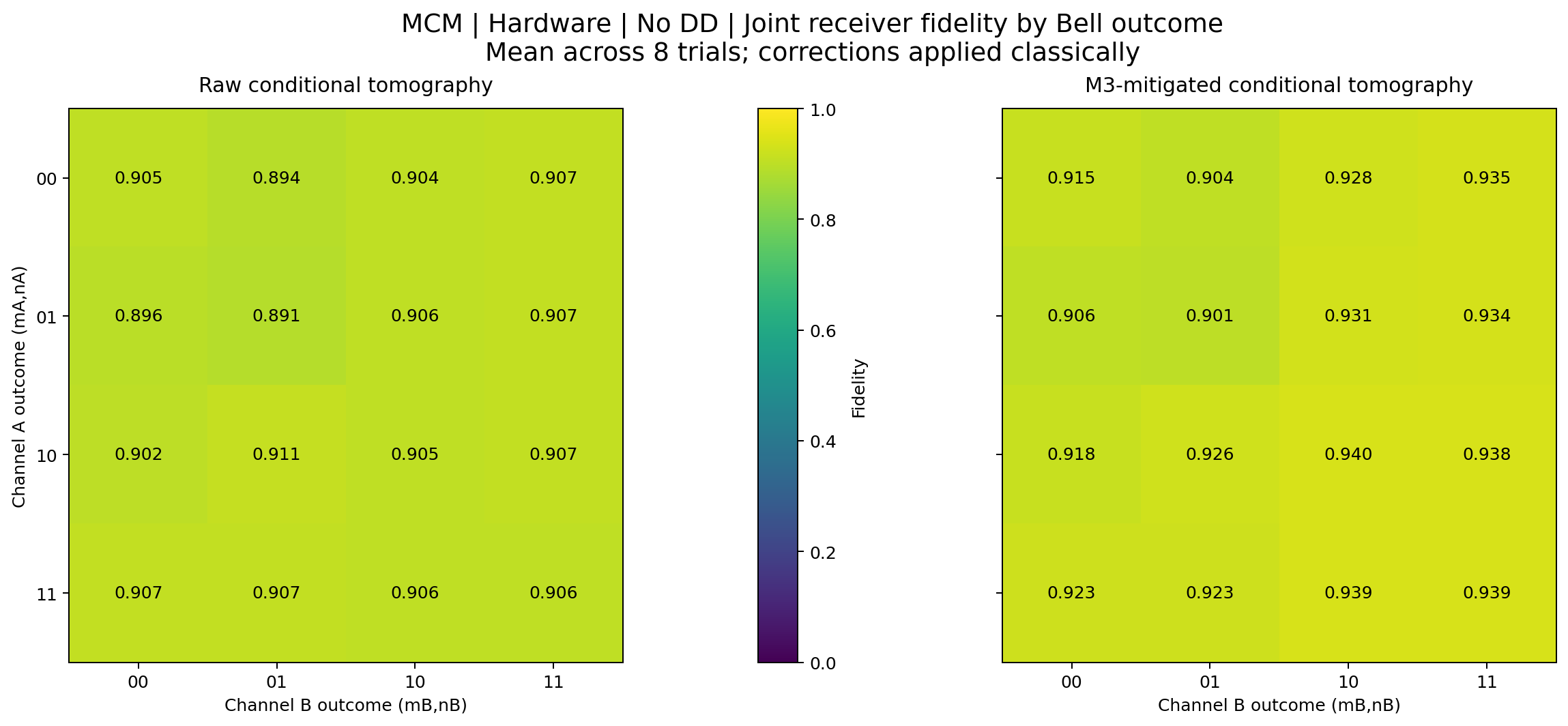}
\caption{Rank-1 Bell outcome conditioned receiver tomography with XY4
(top) and no DD (bottom), with Pauli corrections applied
classically. Heatmaps show raw and M3-mitigated branch means
over eight trials. The slightly higher no-DD means in some
branches illustrate that DD need not improve every output
fidelity. Imperfections in the additional control pulses can
outweigh the benefit of suppressing idle-time
noise~\cite{Souza:2011vcv}. Here, DD pulses fill existing idle
windows without increasing their duration~\cite{QiskitPadDD}.
Pulse errors therefore provide a possible explanation for
these differences.}
\label{supp-fig:stage-branches}
\end{figure}

\clearpage
With XY4, the rank-1 M3 readout-mitigated tomography fidelity means are $0.947$,
$0.959$ and $0.926$ after preparation, decoding and product-state
teleportation, respectively. Generator witnesses provide lower bounds on target-state fidelity,
while SWAP test pass rates report the probability of the test's
acceptance outcome. Neither quantity should be interpreted as a
tomography fidelity. These diagnostics benchmark the state quality achieved after the
resource preparation, decoding and teleportation operations used
in each cascade round. They therefore help interpret how the
gate, measurement and idle-time costs of repeating these operations
can accumulate and reduce the reference-receiver entanglement
fidelity. However, note that the circuits, measured subsystems and target
states differ between stages, so subtracting their fidelities
does not isolate the errors added by individual protocol steps. 

\section{Full hardware entanglement fidelity results via classical post-processing}

This section presents the complete hardware entanglement fidelity
results for the cascade with classical Pauli frame post-processing.
The plots cover both decoders and channels, rounds $R0$--$R3$,
no dynamical decoupling (DD), XX and XY4, and optimisation levels
2 and 3. Raw estimates are shown alongside results corrected using
matrix-free measurement error mitigation (M3), revealing how reference-receiver
entanglement changes as teleportation is repeated.

\begin{figure}[p]
\centering
\includegraphics[width=\textwidth,height=0.81\textheight,keepaspectratio]{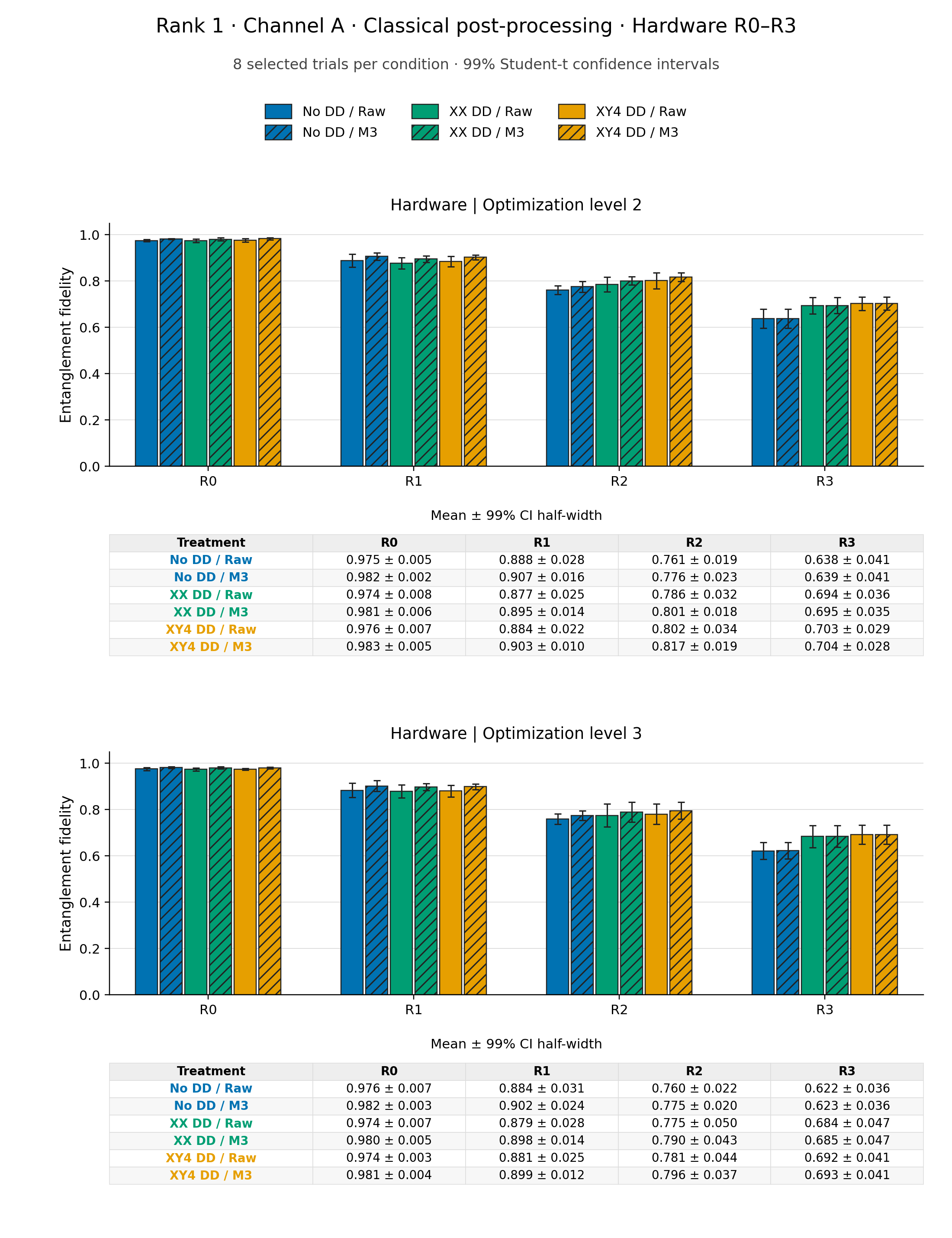}
\caption{Complete classical post-processing hardware results for rank 1, channel $A$.
Optimisation levels 2 and 3 are shown separately. Colours distinguish
no DD, XX and XY4; solid and hatched bars indicate raw and M3-mitigated
estimates. Bars are means over eight selected trials and error bars
are pointwise two-sided $99\%$ confidence intervals. The embedded
tables list each mean and interval half-width.}
\label{supp-fig:full-bars-1}
\end{figure}

\begin{figure}[p]
\centering
\includegraphics[width=\textwidth,height=0.81\textheight,keepaspectratio]{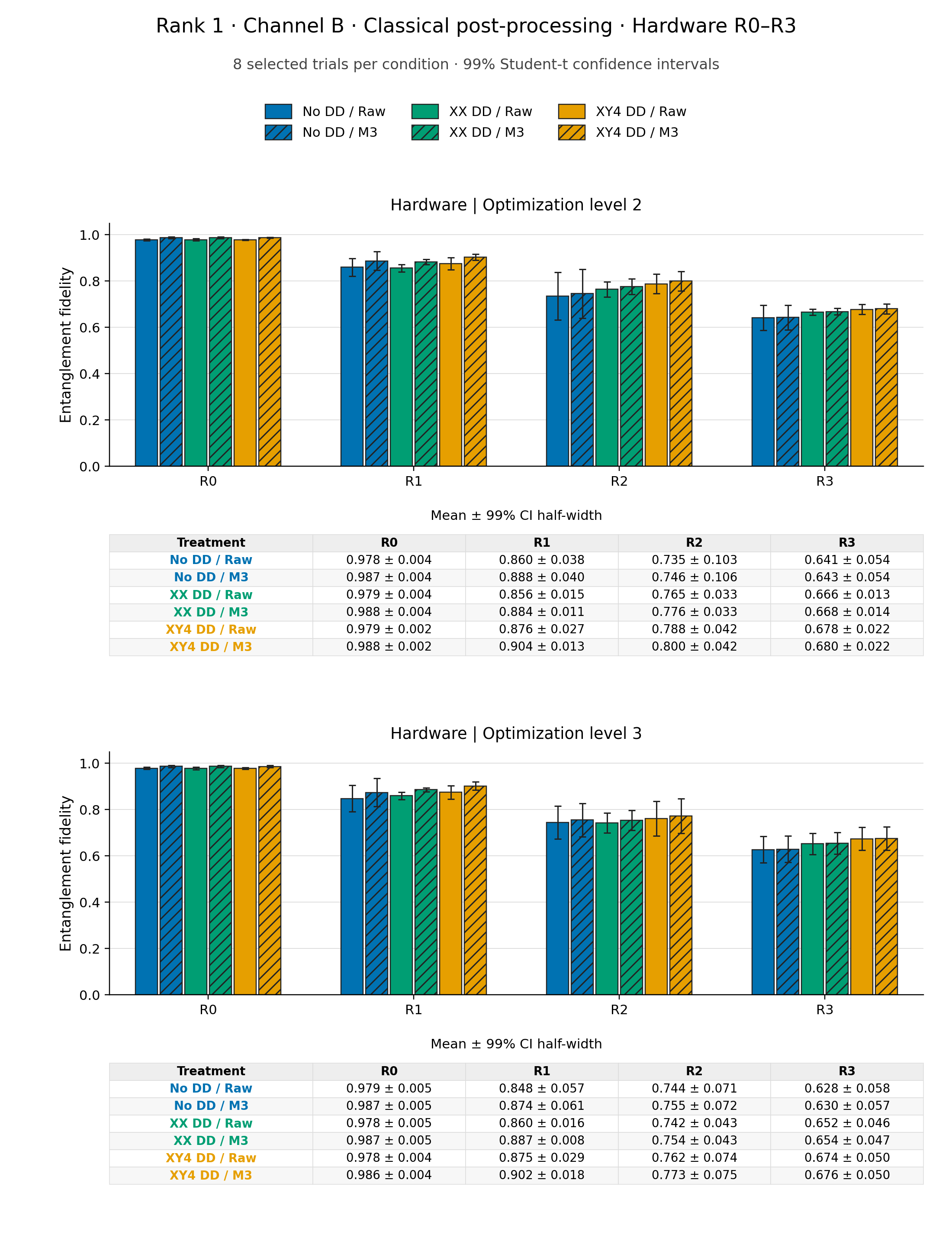}
\caption{Complete classical post-processing hardware results for rank 1, channel $B$.
Optimisation levels 2 and 3 are shown separately. Colours distinguish
no DD, XX and XY4; solid and hatched bars indicate raw and M3-mitigated
estimates. Bars are means over eight selected trials and error bars
are pointwise two-sided $99\%$ confidence intervals. The embedded
tables list each mean and interval half-width.}
\label{supp-fig:full-bars-2}
\end{figure}

\begin{figure}[p]
\centering
\includegraphics[width=\textwidth,height=0.81\textheight,keepaspectratio]{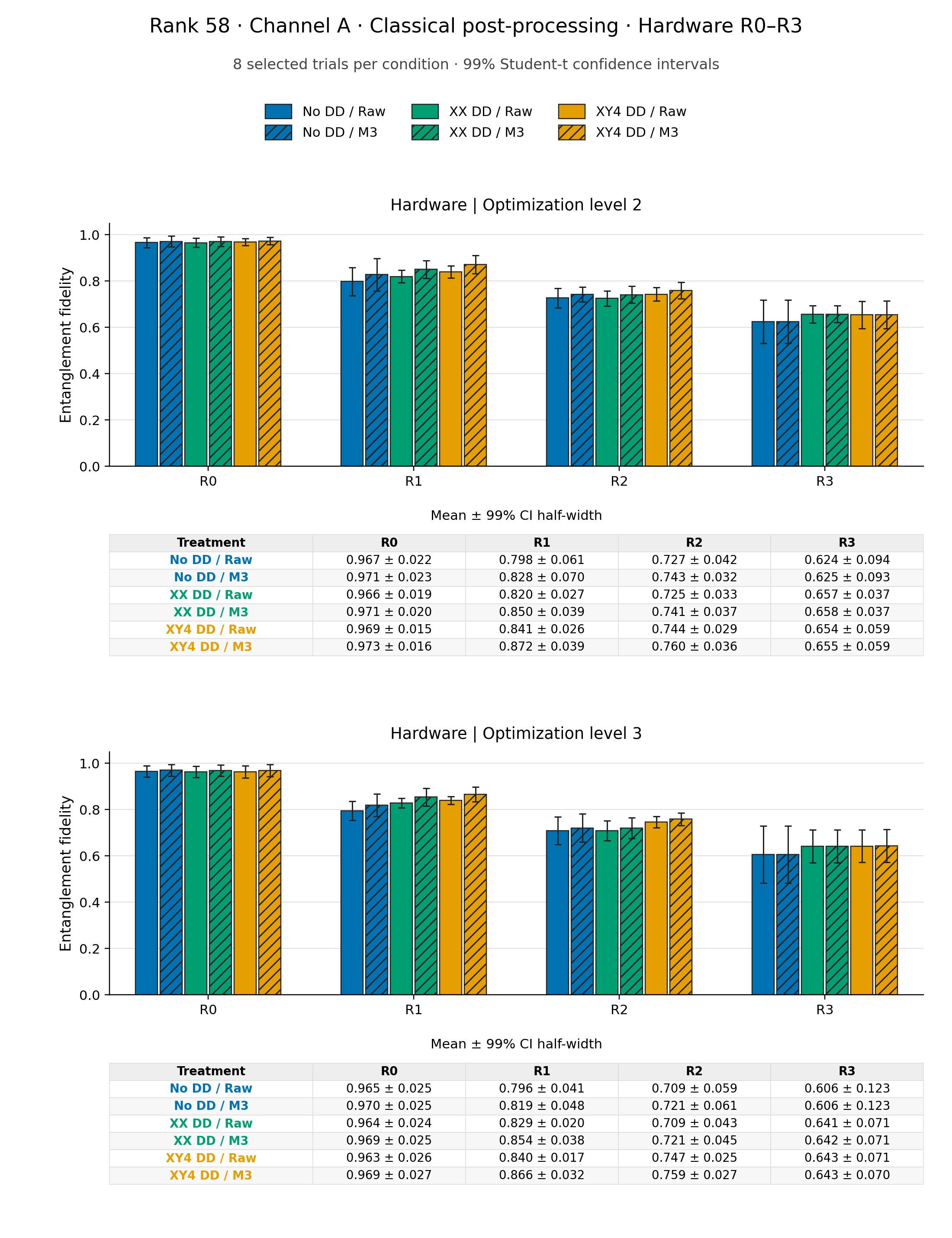}
\caption{Complete classical post-processing hardware results for rank 58, channel $A$.
Optimisation levels 2 and 3 are shown separately. Colours distinguish
no DD, XX and XY4; solid and hatched bars indicate raw and M3-mitigated
estimates. Bars are means over eight selected trials and error bars
are pointwise two-sided $99\%$ confidence intervals. The embedded
tables list each mean and interval half-width.}
\label{supp-fig:full-bars-3}
\end{figure}

\begin{figure}[p]
\centering
\includegraphics[width=\textwidth,height=0.81\textheight,keepaspectratio]{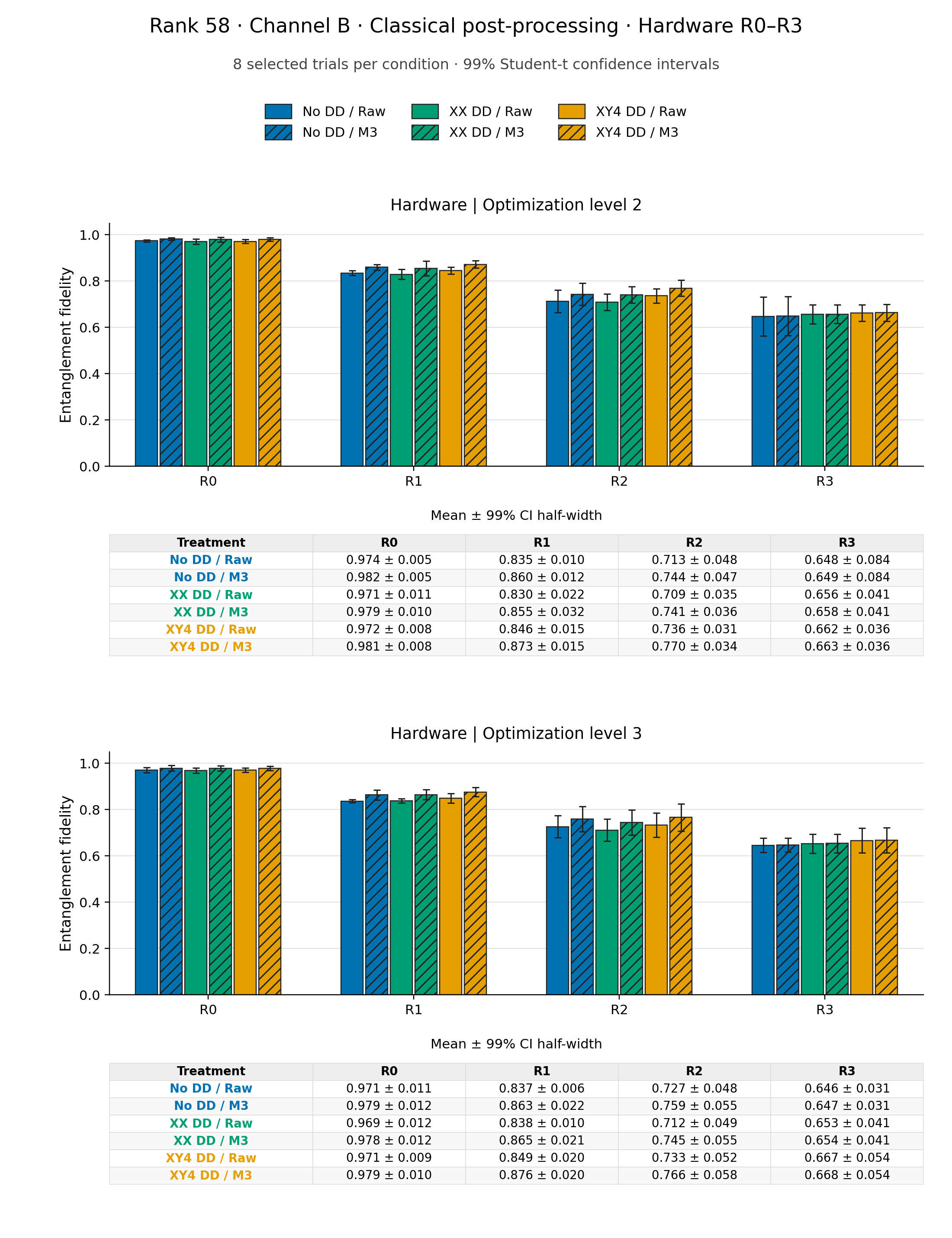}
\caption{Complete classical post-processing hardware results for rank 58, channel $B$.
Optimisation levels 2 and 3 are shown separately. Colours distinguish
no DD, XX and XY4; solid and hatched bars indicate raw and M3-mitigated
estimates. Bars are means over eight selected trials and error bars
are pointwise two-sided $99\%$ confidence intervals. The embedded
tables list each mean and interval half-width.}
\label{supp-fig:full-bars-4}
\end{figure}

Figures~\ref{supp-fig:full-bars-1}-\ref{supp-fig:full-bars-4} provide the full results underlying the summary of the main text, represented as grouped-bar charts.
In the main text, we focus on XY4 dynamical decoupling at otimization level 2.

\clearpage
Although entanglement fidelity decreases over successive rounds, both channels
retain certifiable entanglement after three rounds with XY4 at
optimisation level~2 for both tested decoders. The M3-mitigated
$R3$ means are $(0.704,0.680)$ for rank~1 and $(0.655,0.663)$
for rank~58 on channels $(A,B)$. For each decoder-channel
combination, both the raw and M3-mitigated pointwise $99\%$
confidence intervals lie above $1/2$, so this entanglement survival
does not depend on readout mitigation.

\section{DD, baseline-adjusted retention and Bell correlations of classical post-processing results}
\label{sec:supp-comparisons}
This section compares dynamical decoupling (DD) sequences and decoder
performance in the classical post-processing hardware campaign,
analysing raw and M3 readout-mitigated data separately.
We examine dynamical decoupling (DD) dependence, fidelity changes
relative to each trial's $R0$ baseline, and the Bell-correlation
decomposition using both raw and M3-mitigated hardware results
with classical Pauli frame post-processing. Each comparison is
calculated separately for the two processing methods; raw and
mitigated fidelities are not mixed. The aim is to determine which performance trends persist before
and after readout mitigation.

To assess DD, we compare sequences within each decoder rank at
fixed channel, optimisation level and round. The trial-level
differences are
\begin{align}
 D_i^{\mathrm{XY4-XX}}(R)
 &=100\bigl[F_{i,\mathrm{XY4}}(R)-F_{i,\mathrm{XX}}(R)\bigr],
 \label{supp-eq:xy4-xx-difference}\\
 D_{i,\mathrm{DD}}^{\mathrm{DD-No\,DD}}(R)
 &=100\bigl[F_{i,\mathrm{DD}}(R)-F_{i,\mathrm{No\,DD}}(R)\bigr],
 \label{supp-eq:dd-no-dd-difference}
\end{align}
where $i=1,\ldots,8$, $\mathrm{DD}\in\{\mathrm{XX},\mathrm{XY4}\}$,
and $F$ consistently denotes either raw or M3-mitigated fidelity.
The factor $100$ expresses differences in fidelity percentage
points. Positive values favour XY4 over XX in the first comparison
and the specified DD sequence over no DD in the second.
Throughout this discussion, uncertainties are the half-widths
of pointwise two-sided $99\%$ confidence intervals.

Figures~\ref{supp-fig:dd-optimization}(a,b)
and~\ref{supp-fig:raw-dd-comparisons}(a,b) show that the XY4
advantage is present before readout mitigation. In both processing
methods, XY4 minus XX means are positive in 23 of the 24
comparisons across the two ranks, two channels, optimisation
levels 2 and 3, and rounds $R1$-$R3$. Thus, XY4 generally
achieves higher mean entanglement fidelity than XX and is the
preferred DD sequence for the cascaded experiment. The $R0$
checkpoint contains only input EPR-pair preparation and is
excluded from these post-teleportation comparison counts.

The DD minus no DD comparisons in
Figures~\ref{supp-fig:dd-optimization}(c,d)
and~\ref{supp-fig:raw-dd-comparisons}(c,d) likewise show similar
raw and mitigated trends, with larger mean gains at later rounds,
particularly for rank~1. At optimisation level 2 and $R3$,
rank~1's raw XY4 gains are $6.49\pm6.41$ percentage points on
channel $A$ and $3.66\pm5.98$ on channel $B$, compared with
M3-mitigated gains of $6.51\pm6.39$ and $3.73\pm6.08$,
respectively. For rank~58, the corresponding raw gains are
$2.99\pm6.17$ and $1.42\pm7.66$, compared with
$3.00\pm6.17$ and $1.41\pm7.68$ after M3 mitigation.
Across $R1$-$R3$, M3 changes the DD-minus-no-DD means by
at most $0.211$ percentage points.

Figures~\ref{supp-fig:channel-gap}
and~\ref{supp-fig:raw-channel} compare decoder performance and
channel balance after accounting for the input EPR fidelities.
Writing $\delta F_{r,c}(R)=F_{r,c}(R)-F_{r,c}(R0)$ for the
fidelity change of decoder rank $r$ and channel $c$, the two
comparisons are
\begin{align}
 \Delta F_{1-58,c}(R)
 &=\delta F_{1,c}(R)-\delta F_{58,c}(R),
 \label{supp-eq:baseline-adjusted-rank-gap}\\
 \Delta F_{BA,r}(R)
 &=\delta F_{r,B}(R)-\delta F_{r,A}(R).
 \label{supp-eq:baseline-adjusted-gap}
\end{align}
The left plots show $100\Delta F_{1-58,c}$ for channels $A$
and $B$; the right plots show $100\Delta F_{BA,r}$ for ranks
1 and 58. Each trial's corresponding $R0$ baseline is subtracted
before calculating means and uncertainties, making $R0$ zero
by construction. Positive values indicate better fidelity
retention by rank~1 in the decoder comparison or by channel
$B$ in the channel comparison.

Both processing methods give positive baseline-adjusted
rank 1 minus rank 58 means for XX and XY4 on both channels
throughout $R1$-$R3$. For XY4 at level 2 and $R3$, the raw
rank-1 advantages are $4.08\pm4.66$ percentage points on
channel $A$ and $0.872\pm3.11$ on channel $B$; the
M3-mitigated values are $3.88\pm4.53$ and $0.955\pm3.11$,
respectively. These endpoint intervals include zero, but
the consistently positive means favour rank~1 in both raw
and mitigated results. Mitigation can change the size and
uncertainty of a baseline-adjusted difference because it
changes both the later-round estimate and the subtracted
$R0$ baseline.

For rank~1, both XX and XY4 give negative baseline-adjusted
$B-A$ means at $R1$-$R3$ in both processing methods.
Channel $B$ therefore shows greater mean fidelity loss than
channel $A$ relative to its own input baseline. With XY4 at
$R3$, this difference is $-2.72\pm2.72$ percentage points
in the raw results and $-2.83\pm2.65$ after M3 mitigation.

Figures~\ref{supp-fig:bell-correlations}
and~\ref{supp-fig:raw-correlations} resolve the endpoint
fidelities into signed Bell correlations. Writing
$C=(\langle XX\rangle,-\langle YY\rangle,\langle ZZ\rangle)$,
the XY4, level-2 triples at $R3$ are
\begin{align*}
 C_{1,A}^{\mathrm{raw}}&=(0.593,0.574,0.644),&
 C_{1,A}^{\mathrm{M3}}&=(0.594,0.576,0.646),\\
 C_{58,A}^{\mathrm{raw}}&=(0.591,0.512,0.515),&
 C_{58,A}^{\mathrm{M3}}&=(0.592,0.513,0.516),\\
 C_{1,B}^{\mathrm{raw}}&=(0.616,0.521,0.575),&
 C_{1,B}^{\mathrm{M3}}&=(0.619,0.524,0.578),\\
 C_{58,B}^{\mathrm{raw}}&=(0.566,0.521,0.562),&
 C_{58,B}^{\mathrm{M3}}&=(0.567,0.523,0.563).
\end{align*}
On channel $A$, the $XX$ components are nearly equal,
while rank~1 retains larger $-YY$ and $ZZ$ components.
On channel $B$, the $-YY$ components are nearly equal;
rank~1's advantage comes mainly from $XX$, with a smaller
contribution from $ZZ$. This interpretation holds for both
processing methods: M3 changes each $R3$ component by at
most $0.00310$. These components determine the absolute
endpoint fidelities through
Eq.~\eqref{supp-eq:bell-correlation-fidelity}, whereas the
baseline-adjusted comparisons additionally subtract the
corresponding differences at $R0$.

Residual phase noise during mid-circuit measurements and idle
intervals is a plausible contributor to the correlation loss.
Under local $Z$-dephasing, transverse $X$ and $Y$ components
are attenuated while computational-basis populations remain
unchanged. Measurement-induced phase errors and residual
$ZZ$ crosstalk have been observed in superconducting dynamic
circuits, where suitably designed DD can suppress these
effects~\cite{Shirizly:2024hre}. Neither classical Pauli frame
tracking nor readout mitigation removes the underlying
gate errors or idle-time decoherence.

Altogether, both raw and M3-mitigated hardware results favour
rank~1 over rank~58 and XY4 over XX for cascaded parallel
teleportation with classical Pauli frame post-processing.
These mean trends are therefore not created by readout
mitigation. Rank~1's fewer single-qubit gates and lower
logical decoder depth are consistent with its advantage,
although physical mapping, routing and execution timing
also influence performance. The correlation decomposition
identifies the components contributing to the fidelity
differences, but does not establish a unique underlying
noise mechanism.

\FloatBarrier
\begingroup
\raggedbottom
\renewcommand{\topfraction}{0.95}
\renewcommand{\bottomfraction}{0.95}
\renewcommand{\textfraction}{0.05}
\renewcommand{\floatpagefraction}{0.50}
% Top-align float-only pages instead of vertically centring a short row.
\makeatletter
\setlength{\@fptop}{0pt}
\setlength{\@fpsep}{12pt plus 2pt minus 2pt}
\setlength{\@fpbot}{0pt plus 1fil}
\makeatother

\subsection*{M3-mitigated}

% The two DD rows may fill separate gaps but share one figure number.
\begin{figure}[!htbp]
\centering
\begin{minipage}[t]{0.49\textwidth}
\vspace{0pt}\centering
\includegraphics[width=\linewidth,height=0.34\textheight,keepaspectratio]
 {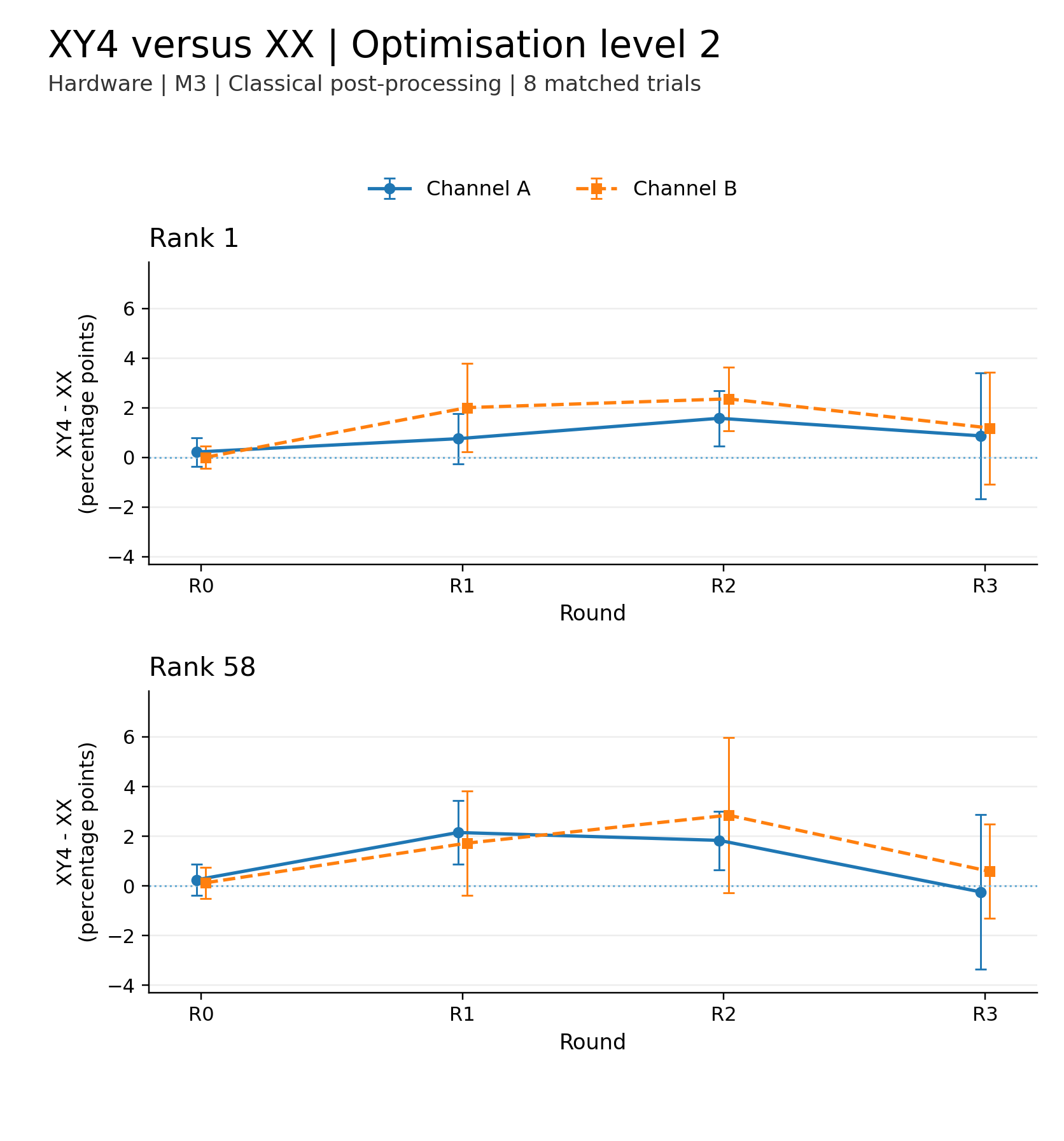}
\par\smallskip (a) XY4 versus XX, level 2.
\end{minipage}\hfill
\begin{minipage}[t]{0.49\textwidth}
\vspace{0pt}\centering
\includegraphics[width=\linewidth,height=0.34\textheight,keepaspectratio]
 {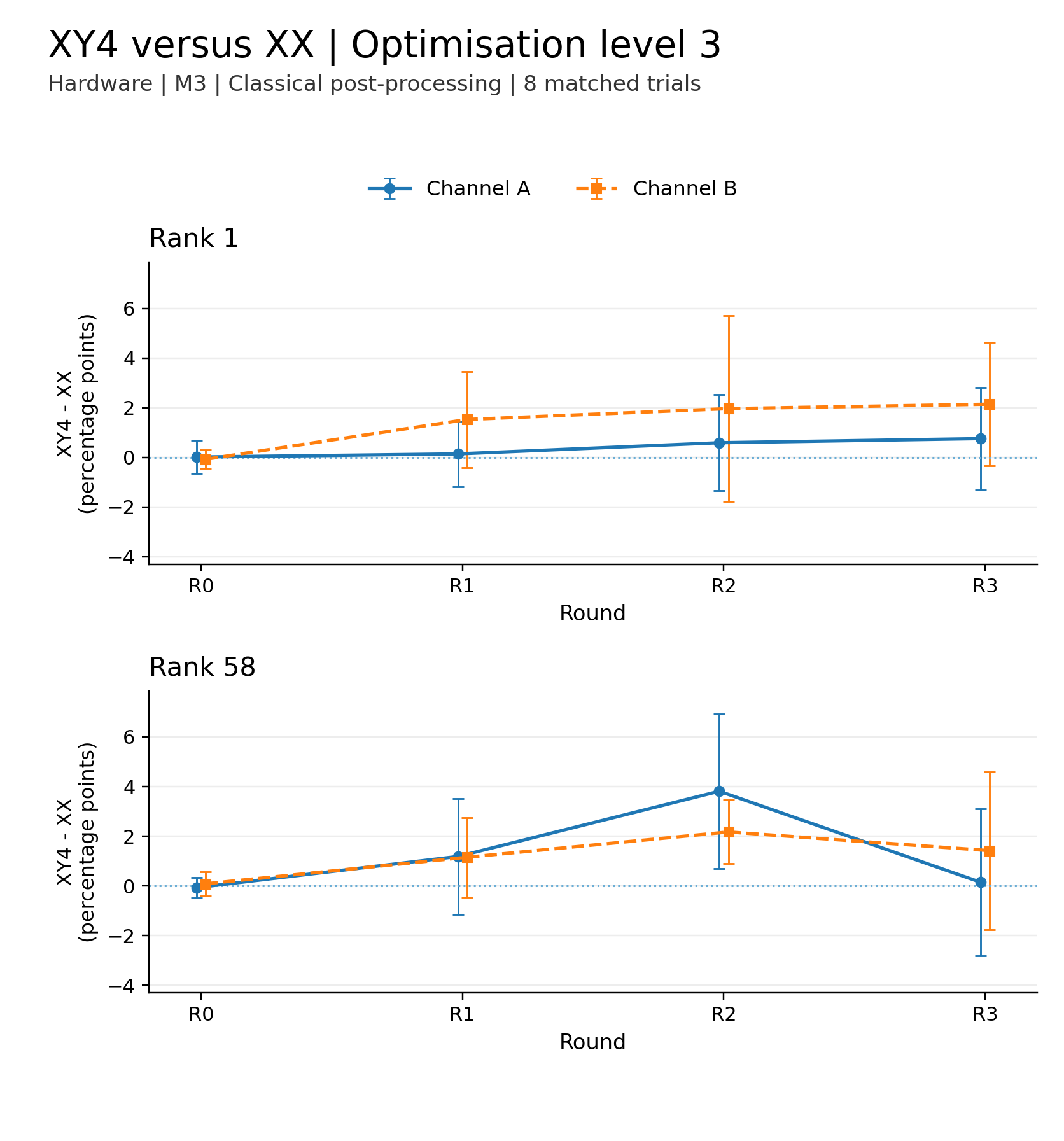}
\par\smallskip (b) XY4 versus XX, level 3.
\end{minipage}
\phantomcaption
\label{supp-fig:dd-optimization}
\end{figure}

\begin{figure}[!htbp]
\ContinuedFloat
\centering
\begin{minipage}[t]{0.49\textwidth}
\vspace{0pt}\centering
\includegraphics[width=\linewidth,height=0.34\textheight,keepaspectratio]
 {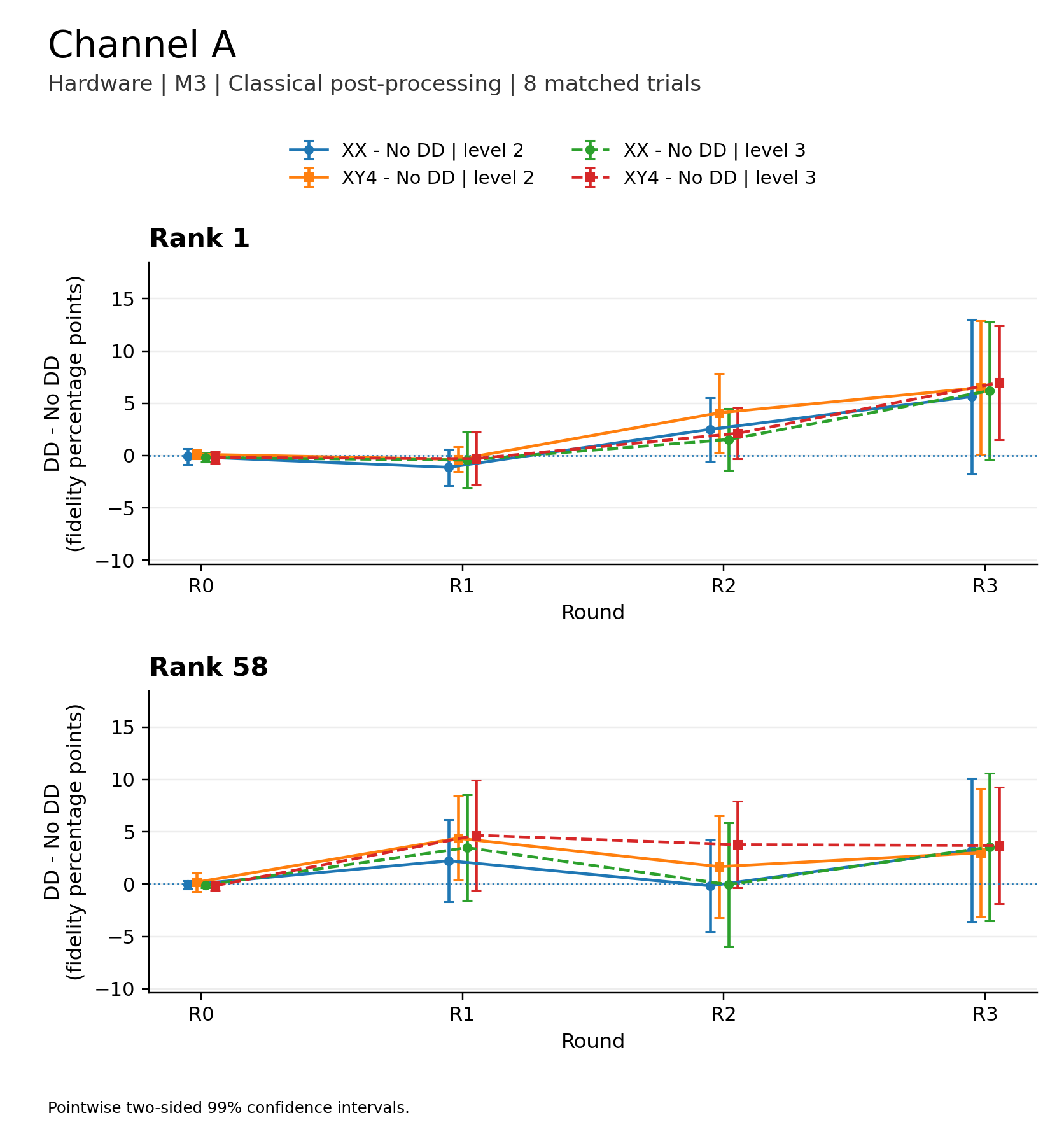}
\par\smallskip (c) DD versus no DD, channel $A$.
\end{minipage}\hfill
\begin{minipage}[t]{0.49\textwidth}
\vspace{0pt}\centering
\includegraphics[width=\linewidth,height=0.34\textheight,keepaspectratio]
 {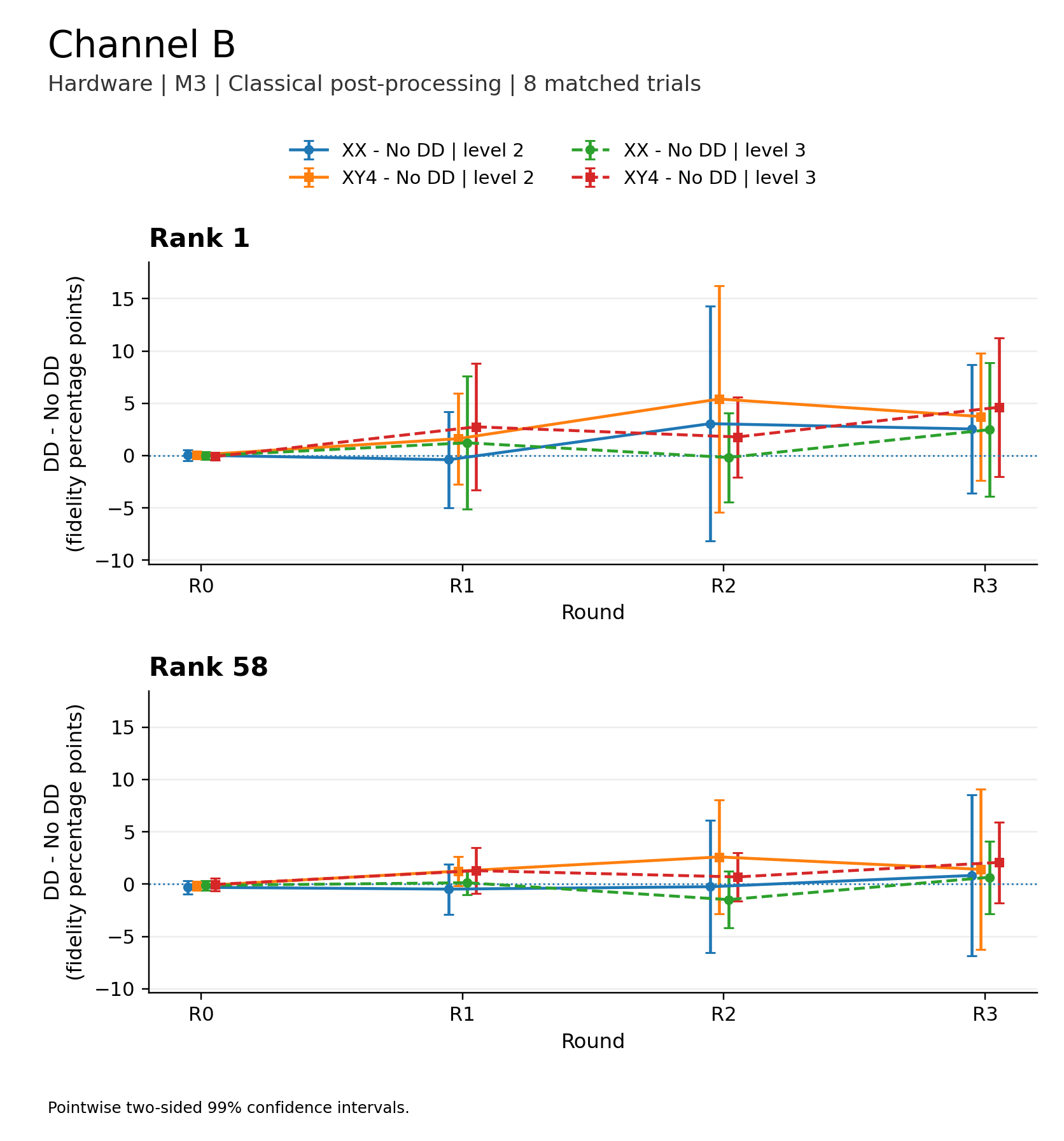}
\par\smallskip (d) DD versus no DD, channel $B$.
\end{minipage}
\captionsetup{font=small}
\caption{M3-mitigated hardware DD comparisons with classical Pauli frame
post-processing. Panels (a) and (b) compare XY4 and XX at optimisation
levels 2 and 3. Panels (c) and (d) compare XX and XY4 with matched
no-DD baselines for channels $A$ and $B$, with rank~1 above
rank~58 and both levels in each subplot. Differences are
fidelity percentage points. Points are means of eight
within-trial differences, and error bars are pointwise
two-sided $99\%$ confidence intervals.}
\end{figure}

% The baseline plot is independent of the preceding DD float.
% Requires graphicx and caption in the preamble.
\newsavebox{\MthreeBaselineCaptionBox}
\newsavebox{\MthreeCorrelationCaptionBox}
\newlength{\MthreePairImageHeight}

\begin{figure}[p]
\centering
\setlength{\parskip}{0pt}

% Measure and store the first caption.
\savebox{\MthreeBaselineCaptionBox}{%
\begin{minipage}{\textwidth}
\captionsetup{font=small}
\caption{Baseline-adjusted M3-mitigated hardware entanglement fidelity
comparisons at optimisation level 2.
Left: rank 1 minus rank 58 retention for channels $A$ and $B$.
Right: channel $B$ minus channel $A$ retention for ranks 1 and 58.
The corresponding within-trial $R0$ difference is subtracted
before forming the comparisons. Positive values favour rank~1
on the left and channel $B$ on the right. Error bars are
pointwise two-sided $99\%$ confidence intervals, with eight
trials per decoder.}
\label{supp-fig:channel-gap}
\end{minipage}%
}

% Measure and store the second caption.
\savebox{\MthreeCorrelationCaptionBox}{%
\begin{minipage}{\textwidth}
\caption{Signed Bell correlations with XY4, optimisation
level 2 and M3 mitigation. The components $XX$, $-YY$ and $ZZ$
determine the entanglement fidelity through
Eq.~\eqref{supp-eq:bell-correlation-fidelity}.
Points are means over eight trials, and error bars are
pointwise two-sided $99\%$ confidence intervals.}
\label{supp-fig:bell-correlations}
\end{minipage}%
}

% Reserve caption space, an 8pt gap and a 4pt safety margin.
% Divide the remaining height equally between the images.
\setlength{\MthreePairImageHeight}{\dimexpr
 \textheight
 -\ht\MthreeBaselineCaptionBox
 -\dp\MthreeBaselineCaptionBox
 -\ht\MthreeCorrelationCaptionBox
 -\dp\MthreeCorrelationCaptionBox
 -12pt\relax}
\setlength{\MthreePairImageHeight}{0.5\MthreePairImageHeight}

\includegraphics[
 width=\textwidth,
 height=\MthreePairImageHeight,
 keepaspectratio
]{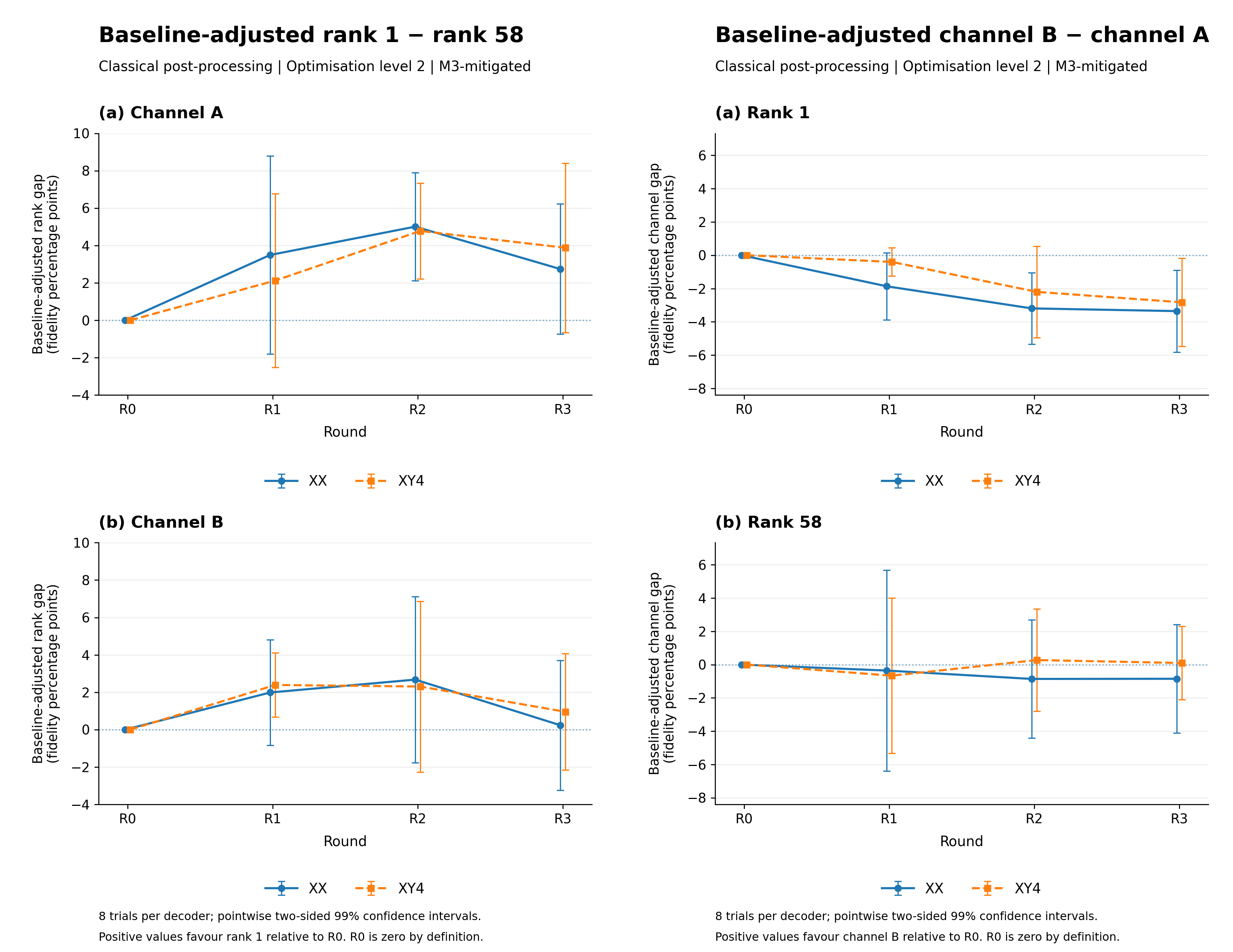}
\par\nointerlineskip
\usebox{\MthreeBaselineCaptionBox}

\par\vspace{8pt}\nointerlineskip

\includegraphics[
 width=\textwidth,
 height=\MthreePairImageHeight,
 keepaspectratio
]{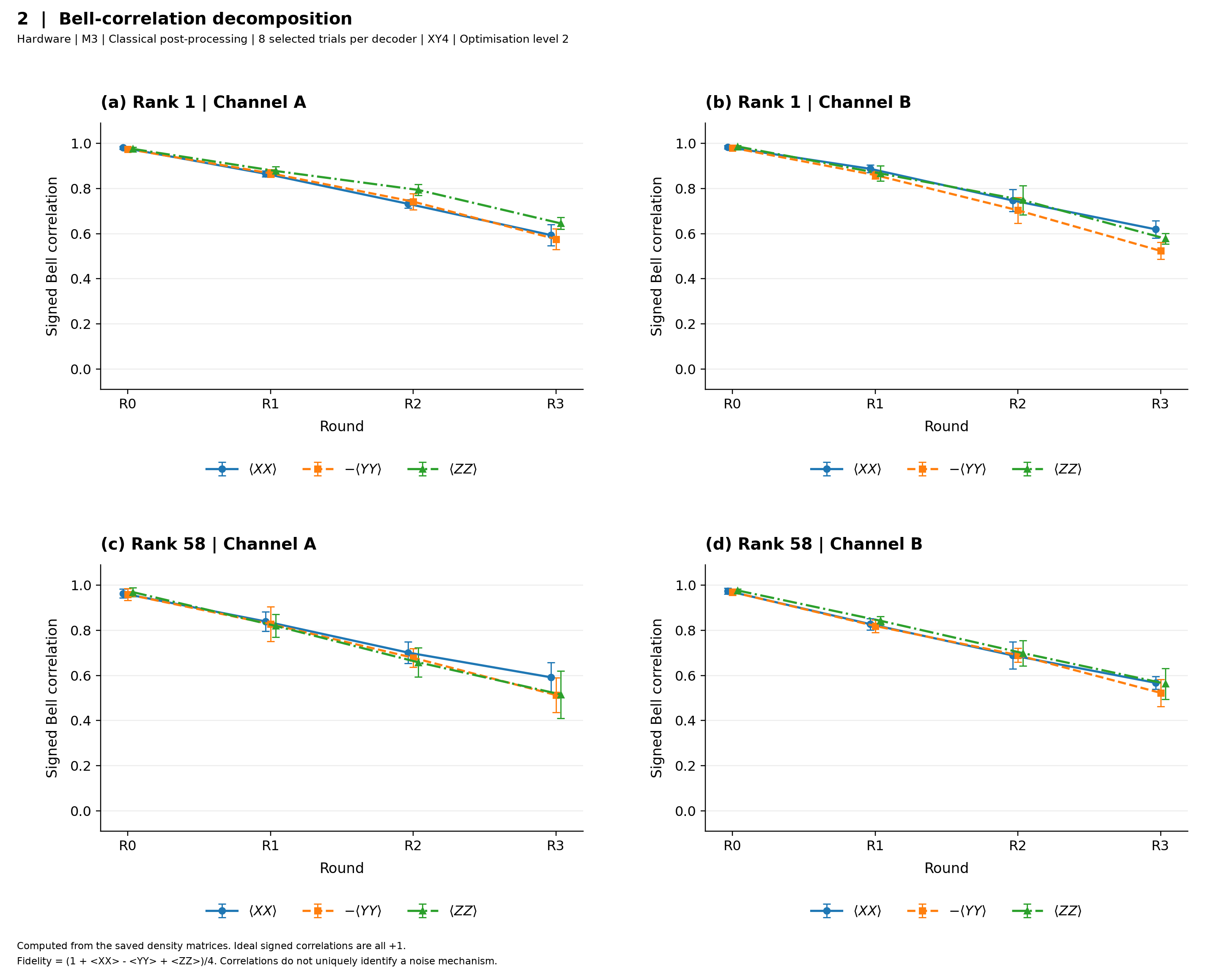}
\par\nointerlineskip
\usebox{\MthreeCorrelationCaptionBox}
\par
\end{figure}

% Finish every M3 float before the raw subsection can begin.
\FloatBarrier
\subsection*{Raw results (unmitigated)}

% Preserve the original effective image sizes, including the 0.95 scale.
\begin{figure}[!htbp]
\centering
\resizebox{0.95\textwidth}{!}{%
\begin{minipage}{\textwidth}
\centering

\begin{minipage}[t]{0.48\linewidth}
 \vspace{0pt}\centering
 \includegraphics[width=\linewidth,height=0.34\textheight,
 keepaspectratio]{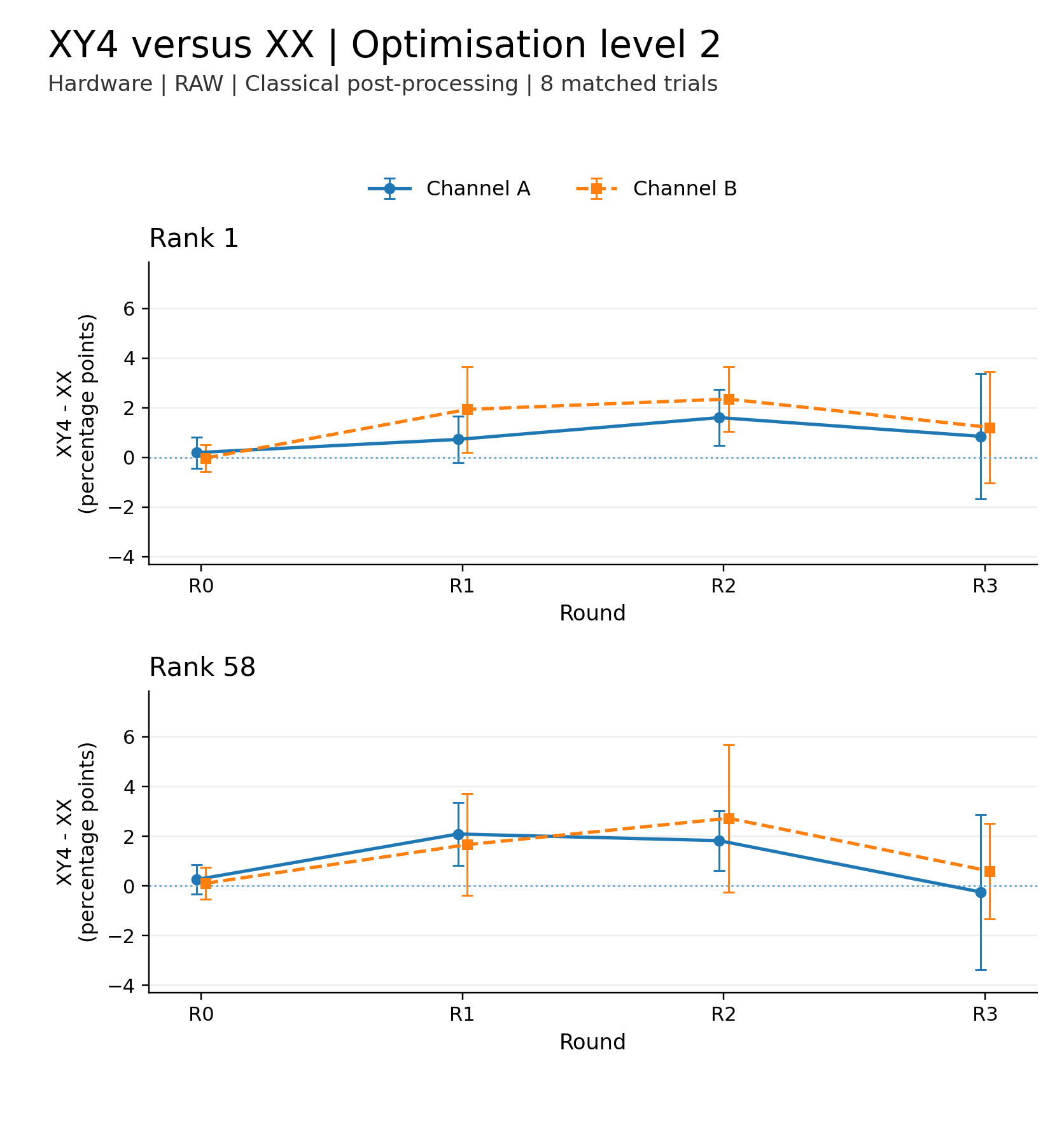}
 \par\smallskip\textbf{(a)} XY4 versus XX, level 2.
\end{minipage}
\hfill
\begin{minipage}[t]{0.48\linewidth}
 \vspace{0pt}\centering
 \includegraphics[width=\linewidth,height=0.34\textheight,
 keepaspectratio]{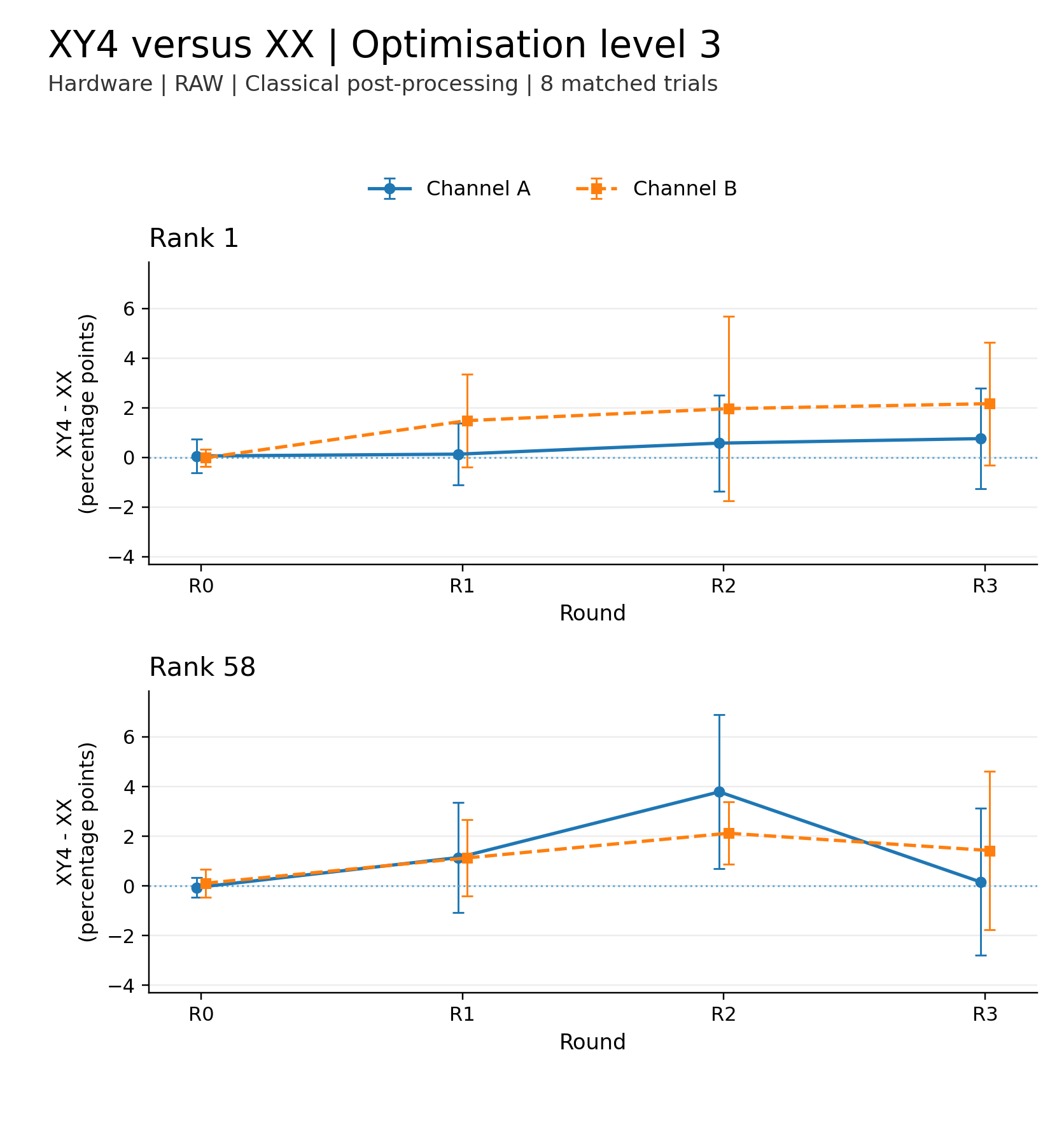}
 \par\smallskip\textbf{(b)} XY4 versus XX, level 3.
\end{minipage}

\end{minipage}%
}
\phantomcaption
\label{supp-fig:raw-dd-comparisons}
\end{figure}

\begin{figure}[!htbp]
\ContinuedFloat
\centering
\resizebox{0.95\textwidth}{!}{%
\begin{minipage}{\textwidth}
\centering

\begin{minipage}[t]{0.48\linewidth}
 \vspace{0pt}\centering
 \includegraphics[width=\linewidth,height=0.34\textheight,
 keepaspectratio]{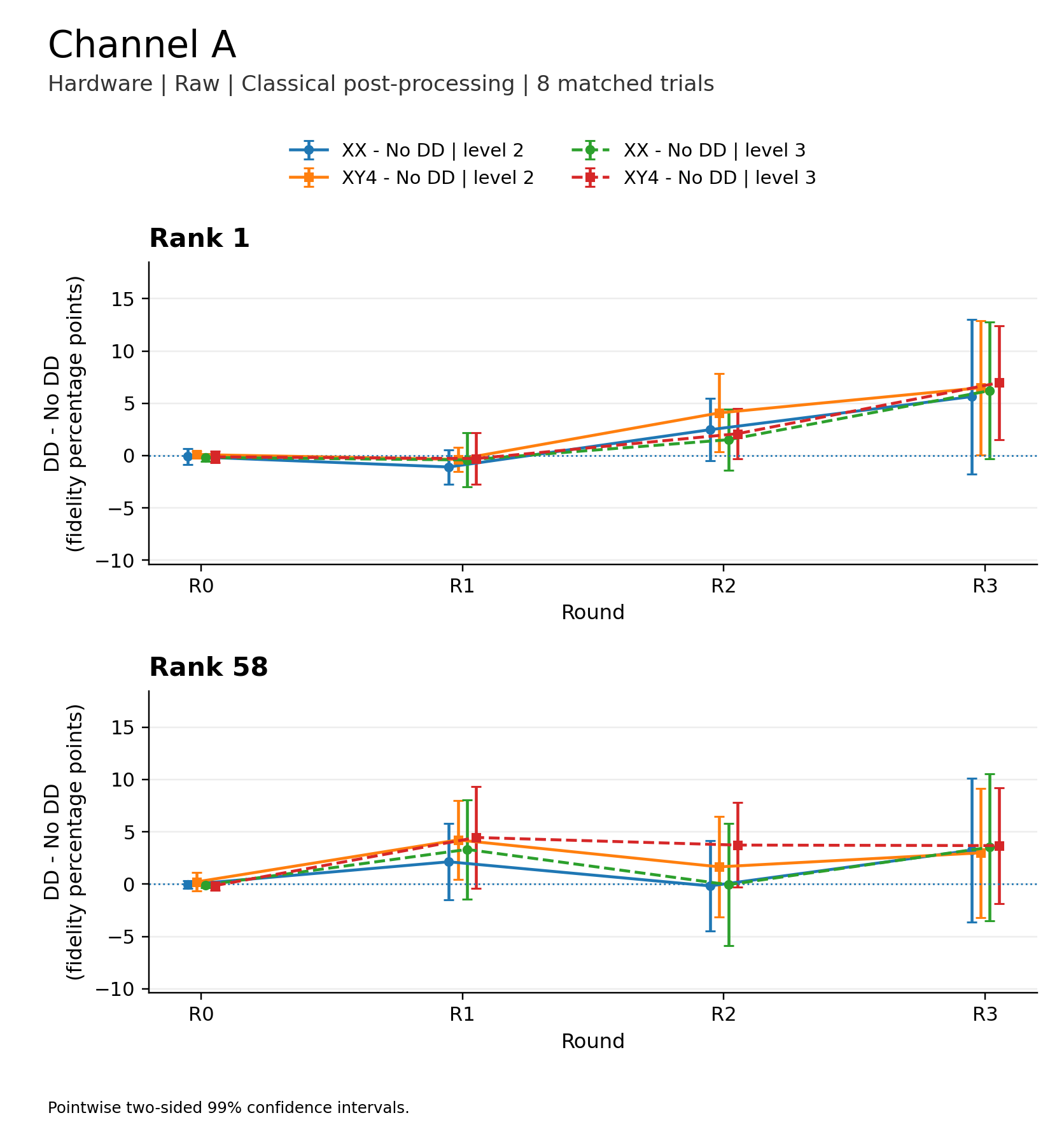}
 \par\smallskip\textbf{(c)} DD versus no DD, channel $A$.
\end{minipage}
\hfill
\begin{minipage}[t]{0.48\linewidth}
 \vspace{0pt}\centering
 \includegraphics[width=\linewidth,height=0.34\textheight,
 keepaspectratio]{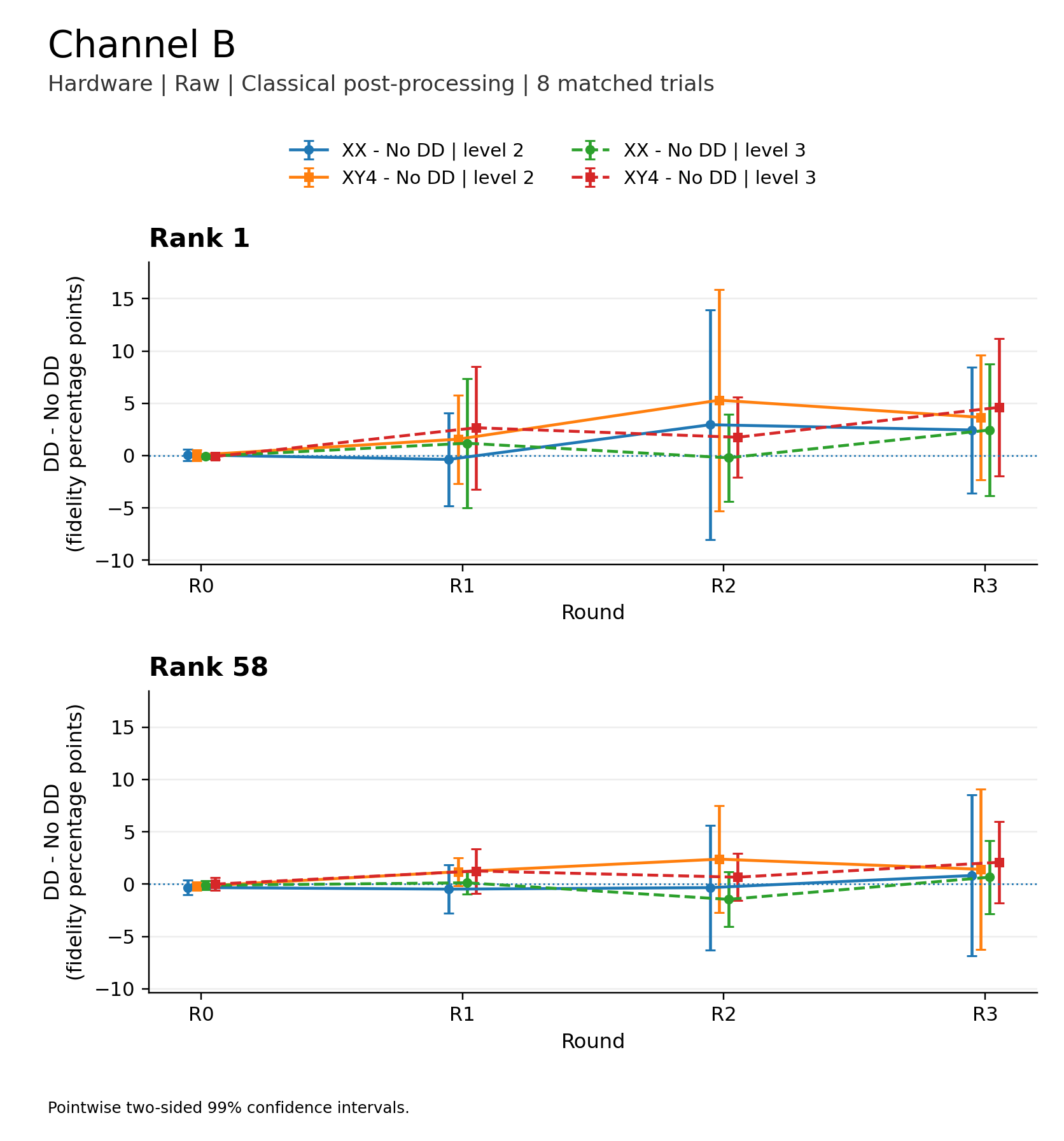}
 \par\smallskip\textbf{(d)} DD versus no DD, channel $B$.
\end{minipage}

\end{minipage}%
}
\captionsetup{font=small}
\caption{Raw hardware entanglement fidelity comparisons with
classical Pauli frame post-processing.
Panels (a) and (b) show XY4-minus-XX differences at
optimisation levels 2 and 3, respectively.
Panels (c) and (d) compare XX and XY4 with matched no-DD
baselines for channels $A$ and $B$, respectively, with
rank~1 above rank~58 and both optimisation levels in each
subplot. Differences are expressed in percentage points.
Positive values favour XY4 in (a) and (b) or the indicated DD
sequence in (c) and (d). Each point is the mean of eight
within-trial differences; error bars denote pointwise
two-sided $99\%$ confidence intervals.}
\end{figure}

\begin{figure}[p]
\centering
\setlength{\parskip}{0pt}

\savebox{\MthreeBaselineCaptionBox}{%
\begin{minipage}{\textwidth}
\captionsetup{font=small}
\caption{Baseline-adjusted raw hardware entanglement fidelity
comparisons at optimisation level 2.
The left panels compare rank 1 minus rank 58 retention for
channels $A$ and $B$. The right panels compare channel $B$
minus channel $A$ retention for ranks 1 and 58.
The corresponding within-trial $R0$ difference is subtracted
before forming the comparisons. Positive values favour rank~1
on the left and channel $B$ on the right. Error bars are
pointwise two-sided $99\%$ confidence intervals, with eight
trials per decoder.}
\label{supp-fig:raw-channel}
\end{minipage}%
}

\savebox{\MthreeCorrelationCaptionBox}{%
\begin{minipage}{\textwidth}
\caption{Raw signed Bell correlations with XY4 and optimisation
level 2. The components $XX$, $-YY$ and $ZZ$ determine the
entanglement fidelity through
Eq.~\eqref{supp-eq:bell-correlation-fidelity}.
Points are means over eight trials, and error bars are
pointwise two-sided $99\%$ confidence intervals.}
\label{supp-fig:raw-correlations}
\end{minipage}%
}

\setlength{\MthreePairImageHeight}{\dimexpr
 \textheight
 -\ht\MthreeBaselineCaptionBox
 -\dp\MthreeBaselineCaptionBox
 -\ht\MthreeCorrelationCaptionBox
 -\dp\MthreeCorrelationCaptionBox
 -12pt\relax}
\setlength{\MthreePairImageHeight}{0.5\MthreePairImageHeight}

\includegraphics[
 width=\textwidth,
 height=\MthreePairImageHeight,
 keepaspectratio
]{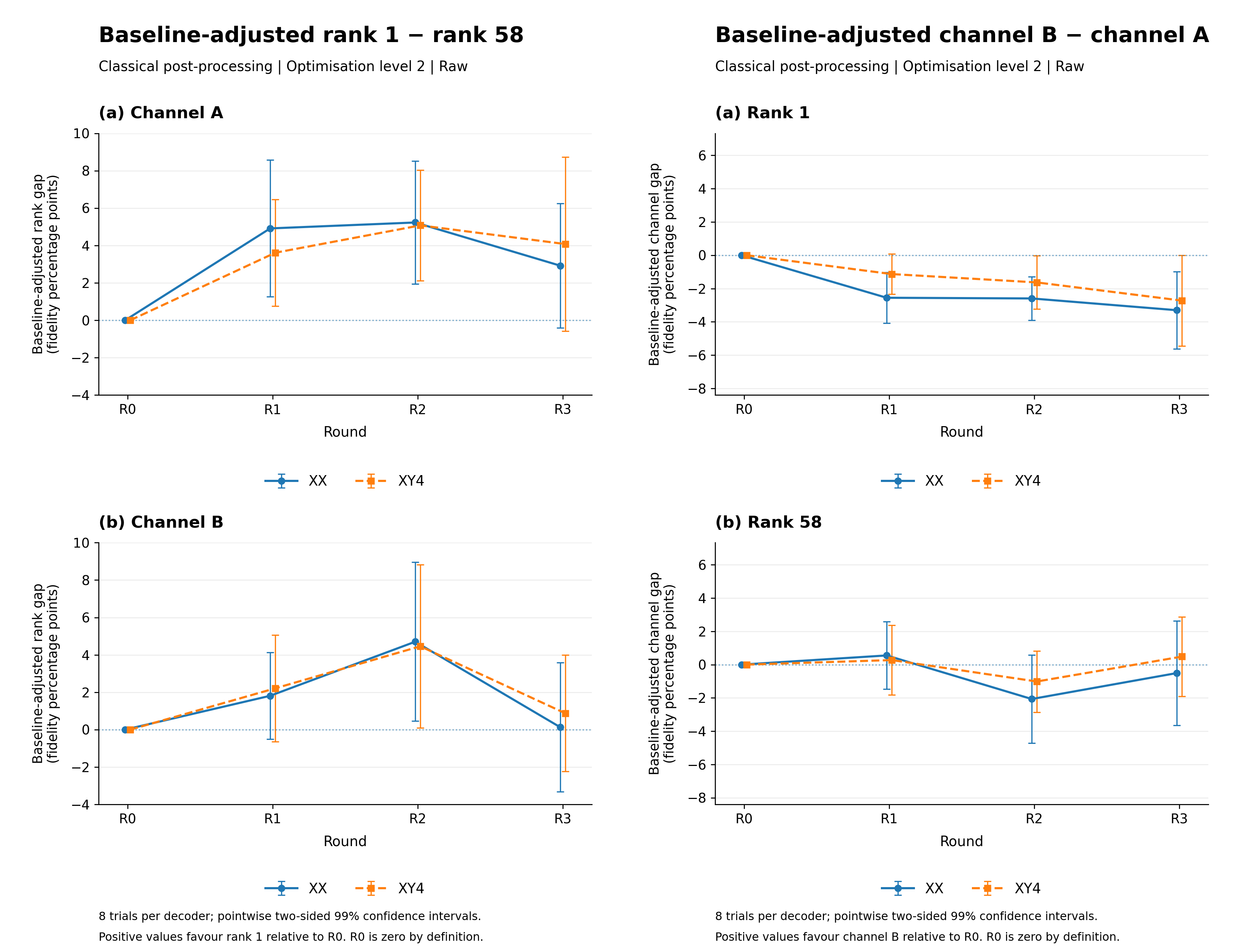}
\par\nointerlineskip
\usebox{\MthreeBaselineCaptionBox}

\par\vspace{8pt}\nointerlineskip

\includegraphics[
 width=\textwidth,
 height=\MthreePairImageHeight,
 keepaspectratio
]{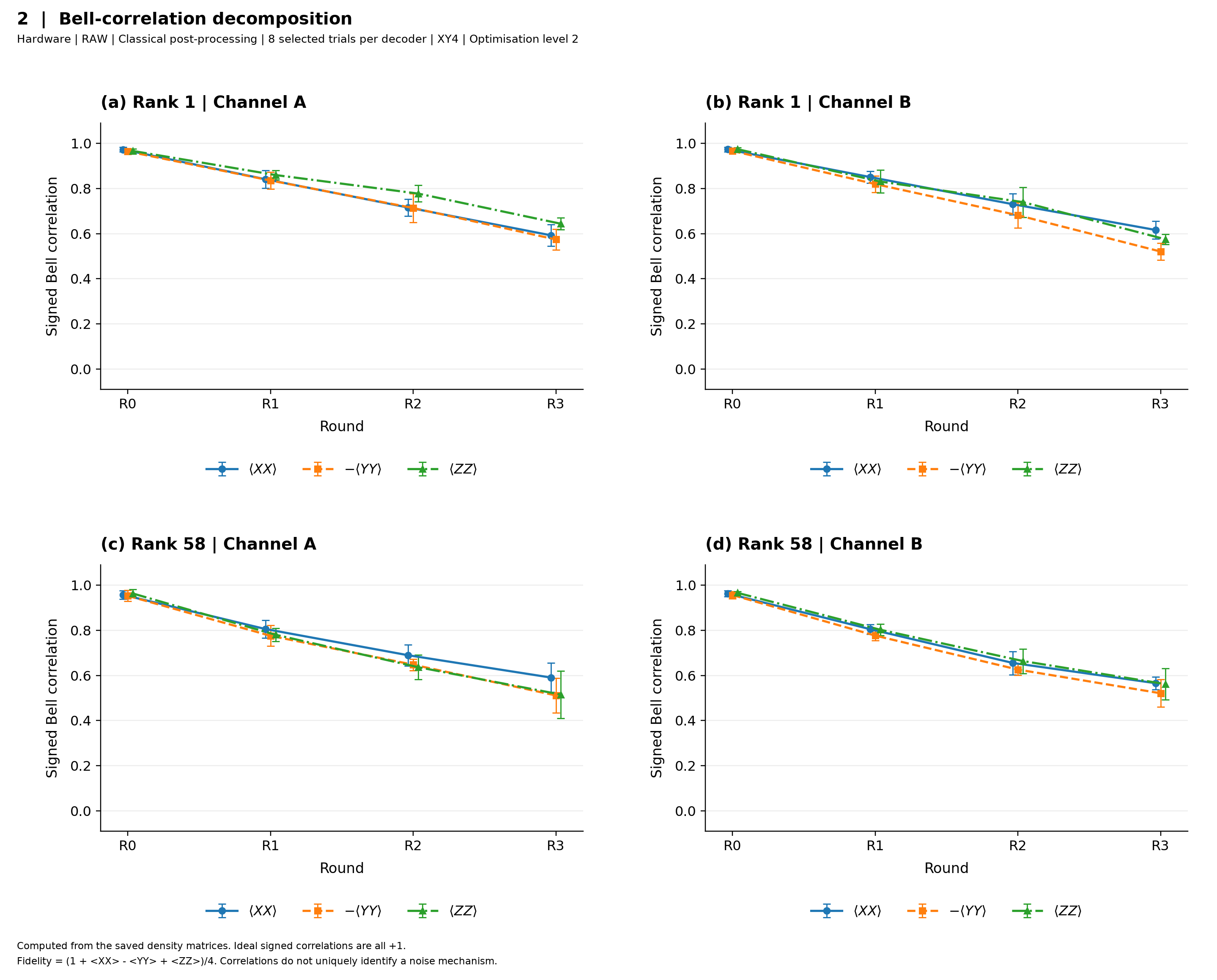}
\par\nointerlineskip
\usebox{\MthreeCorrelationCaptionBox}
\par
\end{figure}

% Do not let raw figures drift into the following manuscript section.
\FloatBarrier
\endgroup

\clearpage

\subsection*{Overall}

% Requires booktabs, already used in the manuscript.
% Figure numbers refer to the supplied manuscript PDF (11).
\begin{table}[!htbp]
\centering
\small
\caption{Raw and M3-mitigated results for the comparisons in
Sec.~VI. Ranges give the minimum and maximum plotted trial means
over $R1$-$R3$, not confidence intervals or pooled estimates.
The maximum shift is the largest absolute M3-minus-raw change
between corresponding points, calculated before rounding.
Each underlying mean uses eight trials per decoder.
All values are given to three significant figures.}
\label{supp-tab:raw-m3-overview}
\setlength{\tabcolsep}{4pt}
\begin{tabular}{@{}llccc@{}}
\toprule
Figures (M3/raw) & Comparison & Raw range & M3 range
& Max. $|\Delta|$ \\
\midrule
\multicolumn{5}{@{}l}{Entanglement-fidelity differences
(percentage points)}\\
S10(a)/S13(a) & XY4 $-$ XX, level 2
& $[-0.244,\,2.72]$ & $[-0.241,\,2.85]$ & $0.124$ \\
S10(b)/S13(b) & XY4 $-$ XX, level 3
& $[0.131,\,3.79]$ & $[0.141,\,3.81]$ & $0.0500$ \\
S10(c)/S13(c) & DD $-$ no DD, channel $A$
& $[-1.09,\,6.97]$ & $[-1.12,\,6.97]$ & $0.210$ \\
S10(d)/S13(d) & DD $-$ no DD, channel $B$
& $[-1.46,\,5.30]$ & $[-1.48,\,5.41]$ & $0.211$ \\
S11/S14, left & Rank 1 $-$ rank 58, baseline-adjusted
& $[0.128,\,5.23]$ & $[0.237,\,5.01]$ & $2.15$ \\
S11/S14, right & Channel $B-A$, baseline-adjusted
& $[-3.30,\,0.550]$ & $[-3.35,\,0.274]$ & $1.29$ \\
\midrule
\multicolumn{5}{@{}l}{Signed Bell correlations
(dimensionless; XY4, level 2)}\\
S12/S15 & $\langle XX\rangle$
& $[0.566,\,0.851]$ & $[0.567,\,0.888]$ & $0.0367$ \\
S12/S15 & $-\langle YY\rangle$
& $[0.512,\,0.836]$ & $[0.513,\,0.868]$ & $0.0641$ \\
S12/S15 & $\langle ZZ\rangle$
& $[0.515,\,0.860]$ & $[0.516,\,0.878]$ & $0.0401$ \\
\bottomrule
\end{tabular}
\par\smallskip
\begin{minipage}{\linewidth}
\footnotesize
The DD minus no DD rows cover XX and XY4, both decoder ranks,
and optimisation levels 2 and 3. The baseline-adjusted rows cover
XX and XY4 at level 2; the rank comparison includes both channels,
and the channel comparison includes both ranks. Correlation ranges
include both ranks and channels. $R0$ is excluded throughout.
Confidence intervals remain those of the individual plotted points.
\end{minipage}
\end{table}

All in all, these analyses test whether the DD and decoder preferences
highlighted in the main text persist without readout mitigation
and whether the decoder advantage remains after accounting for
different measured $R0$ baselines. Both raw and M3-mitigated
results favour XY4 over XX, with positive XY4 minus XX means
in 23 of the 24 comparisons across ranks, channels, optimisation
levels and rounds $R1$-$R3$. They also favour rank~1 in all
12 baseline-adjusted mean comparisons for XX and XY4 at level~2.
These results support the main-text focus on XY4 and rank~1.
Both preferences are present in the raw means, and rank~1 retains
its mean advantage after differences in the measured $R0$
baselines are accounted for. 

\section{Separate hardware method with real-time sequential feed-forward}
\label{sec:supp-feedforward}

In the main text, Pauli byproducts are handled through classical
Pauli frame tracking and postprocessing. In this section, we investigate real
time feedforward in a separate hardware campaign. Bell measurement
outcomes determine the Pauli corrections applied on the processor
in each round, before the receiver becomes the input to the next
round. The additional conditional operations and feedback delays
can introduce gate errors and decoherence, lowering entanglement
fidelity. We compare the two decoders, dynamical decoupling sequences
and optimisation levels using the entanglement fidelity between each
reference and its receiver, evaluated from raw data and with M3
readout mitigation. 

\begin{figure}[!http]
\centering
\includegraphics[width=\textwidth]{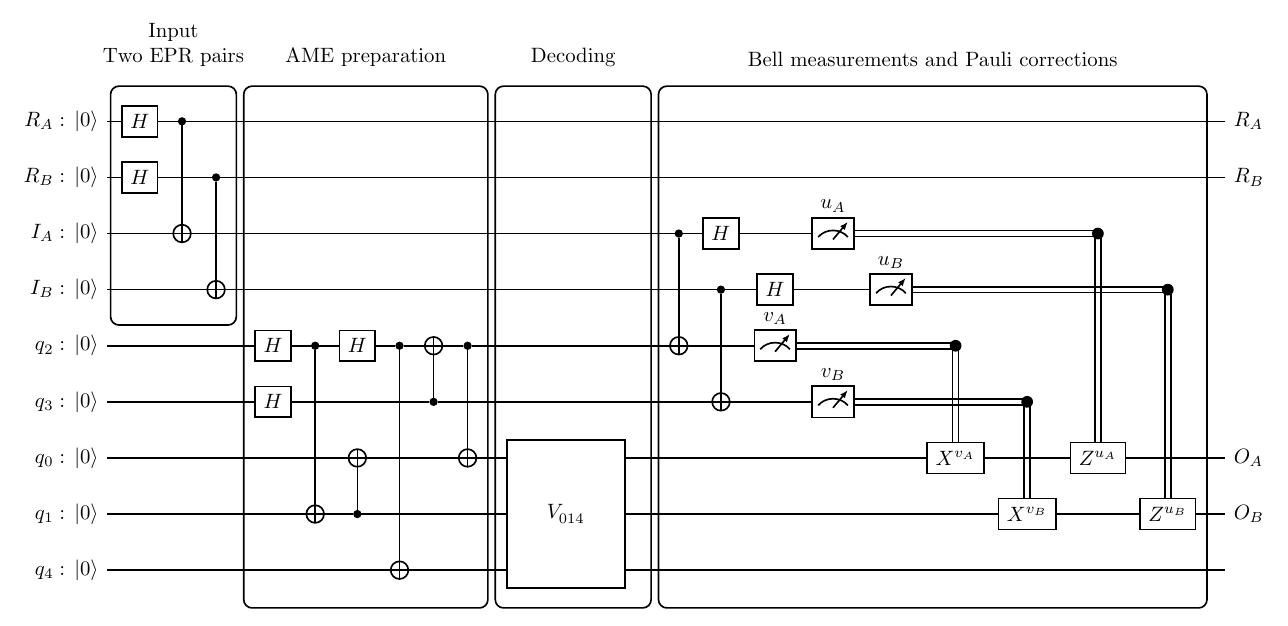}
\caption{Circuit diagram of one logical round of parallel teleportation, using the rank 1 decoder. At $R0$, the $H$-CNOT circuits prepare
$(R_A,I_A)$ and $(R_B,I_B)$ in $\ket{\Phi^+}$. The references
$R_A,R_B$ are retained while $V_{014}$ decodes the AME resource.
Bell outcomes $(u_c,v_c)$ determine the Pauli correction on $O_c$. The sequential
feed-forward implementation executes those corrections on-device;
the classical implementation instead defers them into the analysis
frame.  In the cascade,
$I_c^{(R+1)}\equiv O_c^{(R)}$. The same reference qubits are retained through the cascade.}
\label{supp-fig:single-round-circuit}
\end{figure}

\begin{figure*}[t]
 \centering
 \begin{minipage}[t]{0.49\textwidth}
  \vspace{0pt}\centering
  \makebox[\linewidth][c]{%
   \scalebox{1}[1]{%
    \includegraphics[width=\linewidth]
     {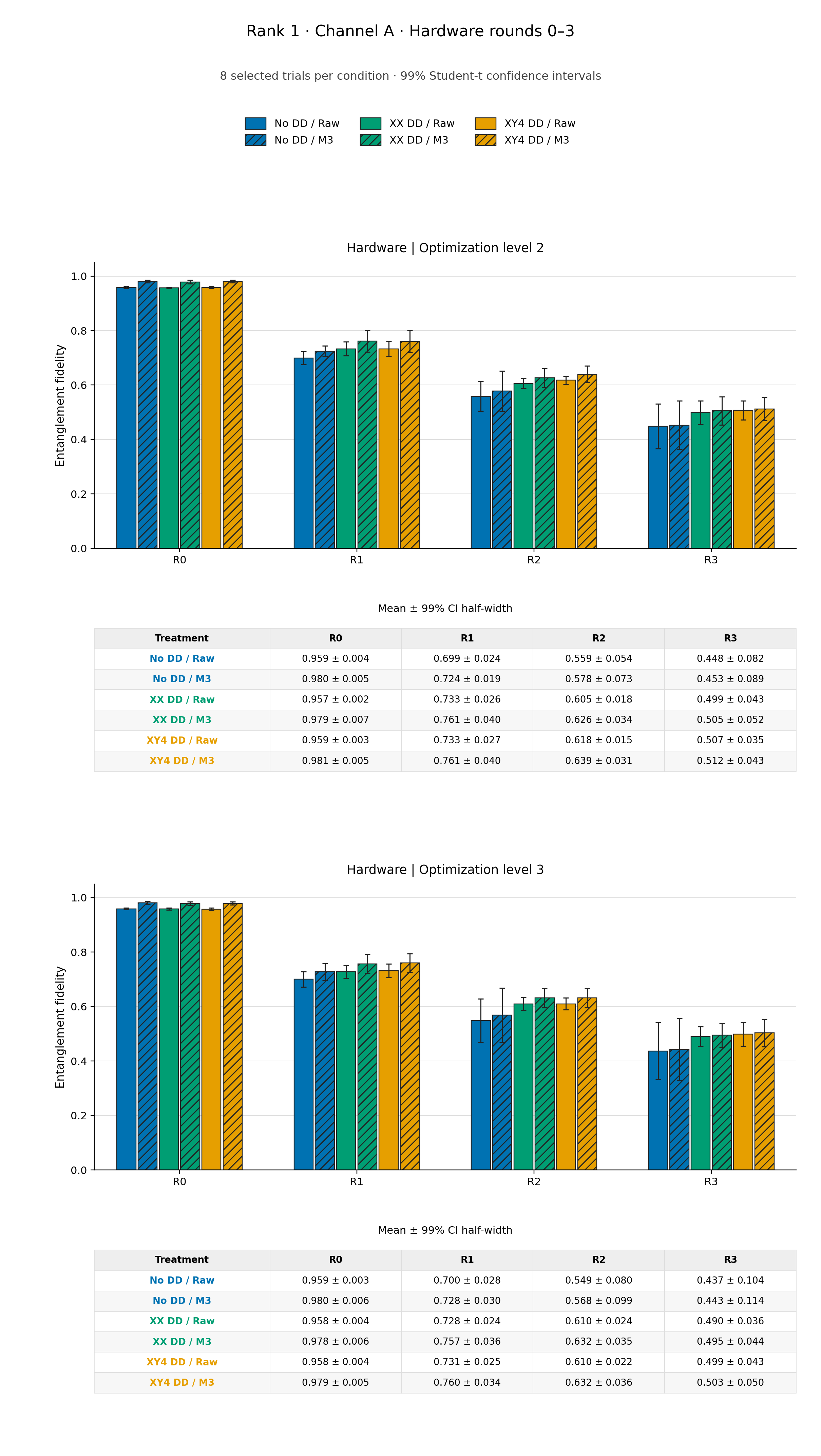}%
   }%
  }
  \par\smallskip\textbf{(a)} Channel \(A\).
 \end{minipage}
 \hfill
 \begin{minipage}[t]{0.49\textwidth}
  \vspace{0pt}\centering
  \makebox[\linewidth][c]{%
   \scalebox{1}[1]{%
    \includegraphics[width=\linewidth]
     {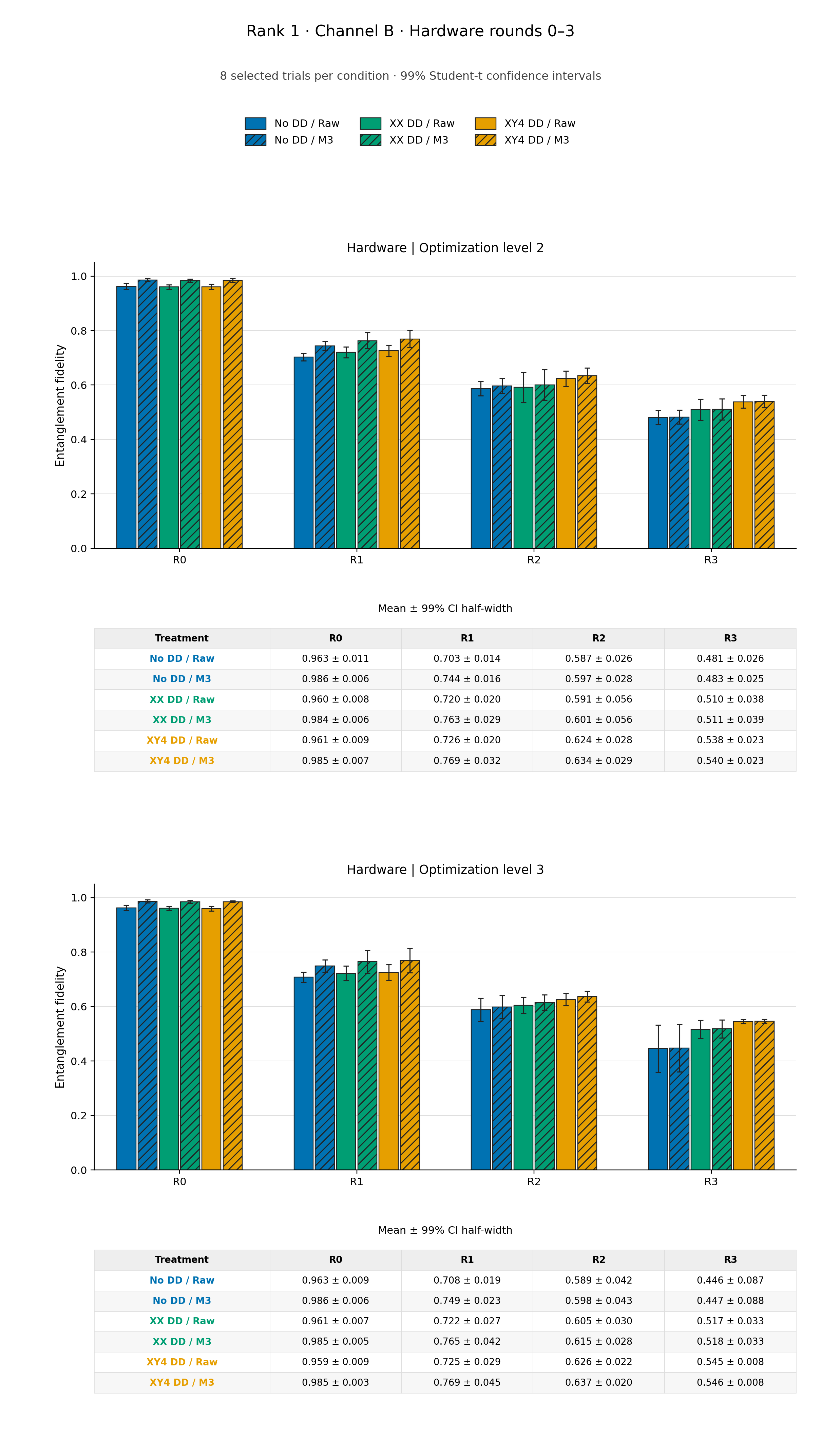}%
   }%
  }
  \par\smallskip\textbf{(b)} Channel \(B\).
 \end{minipage}
 \caption{Hardware feed-forward results for rank~1 through \(R3\).
 Pauli corrections are applied conditionally on the device, and each
 corrected receiver supplies the next-round input. Each panel contains
 optimisation levels \(2\) and \(3\), with no DD, XX and XY4
 distinguished by colour. Solid and hatched bars show raw and
 M3-mitigated fidelities. Values are means over eight selected trials,
 with pointwise two-sided \(99\%\) confidence intervals; the tables
 list means and interval half-widths.}
 \label{supp-fig:supp-cascade-feedforward-rank1}
\end{figure*}

\begin{figure*}[t]
 \centering
 \begin{minipage}[t]{0.49\textwidth}
  \vspace{0pt}\centering
  \makebox[\linewidth][c]{%
   \scalebox{1}[1]{%
    \includegraphics[width=\linewidth]
     {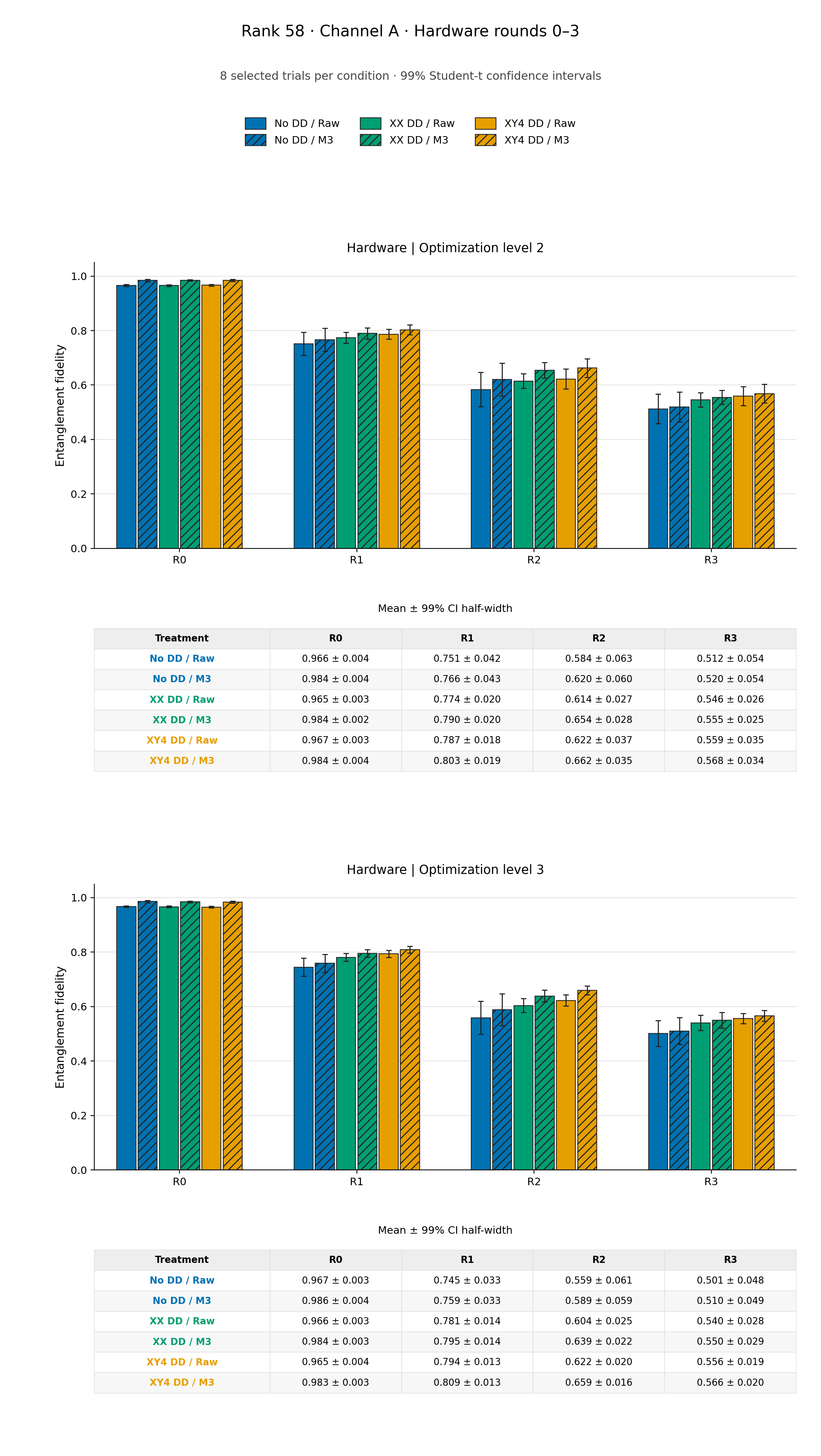}%
   }%
  }
  \par\smallskip\textbf{(a)} Channel \(A\).
 \end{minipage}
 \hfill
 \begin{minipage}[t]{0.49\textwidth}
  \vspace{0pt}\centering
  \makebox[\linewidth][c]{%
   \scalebox{1}[1]{%
    \includegraphics[width=\linewidth]
     {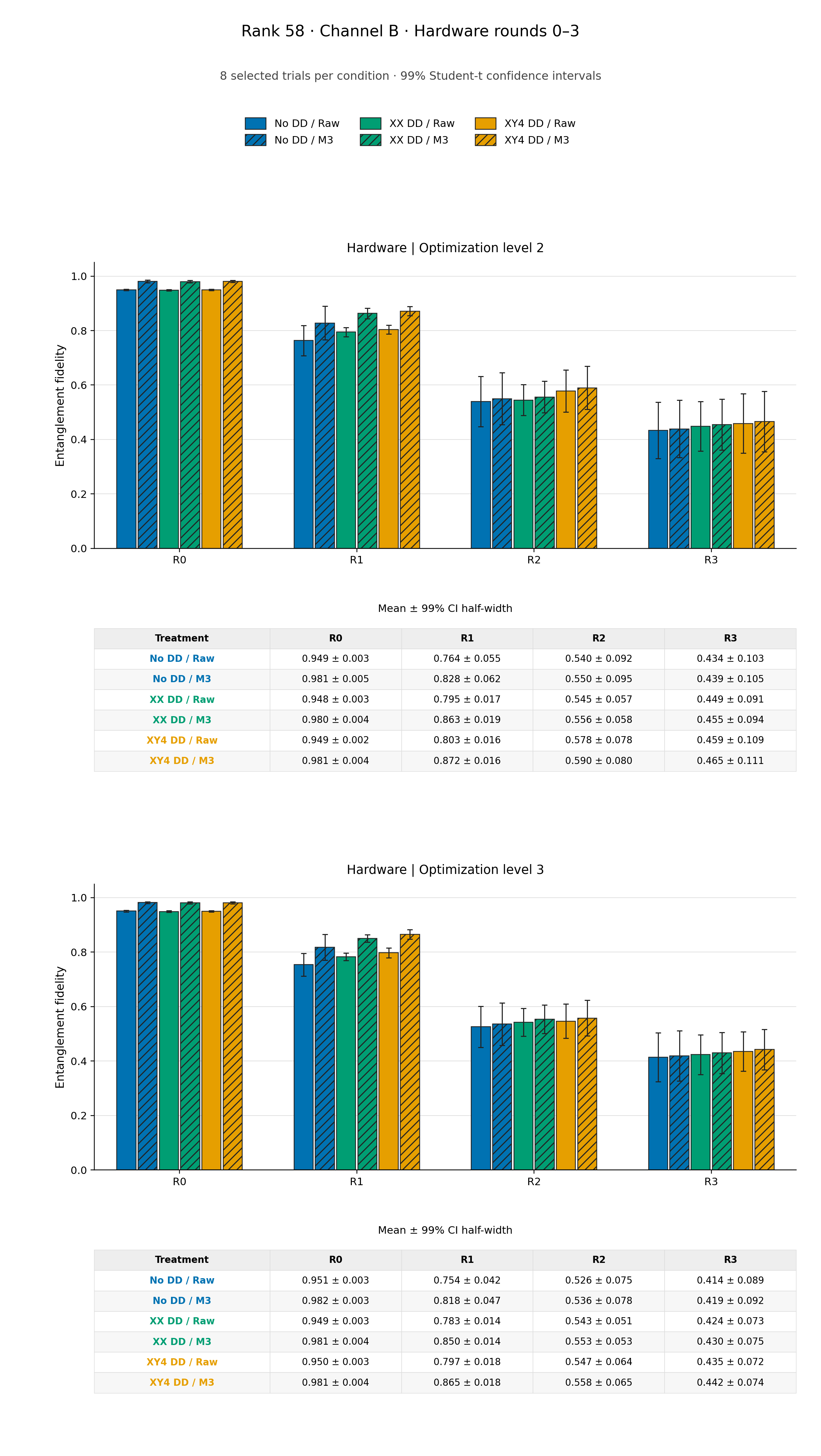}%
   }%
  }
  \par\smallskip\textbf{(b)} Channel \(B\).
 \end{minipage}
 \caption{Hardware feed-forward results for rank~58 through \(R3\).
 Pauli corrections are applied conditionally on the device, and each
 corrected receiver supplies the next-round input. Each panel contains
 optimisation levels \(2\) and \(3\), with no DD, XX and XY4
 distinguished by colour. Solid and hatched bars show raw and
 M3-mitigated fidelities. Values are means over eight selected trials,
 with pointwise two-sided \(99\%\) confidence intervals; the tables
 list means and interval half-widths.}
 \label{supp-fig:supp-cascade-feedforward-rank58}
\end{figure*}

Figure~\ref{supp-fig:single-round-circuit} shows the full
one-round circuit with the rank-1 decoder, including initial
reference-input Bell-pair preparation, AME preparation,
decoding and outcome-dependent Pauli corrections. In the
feed-forward cascade, these corrections are applied on device
before each receiver becomes the next-round input, while the
same reference qubits are retained throughout.
Figures~\ref{supp-fig:supp-cascade-feedforward-rank1} and
\ref{supp-fig:supp-cascade-feedforward-rank58} report the hardware sequential
feed-forward method. In this implementation, each Bell measurement
is followed by physical conditional \(X\) and \(Z\) corrections
before the receiver becomes the input to the next round. The plots
compare no DD, XX and XY4 at optimisation levels \(2\) and \(3\),
with raw and M3-mitigated means over eight selected trials and
pointwise two-sided \(99\%\) confidence intervals.

The entanglement fidelities decrease with the number of teleportation rounds, but
the decoder ordering depends on the channel. With XY4 and optimisation
level \(3\), the rank~58 M3-mitigated channel \(A\) means are
\(0.809\), \(0.659\) and \(0.566\) at \(R1\), \(R2\)
and \(R3\), compared with \(0.760\), \(0.632\) and \(0.503\)
for rank~1. Channel \(B\), however, reverses its mean ordering:
rank~58 gives \(0.865\pm0.018\) at \(R1\), versus
\(0.769\pm0.045\) for rank~1, but falls to
\(0.558\pm0.065\) and \(0.442\pm0.074\) at \(R2\) and
\(R3\), whereas rank~1 gives \(0.637\pm0.020\) and
\(0.546\pm0.008\). The quoted uncertainties are the \(99\%\)
interval half-widths.

At \(R3\) under this setting, the lower confidence bounds remain
above the entanglement fidelity threshold \(1/2\) for rank~58 channel \(A\)
and rank~1 channel \(B\), including in the raw estimates. The
corresponding M3 values for rank~1 channel \(A\)
\((0.503\pm0.050)\) and rank~58 channel \(B\)
\((0.442\pm0.074)\) do not provide that certification. Failure of
this entanglement fidelity criterion does not by itself establish separability; in the sequential feed-forward implementation, decoherence during measurement and feedback delays, conditional gate errors, and
preparation and readout errors can lower the measured fidelity.
XX and XY4 generally improve the later-round means relative to no DD,
but do not remove the channel dependence. In summary, we found experimentally, as expected, that classical post-processing gives a higher entanglement fidelity than real-time Pauli frame correction. Nonetheless, there is a caveat in our conclusion. Since different acquisitions and implementations were performed, the difference in entanglement fidelity cannot be attributed solely to replacing real-time corrections with classical post-processing.

\clearpage
In summary, we found experimentally, as expected, that classical post-processing gives a higher entanglement fidelity than real-time Pauli frame correction. Nonetheless, there is a caveat in our conclusion. Since different acquisitions and implementations were performed, the difference in entanglement fidelity cannot be attributed solely to replacing real-time corrections with classical post-processing.

\bibliographystyle{unsrt}
\bibliography{Ref,additional_refs}